\documentclass[%
 reprint,superscriptaddress,
 amsmath,amssymb,
 aps,
]{revtex4-2}
\usepackage{algorithmic}
\usepackage{graphicx}
\usepackage{dcolumn}
\usepackage{bm}
\usepackage{amssymb}
\usepackage{tabularx}
\usepackage{mathtools,halloweenmath}
\usepackage{wick}
\usepackage{physics}
\usepackage{graphicx}
\usepackage{amsmath}
\usepackage[version=4]{mhchem}
\usepackage{siunitx}
\usepackage{longtable,tabularx}
\usepackage{braket}
\usepackage{wick}
\usepackage{float}
\usepackage{simplewick}
\usepackage{xcolor}
\usepackage{ulem}
\usepackage{comment}
\usepackage{hyperref} 

\usepackage[ruled,lined]{algorithm2e}

\usepackage{mathtools}
\newcommand{\red}[1]{{\color{red}{#1}}}
\newcommand{\rsout}[1]{\red{\sout{#1}}}
\newcommand{\blue}[1]{{\color{blue}{#1}}}

\begin{document}
\preprint{APS/123-QED}
\title[]{Adapting the HHL algorithm to  quantum many-body theory}
\author{Nishanth Baskaran}
\altaffiliation{Contributed equally to the work}
\affiliation{Centre for Quantum Engineering, Research and Education, TCG CREST, Sector V, Salt Lake, Kolkata 700091, India}
\author{Abhishek Singh Rawat}
\altaffiliation{Contributed equally to the work}
\affiliation{Centre for Quantum Engineering, Research and Education, TCG CREST, Sector V, Salt Lake, Kolkata 700091, India}
\author{Akshaya Jayashankar}
\altaffiliation{Contributed equally to the work}
\affiliation{Centre for Quantum Engineering, Research and Education, TCG CREST, Sector V, Salt Lake, Kolkata 700091, India}
\author{Dibyajyoti Chakravarti}
\affiliation{Centre for Quantum Engineering, Research and Education, TCG CREST, Sector V, Salt Lake, Kolkata 700091, India}
\author{K. Sugisaki}
\affiliation{Centre for Quantum Engineering, Research and Education, TCG CREST, Sector V, Salt Lake, Kolkata 700091, India}
\affiliation{Graduate School of Science and Technology, Keio University, 7-1 Shinkawasaki, Saiwai-ku, Kawasaki, Kanagawa 212-0032, Japan}
\affiliation{Quantum Computing Center, Keio University, 3-14-1 Hiyoshi, Kohoku-ku, Yokohama, Kanagawa 223-8522, Japan}
\author{Shibdas Roy}
\affiliation{Centre for Quantum Engineering, Research and Education, TCG CREST, Sector V, Salt Lake, Kolkata 700091, India}
\author{Sudhindu Bikash Mandal}
\affiliation{Centre for Quantum Engineering, Research and Education, TCG CREST, Sector V, Salt Lake, Kolkata 700091, India}
\author{D. Mukherjee}
\affiliation{Centre for Quantum Engineering, Research and Education, TCG CREST, Sector V, Salt Lake, Kolkata 700091, India}
\author{V. S. Prasannaa}
\email{srinivasaprasannaa@gmail.com}
\affiliation{Centre for Quantum Engineering, Research and Education, TCG CREST, Sector V, Salt Lake, Kolkata 700091, India}

\begin{abstract}
Rapid progress in developing near- and long-term quantum algorithms for quantum chemistry has provided us with an impetus to move beyond traditional approaches and explore new ways to apply quantum computing to electronic structure calculations. In this work, we identify the connection between quantum many-body theory and a quantum linear solver, and implement the Harrow-Hassidim-Lloyd (HHL) algorithm to make precise predictions of correlation energies for light molecular systems via the (non-unitary) linearized coupled cluster theory, where the term `light molecular systems' refer to those molecules whose constituent atoms have low atomic number. For the purposes of practical computations, we make suitable changes to the HHL framework. This entails two aspects--- (a) Adapt: prescribing a novel scaling approach that allows one to scale any arbitrary Hermitian matrix, $A$, that in turn dictates the controlled rotation angles without having to precompute the eigenvalues of $A$, and yet achieve a reasonably high precision in $\ket{x}$, and (b) Lite: we devise techniques that reduce the depth of the relevant quantum circuit. In this context, we introduce the following variants of HHL for different eras of quantum computing--- AdaptHHLite in its appropriate forms for noisy intermediate scale quantum (NISQ), late-NISQ, and the early fault-tolerant eras, as well as AdaptHHL for the fault-tolerant quantum computing era. We demonstrate the ability of the NISQ variant of AdaptHHLite to capture correlation energy precisely, while simultaneously being resource-lean, using simulation as well as the 11-qubit IonQ quantum hardware. 
\end{abstract}

\maketitle

\tableofcontents

\section{Introduction}\label{introduction}

Computations in the field of quantum chemistry on quantum computers typically involve evaluating energies by using different variants of either the quantum phase estimation (QPE) algorithm~\cite{Abrams1997SimulationComputer, Aspuru-Guzik2005Chemistry:Energies, Lanyon2010TowardsComputer, Du2010NMRPreparation, Wang2015QuantumRegister, Sugisaki2021BayesianGaps, Sugisaki2021QuantumEvolutions, DipanjaliIQPE_VQE}, which is suited for the fault-tolerant quantum computing (FTQC) era as it involves deep quantum circuits, or the quantum-classical hybrid variational quantum eigensolver (VQE) algorithm, which is noisy intermediate scale quantum (NISQ)-friendly and involves executing shallow circuits for each iteration~\cite{Peruzzo2014AProcessor, Kandala2017Hardware-efficientMagnets, AIQuantum2020Hartree-FockComputer, Grimsley2019AnComputer, Yordanov2021Qubit-excitation-basedEigensolver, RyabinkinQCC2018, BurtonDISCOVQE2022}. In particular, considering that we are currently in the era of NISQ computing, it is important to note that scalable electronic structure calculations using VQE-based approaches can be hampered by two issues: (1) rapid increase of the circuit repetitions in order to suppress the effect of shot noises on the total energy~\cite{Gonthiervqe2022}, and (2) difficulties associated with the variational optimizations of parameters. The latter problem arises because VQE minimizes a single objective function regardless of the number of variational parameters. By contrast, in the traditional quantum chemical calculations on classical computers, the total energy is usually obtained by solving a secular equation, and it works well even for over 100,000  variational parameters. It is, therefore, timely to design approaches that are naturally suited for the current NISQ era to the early fault-tolerant phase, and which could naturally blend into the fault-tolerant era. \\ 

In this work, we identify that the Harrow-Hassidim-Lloyd~(HHL) algorithm~\cite{Harrow2009QuantumEquations}, which is used to solve equations of the form $A\arrowvert x \rangle=\arrowvert b \rangle$, can be employed for quantum chemical calculations. In particular, we solve the linearized coupled cluster~(LCC) equations to calculate correlation energies of light molecular systems with significant precision. The coupled cluster method involves a non-unitary wave operator which assumes the form of an exponential function~\cite{shavitt_bartlett_2009, Bishop1991AnPhysics, Cizek1966OnMethods}. We linearize the coupled cluster ansatz, and appropriately transcribe the problem to a quantum computing framework. In this sense, we go beyond the otherwise traditional but restrictive approach of employing unitary operators for quantum chemistry on quantum computers. It is worth adding at this point that the coupled cluster approach is considered to be the gold standard of electronic structure calculations on traditional computers, due to its ability to make precise predictions of atomic and molecular properties in a wide array of applications ranging from spectroscopy to expanding our understanding of particle physics~\cite{Sahooparitynonconversation2006, SahooRelativisticEDM2018, Prasannaamonohalides2015, NatarajPRA2011, SanghamitraMRCC2010, JerabekPolarizability2018, DibyajyotiUGASSMRPT2021, BishopHeisenberg2019}. Our work may serve as a stepping stone towards the development of quantum algorithms that employ non-linear and non-unitary wave operators for precision quantum chemical studies. \\ 

\begin{figure*}[t]
\includegraphics[width=\textwidth]{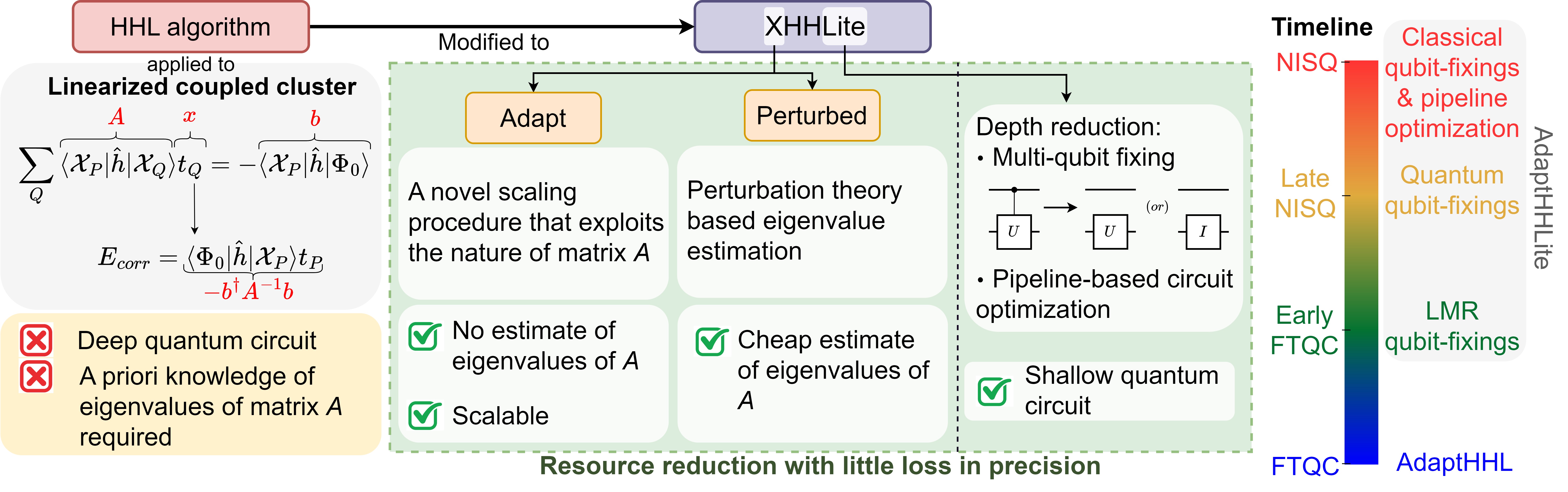}
\caption{\label{fig:FIG1} Figure illustrating the central concepts of the current work. The panel on the left conveys the connection between the HHL algorithm and the linearized coupled cluster equations, which facilitates the computation of correlation energy using the HHL algorithm. The panel also highlights the two limitations associated with the scheme: grappling with a deep quantum circuit, and the requirement of possessing a priori knowledge of the eigenvalues of the matrix, $A$, in order to obtain precise values from the computation. The former challenge is a limitation associated with the NISQ era and can be tackled by introducing classical overheads (cost to algorithm on the classical computing front). These overheads are acceptable for small-scale systems that the NISQ era is equipped to handle. On the other hand, a solution to the latter issue of a priori eigenvalue estimation is necessary for the FTQC era in view of intractable classical computations. However, it is also a welcome and valuable resource-saving addition to the preceding eras. The central panel delineates our solutions to the two issues (with the `X' module addressing eigenvalue computation overhead and the `Lite' component tackling the challenge of quantum circuit depth reduction) for the NISQ era. All of our simulation and hardware results for correlation energies presented in this work employ these variants. The panel to the right outlines the different forms of our recommended AdaptHHLite approach, tailored to meet the demands of each of the quantum computing eras. }
\end{figure*}

After establishing the link between the HHL algorithm and LCC equations, we redirect our focus to the pressing issue of resource reduction while retaining precision in results, in the practical implementation of the algorithm. In the NISQ era, we are limited by the number as well as the quality of qubits. These factors restrict our computations to only small systems, and necessitates quantum resource reductions. This is accompanied by classical computing overheads. Given the reasonably powerful classical computers available today, 
we can accommodate any accompanying cost incurred on the traditional computing side in achieving quantum resource reduction. However, as the NISQ era gradually recedes into the background to usher the FTQC era, one expects to have an abundance of quantum resources and therefore access to large-scale quantum computation. In such calculations that involve large system sizes, any classical computing cost associated intrinsically with a quantum algorithm would be exorbitant. During the in-between transition period, an algorithm needs to be adapted to gradually increase its reliance on quantum resources while diminishing its dependence on classical resources. All the while, it is important to ensure that the precision in results is reasonably good, for predictive sciences such as quantum chemistry, as resource reduction schemes may or may not impact precision. \\

Considering this complicated interplay between resource utilization (classical and quantum) and precision, we introduce revamped versions of the HHL algorithm, which we have termed as the `XHHLite' algorithms. \\ 

We begin by elaborating on the significance of the `X' part. The choice of scaling the input matrix, $A$, has far-reaching effects, as it impacts the QPE evolution time as well as controlled-rotation angles via a constant, `$c$', in the HHL algorithm~\cite{Harrow2009QuantumEquations}. This, in turn, could strongly influence our predictions of correlation energies. However, the current HHL implementations~\cite{kaisHHL2012, richardson, Zhang2022ImprovedQISKIT, Lee2019HybridExperience} rely upon carrying out a classical computation of the eigenvalues of $A$; a necessary yet expensive overhead in order to scale $A$ and choose the constant $c$ efficiently. This cost becomes intractable as system size increases, and negates the speed-up benefit that the HHL algorithm offers. The `X' module is designed to address precisely this issue, and to this end, we introduce the `AdaptHHL' and the `PerturbedHHL' algorithms in this work. Through these variants, we completely circumvent the under-emphasized issue of having to learn the eigenvalues of the input matrix using traditional computation a priori to achieve results with high accuracy, and therefore the associated cost. We discuss how the `AdaptHHL' algorithm can become a powerful framework that obviates compute-intensive eigenvalue estimation, starting from the NISQ and all the way into the FTQC eras. We add at this point that in the `PerturbedHHL' variant, one still estimates the eigenvalues of the matrix, but via perturbation theory in a resource-efficient way. \\ 

We now proceed to cover the `Lite' aspect. The crux of our approach involves studying the eigenvalue distribution from a QPE module, in order to reduce the number of gate applications in the HHL quantum circuit. This strategy is inspired by the improved HHL algorithm~\cite{Zhang2022ImprovedQISKIT}, but is sufficiently different, in that our methodology involves multi-qubit fixing on the QPE ancilla qubits, with the number of fixings controlled by a single tunable parameter, which we refer to as probability threshold. Furthermore, we augment this strategy with our own variant of pipeline-based circuit optimization, which is based on an earlier work~\cite{Kharkov2022ArlineCompilers}. Both of these approaches, which rely upon classical computing resources heavily, are strung together in our HHLite algorithm for the NISQ era. We show how the `Lite' approach can be customized in a resource-efficient way to blend into different timelines of quantum computing. To this end, we propose suitable adaptations to accommodate the needs of different eras. For the late NISQ era, we propose to use a quantum version of multi-qubit fixing~(thereby removing the exponential cost associated with the classical processing of data incurred in studying the outputs of the QPE module), while for the early FTQC era, we put forward a method based on the  Lloyd-Mohseni-Rebentrost~(LMR) algorithm~\cite{Lloyd2014} for achieving multi-qubit fixing. The LMR-based route offers a single-shot solution to identify dominant probabilities from probability distributions, which makes it a viable option for applications beyond quantum chemistry. We envisage that our LMR-based multi-qubit fixing approach could come in handy in real-world situations, where there is a need to frequently identify dominant probabilities from rapidly changing large probability distributions. The algorithm could also be adapted and/or augmented into other quantum algorithms, particularly in those involving quantum phase estimation subroutines such as Shor's~\cite{ThomasM, Lanyon} and quantum state discrimination approaches~\cite{AnthonyC,barnett2009quantum}. Figure \ref{fig:FIG1} showcases the core themes that underpin our work. \\ 

We evaluate our various NISQ era-suited implementations by carrying out calculations of correlation energies of several small molecules. We work with the following molecules: $H_2$, $H_3^+$, $LiH$, $BeH^+$ and $HF$, with each of them considered in five different geometries. Throughout, we keep in mind the applicability of a single reference theory and select appropriate geometries where strong correlation effects do not dominate. We also execute our implementations on the 11-qubit IonQ hardware, to compute the correlation energies of select light molecular systems from our list. It is also worth mentioning that to the best of our knowledge, a few works propose the use of quantum linear solvers to quantum chemistry~\cite{Cai2020QuantumProperties,Wiebe_2021_precond, chiew2023scalable}. We now briefly explain each of these works in literature. Cai \textit{et al}~\cite{Cai2020QuantumProperties} propose a theoretical framework of carrying out molecular response properties using the HHL algorithm. Another theoretical work carried out by Tong \textit{et al}~\cite{Wiebe_2021_precond} proposes employing preconditioned quantum linear solvers to obtain single-particle Green’s functions of quantum many-body systems. It is worth noting that the work utilizes block encoding in realizing the algorithm. The recent work by Kwek \textit{et al}~\cite{chiew2023scalable}(carried out independently around the same time as ours) prescribes the preparation of highly excited eigenstates of physical systems using a combination of HHL and the variational principle (which utilizes classical optimizer subroutines), analyze their applicability for near- and long-term quantum computers, and apply it to the \textit{LiH} system. We now present the salient features of our work: (1) Identifying the connection between LCC and HHL, (2) introducing in view of the key issue of resource reduction the PerturbedHHL and AdaptHHL variants as well as the Lite modules for each quantum computing era (multi-qubit fixing and pipeline-based quantum circuit optimization for the NISQ era, quantum multi-qubit fixing for the late NISQ era, and the LMR-based approach for the early fault-tolerant quantum computing era), and (3) numerical results for several molecules for the NISQ era AdaptHHLite approach (both simulation and on quantum hardware with error mitigation when necessary) in the spirit of quantum chemistry being a predictive science. \\ 

\section{Theory}\label{theory}

\subsection{\label{theory-secA}HHL algorithm and linearized coupled cluster equations: the connection}

In this subsection, we briefly introduce the linearized coupled cluster equations and the HHL algorithm in that order (we delve into some details of the former in the Appendix Section A1). We then proceed to find the connection between the two topics. \\ 

\subsubsection{\label{theory-secA1}Linearized coupled cluster equations}

The coupled cluster (CC) wave function for closed-shell systems is represented as 
\begin{eqnarray} \label{ccansatz}  |\Psi\rangle=\text{e}^{\hat{T}}\left|\Phi_{0}\right\rangle, 
\end{eqnarray}

where ${\hat{T}}={\hat{T}}_{1}+{\hat{T}}_{2} \ldots + {\hat{T}}_{N}$ for an $N$-electron molecule. ${\hat{T}}_{1}$ and ${\hat{T}}_{2}$ are the single and double excitation operators that produce one-hole one-particle and two-hole two-particle states respectively, while $|\Phi_{0}\rangle$ is the Hartree-Fock state. It is worth noting here that e$^{\hat{T}}$ is not a unitary operator. In the spin-free framework, which we adopt in this work, $\hat{T}_1 = \sum_{ia}t_i^a\{\hat{e}_i^a\}$ and $\hat{T}_2 = \sum_{ijab}t_{ij}^{ab}\{\hat{e}_{ij}^{ab}\}$, where the set \textit{i,j,k,$\cdots$} denotes occupied `hole' orbitals and \textit{a,b,c,$\cdots$} signifies the unoccupied `particle' orbitals. $t_i^a$ and $t_{ij}^{ab}$ are the cluster amplitudes (the unknowns) associated with $\hat{T}_1$ and $\hat{T}_2$ respectively, and $\hat{e}$ refers to a spin-free operator that induces the orbital substitution out of the Hartree-Fock reference state. \{\} represents the normal ordering of the elementary excitation operators with respect to the closed shell Hartree-Fock function (which is chosen as the vacuum). \\ 

\begin{figure}[t]
\includegraphics[width = \columnwidth]{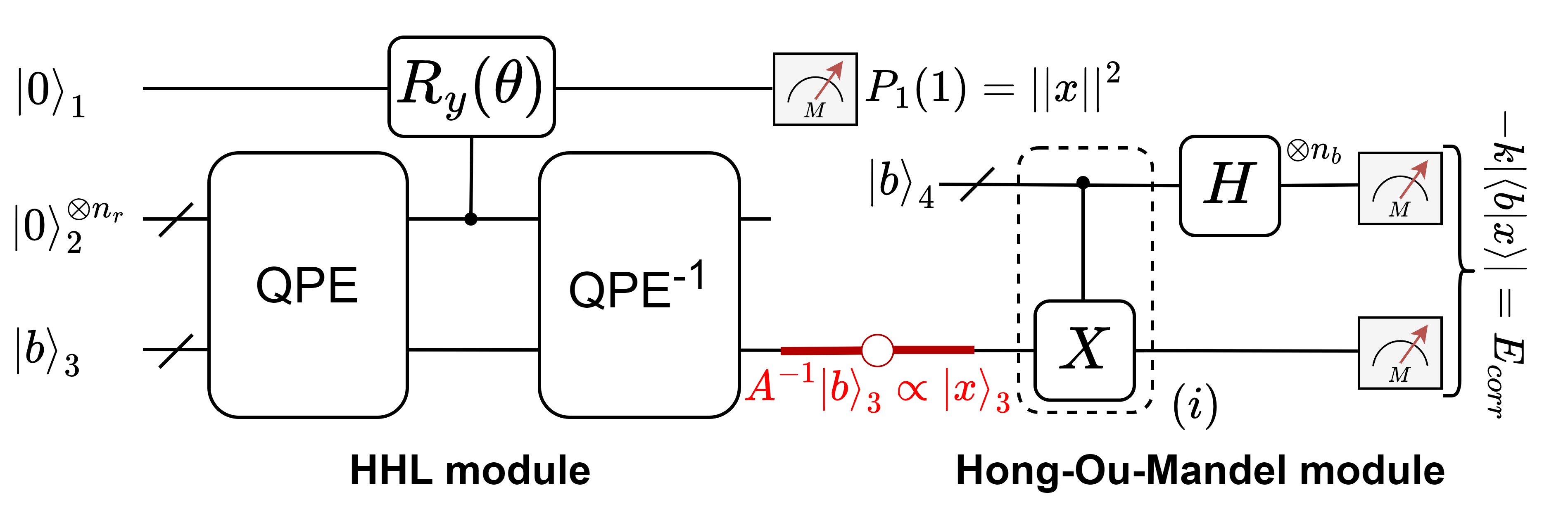}
\caption{\label{fig:FIG2} Quantum circuit schematic of the HHL algorithm, followed by the Hong-Ou-Mandel module, using which one computes the overlap, $|\braket{b|x}|$, and hence the correlation energy, $E_{\rm corr}$, of a quantum many-body system in the LCC framework. In the Hong-Ou-Mandel module, the compact notation of the controlled-X gate with an $(i)$ indicates that the gate is controlled on the $i^{th}$ qubit of the fourth register (marked as $\ket{b}_4$), with the target being the $i^{th}$ qubit of the state register. The details of the notation employed in the figure can be found in Section~\ref{theory-secA3}. }
\end{figure}

In the present work, we demonstrate the potential of a quantum linear solver to solve quantum many-body theoretic equations by considering the linearized CC (LCC) approach, where only the terms linear in $\hat{T}$ are considered from Eq. (\ref{ccansatz}). The form of the LCC amplitude equations can be shown to be 

\begin{equation} \label{LCCtampeqn}
    \bra{\chi_P}\hat{h}\ket{\Phi_0} + \sum_Q \bra{\chi_P} \hat{h} \{\hat{e}_Q\} \ket{\Phi_0}t_Q=0,\ \  \forall P. 
\end{equation}

We refer the reader to Section A1 of our Appendix for details. In the above equation, $t_Q$ refers to the $Q^{th}$ cluster amplitude, while $\hat{h}$ refers to the Hamiltonian operator. The $\ket{\chi}$s denote the excited functions, constructed as $\ket{\chi_P} = \{\hat{e}_P\}\ket{\Phi_0}$. The above equation can be rearranged as \\

\begin{equation} \label{LCCmatrixbraket}
    \boxed{\sum_Q\bra{\chi_P} \hat{h}\ket{\chi_Q}t_Q=-\bra{\chi_P}\hat{h}\ket{\Phi_0} \ \ \forall\ P. }
\end{equation}
\\ 
In this work, we use the HHL algorithm for the inversion of the matrix on the left-hand side of the above equation, to be denoted as \textit{A}. The constant vector on the right-hand side is identified with \textbf{b}, and the vector of unknown cluster amplitudes with \textbf{x}. The right hand side of Eq. \eqref{LCCmatrixbraket} is a vector because the bra index varies over the various possible excited functions while the ket index is fixed to be the Hartree-Fock determinant, and the same holds true for the cluster amplitudes. Thus, we have a matrix-vector equation of the form 

\begin{equation}\label{matrixLCCeqn}
\textit{A}\textbf{x}=-\textbf{b}. 
\end{equation}

It is pertinent to mention at this stage that treating the matrix \textit{A} in the HHL circuit would be simpler if it is Hermitian. This would require that the excited functions forming the basis of this matrix be orthogonal to each other. In other words, the spin-free excitation operators need to have zero overlap within themselves. This condition holds true for the single excitation operators but not for double and higher excitations. For example, the spin-free operators $\hat{e}_{ij}^{ab}$ and $\hat{e}_{ij}^{ba}$ ($i\neq j$ and $a\neq b$) generate excited determinants which are indistinguishable in terms of spatial orbital occupancies and hence will have a non-zero overlap among them. Thus, to have a Hermitian matrix, \textit{A}, one needs to suitably generate a new set of orthogonal excitation operators, which would be linear combinations of the non-orthogonal $\hat{e}_P$'s. We note that the matrices we encounter in the context of closed-shell non-relativistic coupled cluster theory are symmetric positive definite. \\

\subsubsection{\label{theory-secA2}Correlation energy in the LCC framework}

The LCC correlation energy can be expressed as (see Section A1 of the Appendix for details) 

\begin{eqnarray}
    \bra{\Phi_0} \hat{h} \ket{\chi_P}t_P&=&E_{\rm corr} . \label{LCCecorreqn}
\end{eqnarray}
Identifying the term on the left-hand side of the above equation as an inner product between the vector \textbf{b}$^\dagger$ and the cluster amplitudes, i.e., the vector ($-\textit{A}^{-1}\textbf{b}$), one gets the final form for obtaining the LCC correlation energy as 

\begin{equation}
    E_{\rm corr}=-\textbf{b}^\dagger \textit{A}^{-1}\textbf{b}. 
    \label{classical-correlation}
\end{equation}

\begin{figure*}
 \begin{tabular}{c}
\includegraphics[width=\textwidth]{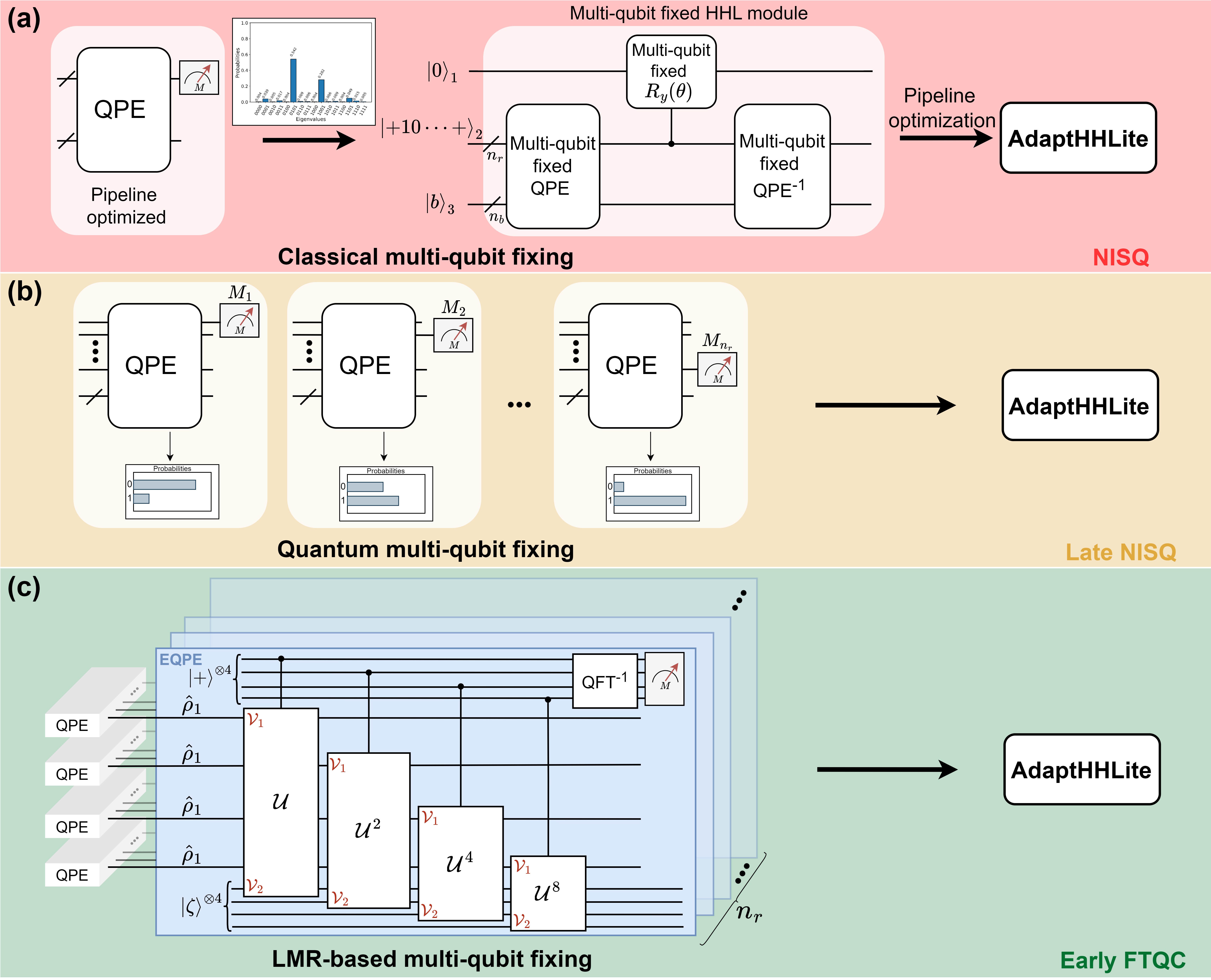}\\
\end{tabular}
\caption{\label{fig:FIG3} Subfigure (a) illustrates the role of the pipeline optimized QPE algorithm, whose output eigenvalue distribution is utilized for depth reduction in each stage of the AdaptHHL circuit via multi-qubit fixing. The resulting multi-qubit fixed HHL circuit is then pipeline optimized to obtain the AdaptHHLite circuit suited for the NISQ era. Subfigure (b) illustrates the workflow of the late-NISQ era AdaptHHLite variant, which utilizes multiple QPE modules to achieve a quantum version of multi-qubit fixing. On the other hand, Subfigure (c) depicts our LMR-based approach to multi-qubit fixing for a simple example of a 4-qubit EQPE ($n_e=4$). The approach, while requiring additional gates, involves substantially fewer measurements as system size increases. $\mathcal{V}_1$ and $\mathcal{V}_2$ refer to the two qubits between which the operation, $\hat{\mathcal{U}}$, occurs. The choice of notation used in this figure can be found in Section \ref{theory-secB1}(e).} 
\end{figure*}

\subsubsection{\label{theory-secA3}The HHL algorithm to calculate the LCC correlation energy} 

The HHL algorithm~\cite{Harrow2009QuantumEquations} is the most widely used quantum linear solver for solving linear systems of equations ($\textit{A}\textbf{x}=\textbf{b}$), and it is known to provide exponential speedup over the best known classical algorithm under certain conditions. The solution to these equations is encoded as amplitudes of the quantum state $\ket{x}$. This algorithm allows one to extract a feature of the solution vector by obtaining the expectation value of a linear operator $M$ on $|x\rangle$, as $\bra{x}M\ket{x}$. It also presumes that there exists a unitary that is capable of preparing vector \textbf{b}, an $n_b$-qubit quantum state, $\arrowvert b \rangle$. We proceed to give a brief overview of the HHL algorithm followed by our module to directly extract correlation energy. Figure~\ref{fig:FIG2} accompanies the description. 

\begin{enumerate}
\item Initialize three quantum registers to $\ket{0}_1\ket{0}_2\ket{b}_3$. The first is the one qubit ancilla register, the second is the $n_r$ qubit clock (or the QPE ancilla) register, and the third being the state register that holds information of the input state, $\ket{b}$. 
\item QPE is applied between the state register and the clock register. After this step, the state is $\ket{0}_1\otimes\sum_{j}\beta_j\ket{\tilde{\lambda_j}}_2\ket{v_j}_3$. We note that $\ket{b} = \sum_{j}\beta_j\ket{v_j}$, and ${\tilde{\lambda_j}}$ is the $n_r$-bit binary approximation of the eigenvalue of $A$ associated with the eigenvector $\ket{v_j}$. 
\item A controlled-rotation is now performed on the ancilla qubit $\ket{0}_1$ conditioned on the state of the clock register qubits, $\ket{\tilde{\lambda_j}}$. Such a rotation encodes the inverted eigenvalues in the amplitudes of the ancilla qubit $\ket{0}_1$ as shown below
\begin{equation}\label{cr}
    \sum_{j}\beta_j\left(\sqrt{1-\frac{c^2}{\tilde{\lambda_j}^2}}\ket{0}+\frac{c}{\tilde{\lambda_j}}\ket{1}\right)_1\ket{\tilde{\lambda_j}}_2\ket{v_j}_3. 
\end{equation} 
\item An inverse QPE module restores the clock register's qubits to $|0\rangle$. The state is now 
\begin{equation}\label{eq8}
    \sum_{j}\beta_j\left(\sqrt{1-\frac{c^2}{\tilde{\lambda_j}^2}}\ket{0}+\frac{c}{\tilde{\lambda_j}}\ket{1}\right)_1\ket{0}_2\ket{v_j}_3. 
\end{equation}
\item Measurement: Finally, the ancilla qubit is measured in the $Z$-basis and a nontrivial measurement outcome would mean that solution $|x\rangle$ is obtained in the third register as shown below 
\begin{equation}
    \sum_{j}\beta_j\frac{c}{\tilde{\lambda_j}}\ket{1}_1\ket{0}_2\ket{v_j}_3 = \ket{1}_1\ket{0}_2\ket{x}_3. 
\end{equation}
However, a trivial outcome would mean a failure in obtaining the solution vector. Therefore, the probability of successfully obtaining the result is the probability of obtaining a nontrivial measurement outcome and is given as 
\begin{equation}
    P(1)= {\| \sum_j\frac{c\beta_j}{\tilde{\lambda}_j} \ket{v_j}\|}^2. 
\end{equation}
Once a nontrivial outcome is detected in the first register, one extracts some feature of the solution vector obtained in the third register by applying suitable operations to it. In our case, $E_{corr}$ can be extracted as the inner product between the solution vector $|x\rangle$ and $|b\rangle$. 
\item From Eq.~\eqref{classical-correlation}, we can write our correlation energy as 
\begin{equation}
    E_{corr}=-k|\braket{b|x}|, 
    \label{Correlation-energy}
\end{equation}
where $k$ is the norm factor corresponding to normalized $\ket{x}$ and $\ket{b}$ (i.e., $k = \norm{x}\norm{b}^2$). $\ket{x}$ is proportional to the solution vector $\textit{A}^{-1}\ket{b}$ obtained in the third register. One now adds an additional fourth register, initialized to $|b\rangle$, which we refer to as $|b\rangle_4$, as shown in Figure~\ref{fig:FIG2}, and thus obtain an overlap between $|x\rangle_3$ and $|b\rangle_4$ using the Hong-Ou-Mandel module~\cite{Garcia-Escartin2013spanEquivalent} to compute the correlation energy $E_{corr}$. 
\item The overlap of  quantum states $\ket{x}_3$ and $\ket{b}_4$ can be computed following Ref.~\cite{Garcia-Escartin2013spanEquivalent} as 
\begin{equation}
  |\braket{b|x}|^2=\sum_{\alpha,\beta\in\{0,1\}^n}(-1)^{\alpha\cdot\beta}P(\alpha \beta), 
\end{equation}
where $\alpha\cdot\beta$ refers to a bit-wise AND of the two $n$-bitstrings, and $P(\alpha\beta)$ refers to the probability of obtaining each of the $2^{n_r}$-bitstrings. 
\end{enumerate}

\subsection{\label{theory-secB}XHHLite Algorithm}

In this subsection, we introduce our variants of the HHL algorithm that suit the needs of different quantum computing eras in detail. We begin with the `Lite' modules and proceed to introduce the `X' parts, and then integrate the two. Lastly, we comment briefly on the time complexity of our algorithm. \\

\subsubsection{\label{theory-secB1}The Lite module}

The `HHLite' module, whose backronym is `HHL implementation using truncated eigenspace', aims to `light'en the quantum resource consumption, by encompassing two key elements of depth reduction, the first of which stems from studying the eigenvalue distribution from a QPE module to fix the qubits, while the other aspect involves optimizing the quantum circuit by executing a sequence of operations in tandem, as we shall describe below. \\

\noindent \paragraph{\textbf{Classical multi-qubit fixing:}} A QPE module with $n_r$ ancilla qubits can capture a total of $2^{n_r}$ eigenvalues. The goal of our multi-qubit fixing strategy is to identify the dominant bits from these $2^{n_r}$ bit strings so that one can markedly reduce the controlled-unitaries that occur subsequently in the HHL circuit. We now enumerate the steps involved in this strategy below: 
\begin{itemize}
\item[1.] We first measure all the $n_r$ qubits in $Z$ basis and obtain their probability distributions. For each of the $n_r$ qubits, we say that an outcome $0$ or $1$ is dominant when its probability of occurrence is greater than or equal to a fixed probability threshold value, $P_{\rm th}$.
\item[2.] Note that classical multi-qubit fixing only requires classical post-processing of the probability distributions of $2^{n_r}$ bit outcomes that one obtains by measuring $n_r$ qubits in $Z$ basis. We fix the maximum among the dominant bit outcomes based on step[1.]. This defines $n_f=1$, that is, one qubit is said to be fixed. We note that if we identify no dominant outcome, that is, there are no probabilities of occurrence greater than or equal to 0.8, we stop. We now proceed to obtain the marginal probability distribution for each of the remaining ${2^{n_r-1}}$ bits to decide the maximum of the dominant bit outcomes again. This defines $n_f=2$. 
\item[3.] We recursively perform step [2.] after fixing the dominant bit outcome each time.
\item[4.] After thus obtaining the dominant bit outcomes for the $n_r$ qubits, we are ready to fix the respective qubits in the subsequent HHL run. For instance, for a given qubit, if we identify that $0 (1)$ is the dominant bit outcome after $Z$ measurement then we fix that particular qubit to $|0\rangle(|1\rangle)$ in the subsequent HHL run. 
The reduction in depth is achieved in this framework as follows: if the control qubit of the clock register is fixed to $|0\rangle$, one performs nothing on the target, and if the control qubit is fixed to $|1\rangle$, one executes the unitary, $U^{2^m}$, where $m \in \{0,1,2,\dots, n_r-1\}$. 
\end{itemize}
Thus, our procedure enables fixing multiple qubits $n_f >1 $, which we refer to as multiple qubit fixing (or multi-qubit fixing). Henceforth, we consistently use the notation $n_f$ to indicate the number of qubit fixings. We note at this juncture that through the process of multi-qubit fixing, several controlled unitaries that occur in HHL quantum circuit either reduce to the corresponding unitaries or to simply identity gates. \\

\noindent Note that this process is controlled by a single tunable parameter, which is a probability threshold, $P_{\rm th}$. It is important to stress that the idea of a probability threshold avoids the need to carry out multiple calculations at different values of $n_f$ to find the optimal fixing number for a given molecule. Instead, this single parameter decides $n_f$ and hence dictates the precision of our results. In other words, the precision in the correlation energies may vary for a given $P_{\rm th}$, depending on the choice of the molecule. In this work, we have set $P_{\rm th}$ to 0.8 throughout, drawing from our background in applying traditional many-body techniques on such molecular systems. It is to be noted that while the set threshold is found to work well in general, in case one seeks to systematically find the threshold in order to obtain a desired precision, the process is accompanied by an associated classical overhead. A simple illustration of multi-qubit fixing for the NISQ era is presented in Figure \ref{fig:FIG3}(a). \\ 

\noindent \paragraph{\textbf{\label{pipeline-basedopti}Pipeline-based quantum circuit optimization:}}
Current-day quantum hardware has limited operational qubits that are noisy and gates that are not yet sufficiently robust for calculations involving deep circuits. Therefore, it is necessary to efficiently decompose quantum circuits~\cite{nam_automated_2018,Mizuta,filip_reducing_2022}. Particularly, implementing the HHL algorithm which has two QPE modules is challenging even for light molecules. Therefore, we adopt the idea of quantum circuit optimization, which reduces the depth of the overall circuit. This makes it amenable to implementation on a quantum device, but accompanied by a significant amount of classical overheads. Therefore, we note that this technique could only be employed for the NISQ era of quantum computing. \\

Specifically, we opt for a pipeline-based optimization strategy, inspired by Ref.~\cite{Kharkov2022ArlineCompilers}. The underlying idea is to be able to carry out a sequence of several optimization procedures (such as peephole optimization based on KAK Decomposition~\cite{KHANEJA200111}, removal of redundant gates by combining gate-inverse pairs or by using commutation relations, etc) thereby compressing the circuits. A ZX-calculus-based approach may not lead to noticeable gains, since most of the gates that we have are non-Clifford in nature due to arbitrary rotation gates. It is important to note that a given optimization subroutine can be used again in the pipeline, as long as they are not successive in their occurrence. We employ a brute-force search to narrow down the best-performing pipeline, and for our purposes, we find that the Qiskit-Pytket-Qiskit~\cite{qiskit, pytket} pipeline (a note on notation: the name of each software development kit here actually refers to the optimization routines employed within its framework) gives the best reduction in depth. It is worth noting that we have a specific quantum hardware, the 11-qubit IonQ Harmony device, in mind for carrying out the calculations. Hence, we choose for our optimization strategies a supporting gate set that is sufficiently close to the native gate set of the hardware~\cite{Maslov_2017}. We expect that the same pipeline-based procedure should work efficiently for other hardware devices too with optimization carried out in their native gate sets. A part of Figure \ref{fig:FIG3}(a) illustrates the `Lite' workflow for the NISQ era. \\ 

We now proceed to assess the degree of compression in the circuit due to optimization in terms of reduction in circuit depth, involving two-qubit and single-qubit gates from the IonQ-supported gate set-- $(R_x,R_y,R_z$ and $R_{xx})$. The metric that we use is termed as depth compression (expressed in percentage), which is defined as 
    \begin{equation}
      \text{Depth compression} = \frac{\left|D_{out}-D_{in}\right|}{D_{in}} \times 100, 
 \end{equation}

\noindent where $D_{in}$-Depth of the input circuit (that is, the unoptimized circuit), and \\
$D_{out}$-Depth of the output circuit (that is, the pipeline optimized circuit). \\

\noindent \paragraph{\textbf{\label{opt_ver}Optimization Verification:}} We now discuss the verification process to check for the equivalence of circuits with and without optimization. 
We concatenate the unoptimized circuit with the conjugate transpose of the optimized circuit and observe that it yields the identity matrix up to a global phase: 
 \begin{equation}
 U_{\rm unopt}U_{\rm opt}^\dagger= e^{i\phi}I, 
 \end{equation}
 where $U_{\rm unopt (opt)}$ is the unitary before (after) optimization and $\phi \in [0,2\pi]$. Furthermore, we also study the classical fidelity~\cite{Kharkov2022ArlineCompilers} between the two output probability distributions obtained from optimized and unoptimized circuits, given as,
\begin{equation}
     \mathcal{F}_{cl}= \sum_{\gamma\in\{0,1\}^n}\sqrt{p_{\rm unopt}\left(\gamma\right)p_{\rm opt}\left(\gamma\right)}, 
\end{equation}
where $p_{\rm unopt (opt)}\left(\gamma\right)$ is the probability distribution of the measured bitstring $\gamma$ of length \textit{n}, in the unoptimized (optimized) circuit.
If the output probability distributions of the circuits exactly coincide, then the measured classical fidelity reaches $1$, implying the equivalence of the original and optimized circuits. The classical fidelity, in our case, is sufficiently close to 1, at 0.99998669. \\

\noindent \paragraph{\textbf{Quantum multi-qubit fixing:}} The cost of a classical multi-qubit fixing algorithm grows exponentially with the system size. Therefore, it becomes necessary to develop a quantum variant of our multi-qubit fixing algorithm. 
We prepare a QPE circuit with $n_r$ ancilla qubits and measure only a subsystem in the $Z$ basis to obtain its single-bit probability distribution. This process is repeated for each of the $n_r$ ancilla qubits. From the $n_r$ histograms thus obtained, we only fix those ancilla qubits to $|0\rangle (|1\rangle)$ in the subsequent HHL run, depending on which of those exceeds a threshold probability $P_{\rm th}$ of $0.80$ for obtaining bit $0 (1)$. 
Such an algorithm could be thought of as more suitable for the late NISQ era where one has access to more quantum resources, and where the reliance upon classical resources begins to wane. A schematic of the quantum multi-qubit fixing process is given in Figure \ref{fig:FIG3}(b). \\ 

\begin{figure}[t]
\begin{tabular}{c}
\includegraphics[width = \columnwidth]{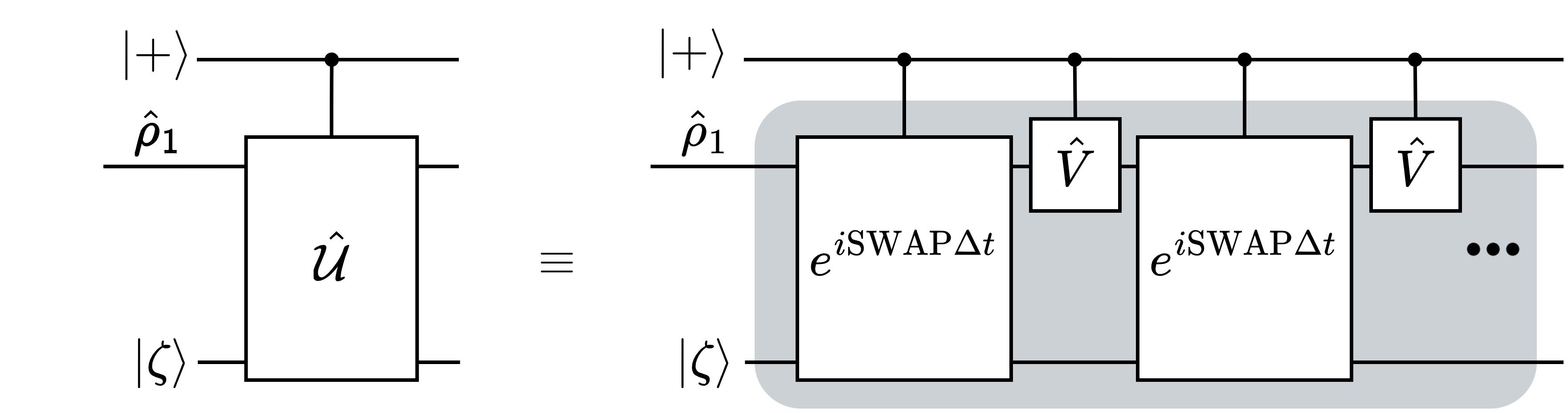} \\
(a)  \\
\includegraphics[scale = 0.5]{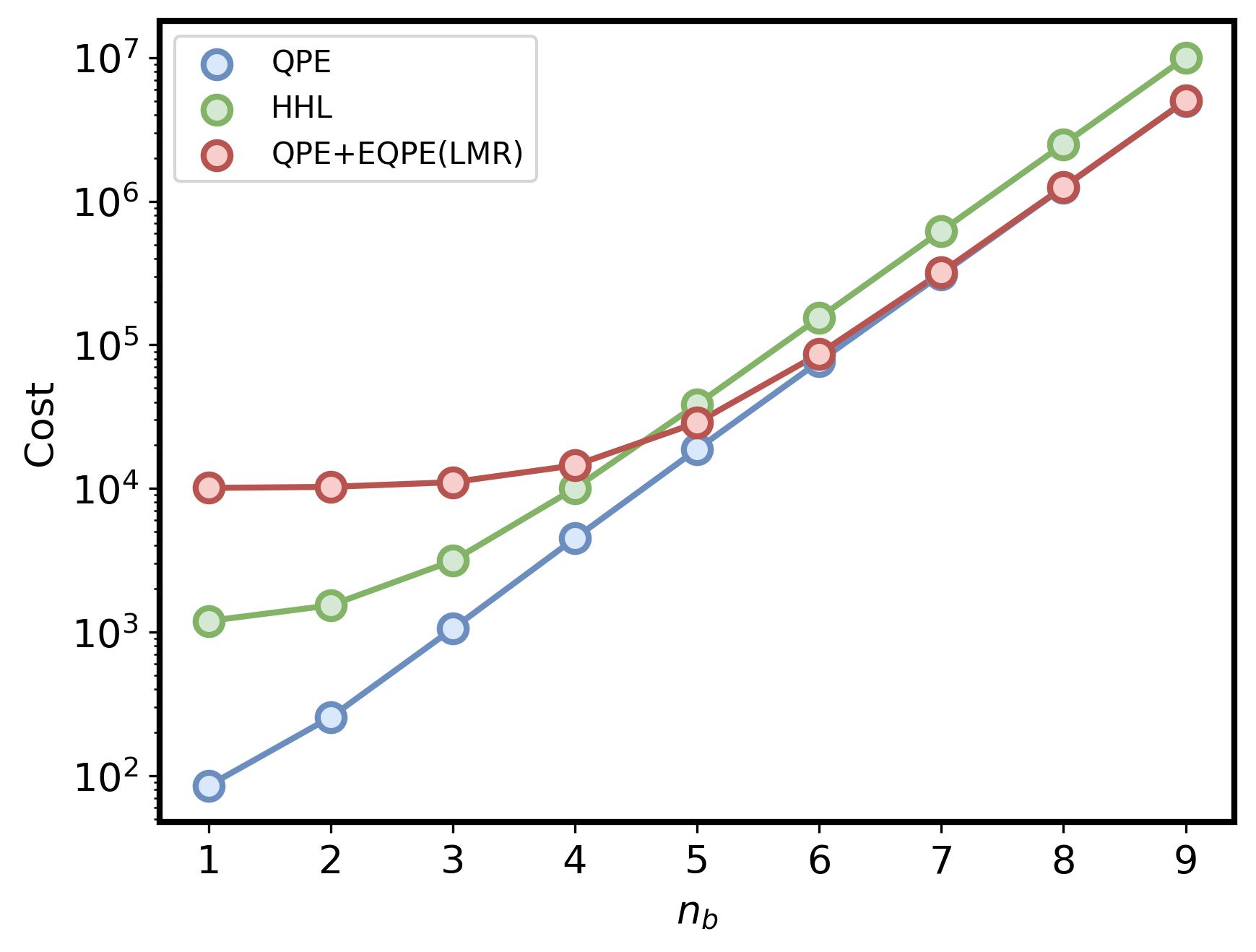} \\
(b)  \\
\end{tabular}
\caption{\label{fig:FIG4} (a) Figure depicting the EQPE module, where $|\zeta\rangle$ $\in$ $\{|0\rangle,|1\rangle\}$ evolves with respect to the subsystem $\hat{\rho}_1$ via the LMR circuit~\cite{Lloyd2014}, based on the control qubit. We note that this illustration is for one EQPE ancilla qubit. (b) The subfigure shows the cost estimate in terms of total number of two-qubit gates required, relative to input state size, $n_b$, for the QPE, HHL, and QPE$+$EQPE approaches. }
\end{figure}

\noindent \paragraph{\textbf{LMR-based multi-qubit fixing:}} 
We now present another variant of our multi-qubit fixing algorithm, which we call the LMR-based scheme for multi-qubit fixing.  
In this variant, as illustrated in  Figure \ref{fig:FIG3}(c), one performs an additional set of QPE computations (referred to as extended QPE (EQPE)) on each of the $n_r$ eigenvalue qubits at the output of the initial QPE modules, to extract the dominant bit outcome for each of the $n_r$ subsystems. We constructed the circuits based on the idea illustrated in Ref.~\cite{Lloyd2014}. The goal of an EQPE module is to extract the expectation values of each QPE ancilla qubit density matrix in $Z$ eigenbasis, in a single shot. In doing so, we obviate the need of having to prepare and measure a QPE module multiple times. We rely on improved qubit quality, better coherence times, and improved gate fidelities to achieve our desired outcomes much faster, by performing EQPE computations in parallel, on $n_r$ qubits. \\ 

We now present the steps involved in the implementation of the EQPE module. In order to extract the expectation values of interest to us for each QPE ancilla qubit whose density matrix could be given as $\hat{\rho}$, via EQPE, one has to perform a controlled-unitary operation- $\{\hat{C\mathcal{U}}, \hat{C\mathcal{U}^2}, \hat{C\mathcal{U}^4}, \cdots\}$, where the unitary $\hat{\mathcal{U}}$ is implemented as a sequence of  $e^{i {\rm SWAP} \Delta t}$ repeatedly on $\hat{\rho}$ and an ancilla  $|\zeta\rangle$ $\in$ $\mathcal{C}^2$, as shown in Figure~\ref{fig:FIG4}(a), such that for $\Delta t$ time step of evolution, the following relation holds 

\begin{eqnarray}\label{eq:swap}
&&e^{-i \rm {\rm SWAP} \Delta t}(\hat{\rho} \otimes \hat{\sigma})e^{i \rm {\rm SWAP} \Delta t}\nonumber \\ \nonumber 
&=& e^{-i \hat{\sigma} \Delta t } (\hat{\rho})  e^{i \hat{\sigma} \Delta t } \otimes e^{-i \hat{\rho} \Delta t } (\hat{\sigma})  e^{i \hat{\rho}\Delta t } +O(\Delta t ^2), \nonumber \\ \nonumber
\end{eqnarray} 

where $\hat{\sigma}=|\zeta\rangle\langle \zeta|$ and $|\zeta\rangle$ $\in$ $\{|0\rangle,|1\rangle\}$ satisfying, 

\begin{equation}
    e^{i\hat{\rho} t}|\zeta\rangle= e^{i p_1 t} |\zeta\rangle, 
\end{equation}

\noindent where $p_1$ is the expectation value of  $\hat{\rho}$ in $|\zeta\rangle$. Once $p_1$ is known, one can easily obtain $p_2=1-p_1$ for the complementary input state to determine which among the two is dominant. Each QPE ancilla density matrix $\hat{\rho}$ could be made diagonal by entangling it with an ancilla in $|0\rangle$ via a CNOT. This step is crucial since it is only then that one can have a diagonal $\hat{\rho}$, such that its eigenvectors are $\{|0\rangle,|1\rangle\}$.
In order to execute an EQPE module, we prepare $n_e$ copies of $\hat{\rho}$ by performing QPE $n_e$ times and initialize $n_e$ ancilla qubits to $|\zeta\rangle$ $\in$ $\{ |0\rangle, |1\rangle\}$. One then performs controlled-unitary operations connecting each copy of a $\hat{\rho}$ with one input state $|\zeta\rangle$, as shown in Figure~\ref{fig:FIG3}(c), where each $\hat{C\mathcal{U}}$ operation is executed as shown in Figure~\ref{fig:FIG4}(a) by applying a series of controlled-$e^{i\rm {SWAP}\Delta t}$ shown in Eq.~\eqref{eq:swap} interspersed with a unitary $\hat{V}= e^{i \hat{\sigma} \Delta t}$ impacting an evolution on $|\zeta\rangle$, as $e^{-i\hat{\rho} t}|\zeta\rangle$. Thus, we perform EQPEs in parallel across each QPE ancilla as illustrated in Figure~\ref{fig:FIG3}(c). The notation $n_e$ here refers to the precision with which one wishes to compute the expectation values for a QPE ancilla qubit $\hat{\rho}$. \\

Figure~\ref{fig:FIG4}(b) demonstrates the scaling of the cost associated with various modules (in terms of the number of two-qubit gates~\cite{Shende2004}) with the system size. The cost associated with QPE and QPE-LMR modules becomes comparable with the increase in system size and is nearly half of the total cost associated with executing an HHL module. This point emphasizes the fact that our LMR-based approach is faster and more efficient when compared to the previous approach (quantum multi-qubit fixing) of having to perform $n_rM$ measurements (for $M$ number of shots) to extract the probabilities, whereas the LMR-based approach requires $n_rn_e$ number of measurements, with $n_e > M$, as Figure~\ref{fig:FIG3} shows. \\ 

\subsubsection{\label{theory-secB2}The X module and integration with the Lite segment} 

In this section, we delve into the workings of our XHHLite algorithm. In order to solve a system of linear equations of the form $A\arrowvert x\rangle = \arrowvert b \rangle$, where the elements of the matrix are arbitrary, one must first scale $A$ appropriately to be able to capture its inverse eigenvalues with sufficiently good precision. One often scales according to the condition number $\kappa=\lambda_{\max}/\lambda_{\min}$, where $\lambda_{\max}$ and $ \lambda_{\min}$ refers to the largest and the smallest eigenvalues of $A$ respectively, in order to maximize the probability of obtaining the solution vector. However, this incurs  $\mathcal{O}(N^3)$ cost that stems from diagonalizing an $(N \times N)$ matrix. On the other hand, employing an arbitrary approach to scaling $A$ could lead to a severe loss in precision. In our work, we circumvent the problem of having to learn the eigenvalues of $A$. We propose two distinct methods to scale down the input matrix $A$ efficiently, both in terms of cost and precision. The letter `X' in our algorithm refers to these two variants, namely, a) AdaptHHL and b) PerturbedHHL. We elaborate on each one of these techniques in the following subsections. \\

\noindent \paragraph{\textbf{AdaptHHL:}} When the size of the input matrix $A$ grows, calculating its eigenvalues becomes intractable. In this variant, we address the two-fold connected problem of efficiently scaling down an arbitrary Hermitian matrix and picking a close-to-optimum value for the coefficient $c$ in Eq.~\ref{eq8}, without having to learn its eigenvalues. This allows us to capture the solution, $\textbf{x}=A^{-1}\textbf{b}$ with arbitrarily high accuracy. We scale the input Hermitian matrix $A$ as $sA$, where the scaling parameter can be given as $s=2^{-n_r}/\tilde{d}_{min}$, adapted to the input matrix $A$, such that $n_r$ is the number of qubits in the clock register of the QPE module, $\tilde{d}_{min}$ is what we call estimate-by-inspection value chosen carefully, as we describe below. Consider a Hermitian matrix given as 
\begin{equation}
    A =
  \left[ {\begin{array}{cc}
    d_{max} & d_{12} \\
    d_{21} & d_{min} \\
  \end{array} } \right]. 
  \end{equation}
   Here $A=A^{\dagger}$, $d_{max}$ is the largest diagonal element, $d_{min}$ is the smallest diagonal element of $A$, $d_{12}=d_{21}^{*}$ are the off-diagonals. We demand that our scaling factor $s=2^{-n_r}/\tilde{d}_{min}$ satisfy the following conditions
\begin{itemize}
    \item For a given $n_r$, $\tilde{d}_{min}$ is chosen to scale $A$ such that the diagonal entries are lesser than $1$, so that QPE captures its eigenvalues using $n_r$ ancillas. For this reason, the largest diagonal entry of the scaled matrix $s A$ has to satisfy,
    \begin{eqnarray} \label{eq:adapt1}
        2^{-n_r} d_{max}/ \tilde{d}_{min} < 1 
    \end{eqnarray}
    The left-hand side has to be strictly less than $1$ so that the largest eigenvalue is also less than $1$  and is captured appropriately.
    \item The choice of $\tilde{d}_{min}$ decides the representability of the eigenvalues of a given matrix. 
    If $A$ has only positive eigenvalues or only negative eigenvalues, then its  largest eigenvalue $\lambda_{max}$ and smallest eigenvalue $\lambda_{min}$ satisfy the following criterion
    \begin{equation}    
    |\lambda_{max}|\geq d_{max};|\lambda_{min}|\leq d_{min}
    \end{equation}
   Therefore, one then scales A to be able to represent its smallest eigenvalue within $n_r$ qubits, this leads to the condition:
   \begin{equation}\label{eq:adapt2}
   2^{-n_r}d_{min}/\tilde{d}_{min} \geq 2^{-n_r}
   \end{equation}
Combining Eq.~\eqref{eq:adapt1} and Eq.~\eqref{eq:adapt2}, we have,
 \begin{equation}\label{eq:adapt3}
 d_{min}\geq \tilde{d}_{min}>2^{-n_r} d_{max}
 \end{equation}
 
 \item One can also choose $\tilde{d}_{min}$ depending on the nature of the matrix $A$. For a matrix that is  diagonally dominant or a matrix that satisfies, $|\lambda_{min}|\geq d_{min}$, one could choose $\tilde{d}_{min}=d_{min}$. This would just mean that all the eigenvalues are well represented in $n_r$ qubits. We now motivate this through the following example. Consider a matrix $A$, as given below: 
\begin{equation}
    A_{2\times2} =
  \left[ {\begin{array}{cc}
    1.5 & .1 \\
    .1 & .75 \\
  \end{array} } \right]. 
\end{equation}
If we choose a scaling $s$= $2^{-3}/.75$, where $n_r=3$, the matrix scales down as 
\begin{equation}
    A_{2\times2} =
  \left[ {\begin{array}{cc}
    .25 & .0166 \\
    .0166 & .125\\
  \end{array} } \right]. 
\end{equation}
The eigenvalues of this matrix are $\{0.122815,0.252184\}$, which are captured using three qubits in the clock register with almost $1.7$ percent loss in the precision for the input state $|b\rangle= |1\rangle$ which is a $-1$ eigenvalue eigenstate of Pauli $Z$ operator. \\ 

We now discuss another example, and pick a system from the set of molecules that we considered in our results section, namely the $H_2$ molecule in its equilibrium bond length of 1.40 bohr. Its $A$ matrix takes the form \\

    $A_{4\times4} =
  \left[ {\begin{array}{cccc}
    1.12854 & 0 & 0 &0.03593\\
    0 &  1.44616 &  0.08368 &0 \\
    0& 0.08368  & 1.44616 &0 \\
    0.03593 &0  &0 & 1.94607
  \end{array} } \right]$. 

By choosing the scaling $s$= $2^{-3}/1.12854$, where $n_r=3$ and $\tilde{d}_{min}=d_{min}$, the matrix scales down as \\ 

$A_{4\times4} =
  \left[ {\begin{array}{cccc}
    0.01562 & 0 & 0 & 0.00050\\
    0 & 0.02002 &  0.00116 &0 \\
    0&  0.00116  & 0.02002 &0 \\
    0.00050 &0  &0 & 0.02694\\
  \end{array} } \right]$. 
  
The eigenvalues of this matrix are $\{0.01560,0.02697,0.02118,0.01886\}$, which are captured using three qubits in the clock register with almost $4$ percent loss in the precision for an input state $|b\rangle = |11\rangle$, where $|1\rangle$ is the $-1$ eigenstate of Pauli Z operator. 
\item As a consequence of such a scaling we obtain a fixed set of controlled-rotation angles dictated only by $n_r$. We construct a conditional rotation module, with the rotation angle $\theta_i= \rm 2 arc sin(c/\tilde{\lambda}_i)$, where $c=2^{-n_r}$ and the eigenvalue $\tilde{\lambda_i}= i/2^{n_r}$ leading to $c/\tilde{\lambda}_i = 1/i$, $i \in \{1, 2,\dots, 2^{n_r}-1\}.$ The angle $\theta_i$= $0$, when $i=0$.  Importantly, $2^{-n_r}$ is the smallest value that can be chosen for $c$ so that  $0 \leq (c/\tilde{\lambda}_i) \leq 1$.

\item  
The norm of obtaining the solution vector $|x\rangle$ could be written as
\begin{eqnarray}\label{eq:prob}
\lVert \ket{x} \rVert &=&\sqrt{\sum_i \bigg| \frac{b_i s}{\tilde{\lambda}_i}  \bigg|^2} \nonumber \\ \nonumber
 &=&\sqrt{\sum_i \bigg| b_i \frac{2^{-n_r}}{\tilde{d}_{min} \tilde{\lambda}_i} \bigg|^2} \nonumber \\ \nonumber
 &=& \sqrt{P(1)}/ \tilde{d}_{min}, 
\end{eqnarray}
where P(1) is the probability of obtaining $1$ in the ancilla register and the coefficients of the input state $|b\rangle$ satisfy $\sum_i|b_i|^2$ =1, $\{\tilde{\lambda}_i\}s$ are the eigenvalues of the scaled matrix $sA$, and $\tilde{d}_{min}$ is an estimate-by-inspection value of the smallest eigenvalue of $A$ (matrix before scaling). 
\item Our systematic way of looking for $\tilde{d}_{min}$ and $n_r$ to represent all the eigenvalues of the matrix with good precision, as shown in Eq.~\eqref{eq:adapt1}-Eq.~\eqref{eq:adapt3}, would ensure results with arbitrarily high accuracy. We now present the performance of few molecules in the Appendix where we plot percentage fraction differences with LCC on a traditional computer (PFDs) vs $\tilde{d}_{min}$  and $P(1)$ vs $\tilde{d}_{min}$. 

\end{itemize}
It is relevant to mention at this point that for the purposes of this work on applying AdaptHHL to LCC problem, we choose $\tilde{d}_{min} = d_{min}$ for all our numerical results. We further note that the `Adapt' in AdaptHHL is not in the same vein as ADAPT-VQE~\cite{Grimsley2019AnComputer}. It is important to stress that the AdaptHHL approach is relevant for all quantum computing eras, in view of its salient feature of bypassing an otherwise expensive classical overhead that scales as the third power of system size. \\ 


\noindent \paragraph{\textbf{AdaptHHLite:}} In AdaptHHLite, we integrate the two ideas, where we combine our way of scaling down the matrix with the `Lite' module described in early parts of Sec.~\ref{theory-secB}. 
We observe that performing AdaptHHLite could lead to intriguing textbook-like examples, where some molecules in specific geometries accommodate all-fixing, thus effectively reducing an AdaptHHLite computation to a one-qubit calculation. This stems from the fact that when all the qubits are already fixed, there are no controlled operations coming out of the clock register in practice. This leads to a peculiar situation, where $|\langle b|x \rangle|$ is one, as we really extract $|b \rangle$ at the end of the AdaptHHLite procedure on the state register. Therefore, for a practical computation, there is no need for a Hong-Ou-Mandel module, or for that matter, a state register. 
This leaves us with a one-qubit computation, where $|0 \rangle$ is merely rotated and measured to extract $||x||^2$, and hence $E_{corr}$. It is important to appreciate that this trivial single-qubit computation is preceded by a pipeline-optimized QPE calculation to determine $n_f$. \\

For our hardware computations that we will discuss in our Section \ref{results-secD}, we choose the special cases mentioned in the above paragraphs. We note that the pipeline-optimized QPE itself is carried out on a classical computer in view of limitations in current-day quantum resources, but it is conceivable that this be executed on a quantum computer as we gradually tune out of the NISQ era. The AdaptHHLite computation is carried out on the IonQ Harmony device. \\ 

\noindent \paragraph{\textbf{PerturbedHHL:}} This variant allows us to scale a matrix by estimating the condition number, $\kappa=\lambda_{max} / \lambda_{min}$ in a way that is much cheaper than classically diagonalizing a matrix. In this method, we examine the diagonal elements of matrix $A$ to obtain the minimum diagonal element, $d_{min}$, and maximum diagonal element, $d_{max}$, to determine whether they are degenerate or non-degenerate, resulting in two potential scenarios: \\ 
\begin{itemize}
    \item If $d_{min}$ ($d_{max}$) is degenerate, we apply a level shift to the repeating $d_{min}$ ($d_{max}$) diagonal elements of $A$, by replacing them with $d_{ii}-m\xi$, where $m \in \{1,2,3.....\}$ represents the instance of repetition of that element, and $\xi \in (0,1]$. We call this new matrix $B$. Then, we find the resulting eigenvalue by using the equation
    \begin{equation}\label{Perturb_B}
        \Tilde{\lambda}= d_{ii}+ \sum_{j\neq i}^{N}\frac{\abs{b_{ij}}^2}{b_{ii}-b_{jj}}, 
    \end{equation}   
    where $i$ corresponds to the index of $d_{min}$ ($d_{max}$), resulting in an approximate value $\Tilde{\lambda}_{min}$ ($\Tilde{\lambda}_{max}$). 
   
    \item If $d_{min}$ ($d_{max}$) is non-degenerate, we apply perturbation to the $d_{min}$ ($d_{max}$) element of matrix $A$ using Eq.~\eqref{Perturb_B}, resulting in an approximate eigenvalue $\Tilde{\lambda}_{min}$ ($\Tilde{\lambda}_{max}$). 
\end{itemize}

Thus, we utilize the information of the minimum and maximum diagonal elements to arrive at $\Tilde{\lambda}_{min}$ and $\Tilde{\lambda}_{max}$ through perturbation on level shifted $d_{min}$ ($d_{max}$), which has a computational complexity of $\mathcal{O}(N)$. This approach yields correlation energies with reasonable precision, as shown in Table A2. The pseudocode for the algorithm is given below: 

\normalem
\begin{algorithm}
\caption{Algorithm for approximate maximum and minimum eigenvalue computation}\label{alg:cap}
\SetKwInOut{Input}{Input}
\SetKwInOut{Output}{Output}
\Input{$A \in (N \times N), \xi = 1$ }
\Output{$\Tilde{\lambda}_{min}, \Tilde{\lambda}_{max}\rightarrow $ determines $c, \theta$ and $t$.} 
\SetKwBlock{Beginn}{beginn}{ende}
\Begin{
Identify $d_{ii}^{max}$ and $d_{ii}^{min}$ \\
Construct $B \ni b_{ij}=d_{ij}$ \\
\If{$d_{ii}^{max/min}$ repeats along the diagonal of A}
{$b_{ii} = d_{ii} - m\xi$ for $m^{th}$ repetition of $d_{ii}$}
\Else  
{$b_{ii} = d_{ii}$}
Calculate $\Tilde{\lambda}_{min/max}$  using perturbation theory. $\Tilde{\lambda}_{min/max} = d_{ii}^{min/max} + \sum_{j\neq i}^{N}\frac{\abs{b_{ij}}^2}{b_{ii}-b_{jj}}$
}
\end{algorithm}
\ULforem

Through a careful examination of various values of $\xi$, we found $\xi=1$ to be a desirable value (see Figure A2 of the Appendix), resulting in a maximum PFD of about 7 percent and a minimum $-$0.06 percent for the set of molecules considered. Figure A5 of the Appendix shows the behaviour of PFD with $n_r$. We observe anomalous behaviour in $LiH$ for $n_r \geq 9$. We reason that while the focus was to evaluate eigenvalues of matrix $A$ approximately, there was no in-built rule to constrain the scaled $A$ between 0 and 1. As $n_r$ increases, the precision in the $E_{corr}$ of $LiH$ is severely affected. Curing this problem is beyond the scope of the current study, and we defer it to future work. \\ 

\noindent \paragraph{\textbf{PerturbedHHLite}} is a straightforward  integration of PerturbedHHL way of scaling the input matrix and the Lite module described in early paragraphs in Sec.~\ref{theory-secB}. \\ 

\subsubsection{\label{theory-secB3}The complexity of AdaptHHL algorithm} 

In this section, we briefly describe the complexity of our AdaptHHL strategy. We carry out the analysis for $\tilde{d}_{min}$ set to $d_{min}$, as mentioned in section \ref{theory-secB2}(a). In determining the complexity, we exclude state preparation ($\ket{b}$) as well as the Hong-Ou-Mandel and the eigenvalue inversion parts, as the QPE modules significantly outweigh these.\\

Now, recall that the complexity of the original HHL algorithm is $\mathcal{O}(\log_2(N)s^2t)$, where $s$ refers to the sparsity of the input $(N\times N)$ matrix. QPE requires the time variable to be $t=\mathcal{O}(\kappa/\epsilon)$. This is because we have $n_r=\mathcal{O}(\log_2(\kappa/\epsilon))$ without amplitude amplification (see Ref.~\cite{richardson}), so that $t=\mathcal{O}(2^{n_r})=\mathcal{O}(\kappa/\epsilon)$. Here, $n_r$ is the number of qubits in the clock register as used in the earlier sections, $\kappa$ is the condition number of the matrix and is the ratio of the largest to the smallest eigenvalue of the matrix, i.e.~$\lambda_{max}/\lambda_{min}$, and $\epsilon/2$ is the precision error in trace distance of the phase estimation process. Taking into account $\mathcal{O}(\kappa)$ amplitude amplification iterations, the total complexity of the HHL algorithm turns out to be $\mathcal{O}(\log_2(N)s^2\kappa^2/\epsilon)$.\\

Notice from Sec.~\ref{theory-secB} that in the AdaptHHL algorithm, the time variable $t$ gets scaled to $t 2^{-n_r}/d_{min}$. Here, as used in the earlier sections, $d_{min}$ is the minimum diagonal element of the matrix. 
The original HHL algorithm used $c=\mathcal{O}(1/\kappa)$ to yield $t=\mathcal{O}(\kappa/\epsilon)$ without amplitude amplification. By contrast, here we have $c=2^{-n_r}=\mathcal{O}(\epsilon/\kappa)$, which, in turn, yields $t=\mathcal{O}(\kappa/\epsilon^2)$, that also evidently subsumes an extra complexity of $\mathcal{O}(1/\epsilon)$ for amplitude amplification steps, that we need not actually perform as mentioned earlier. Because of the additional scaling of $1/d_{min}$ in the scaling factor $s$, we get $t=\mathcal{O}(\kappa d_{min}/\epsilon^2)\approx \mathcal{O}(\lambda_{max}/\epsilon^2)$. Thus, the total complexity of AdaptHHL turns out to be $\mathcal{O}(\log_2(N)s^2\kappa d_{min}/\epsilon^2)\approx \mathcal{O}(\log_2(N)s^2\lambda_{max}/\epsilon^2)$.\\

In the case of general positive definite matrices, i.e.~without the assumption of the matrix being diagonally dominant, we have: $d_{min}\geq\lambda_{min}$ and $d_{max}\leq\lambda_{max}$, where $d_{min}$ and $d_{max}$ are the smallest and the largest diagonal elements of the matrix, respectively, and $\lambda_{min}$ and $\lambda_{max}$ are the smallest and the largest eigenvalues of the matrix, respectively. Thus, the complexity of our AdaptHHL algorithm is given by $\mathcal{O}(\log_2(N)s^2\kappa d_{min}/\epsilon^2)\geq \mathcal{O}(\log_2(N)s^2\lambda_{max}/\epsilon^2)$, where the equality holds when the matrix is a diagonal matrix. Notice that for matrices with $d_{min}<1$, our AdaptHHL algorithm is a sub-$\kappa$ algorithm, since $\kappa d_{min}<\kappa$, although the cost in $\epsilon$ increases from $\mathcal{O}(1/\epsilon)$ to $\mathcal{O}(1/\epsilon^2)$, when compared to the original HHL algorithm.

\begin{figure*} 
\begin{tabular}{c c c}
\includegraphics[scale=0.40]{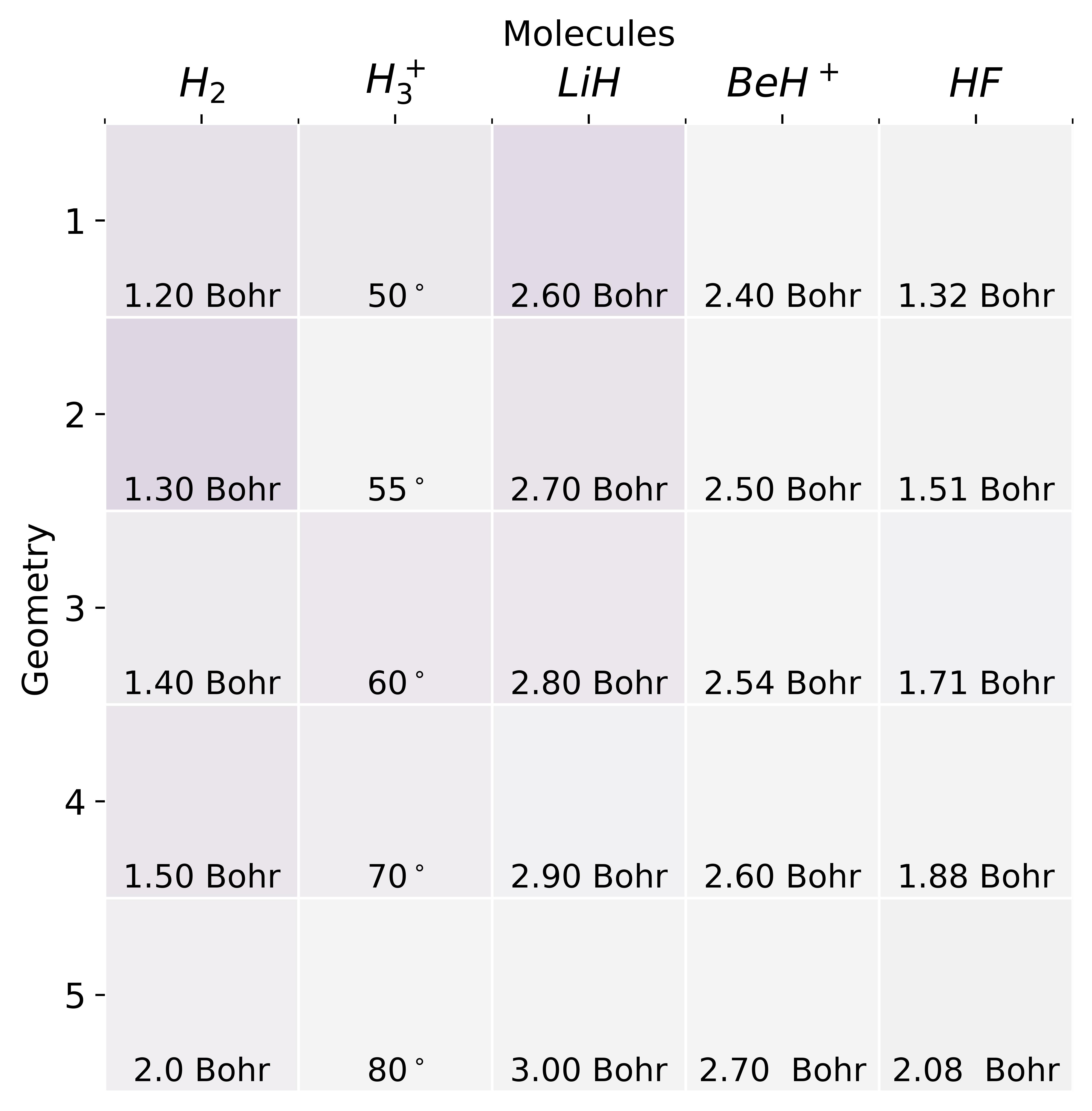} & \includegraphics[scale=0.40]{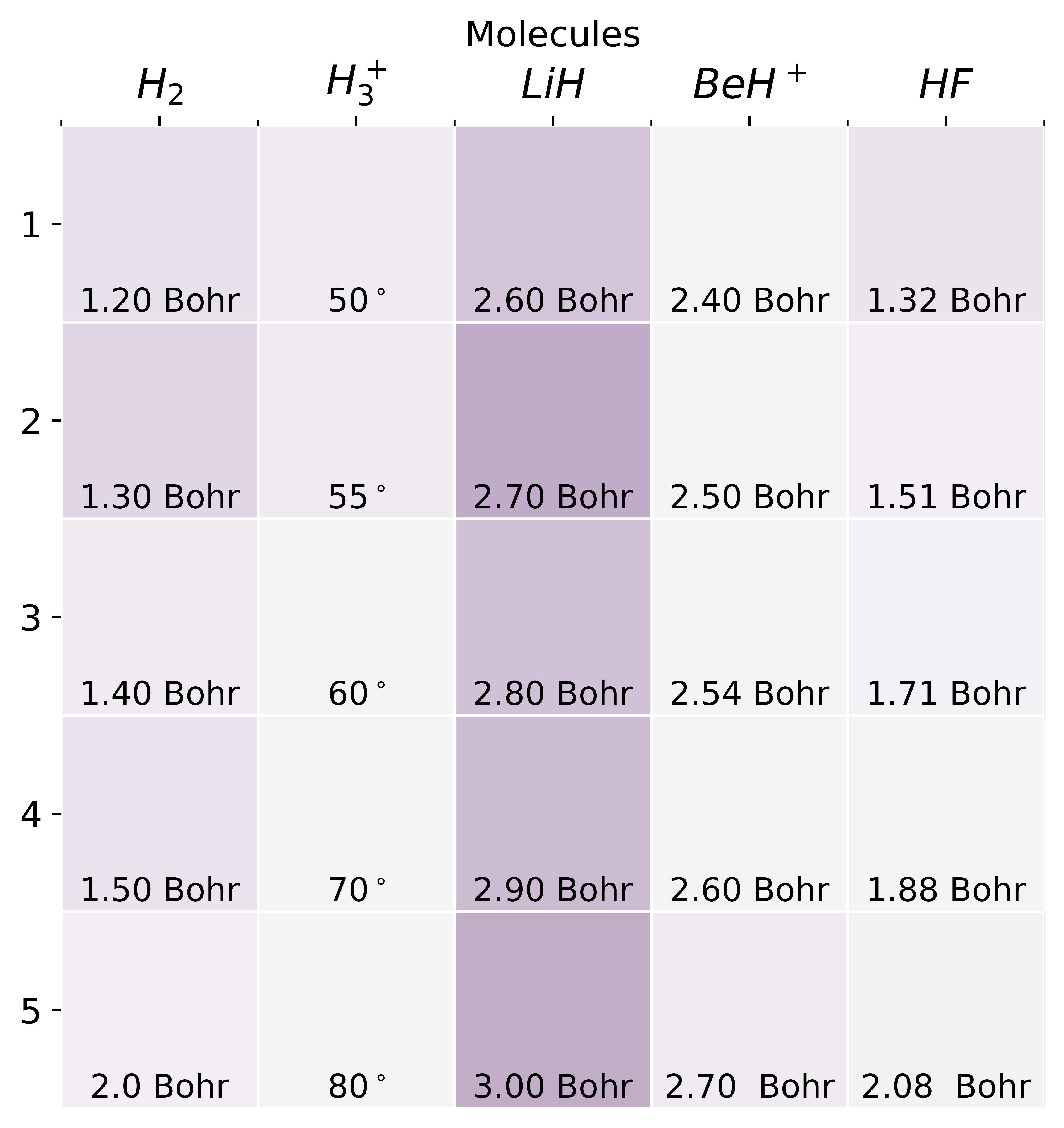} & \includegraphics[scale=0.40]{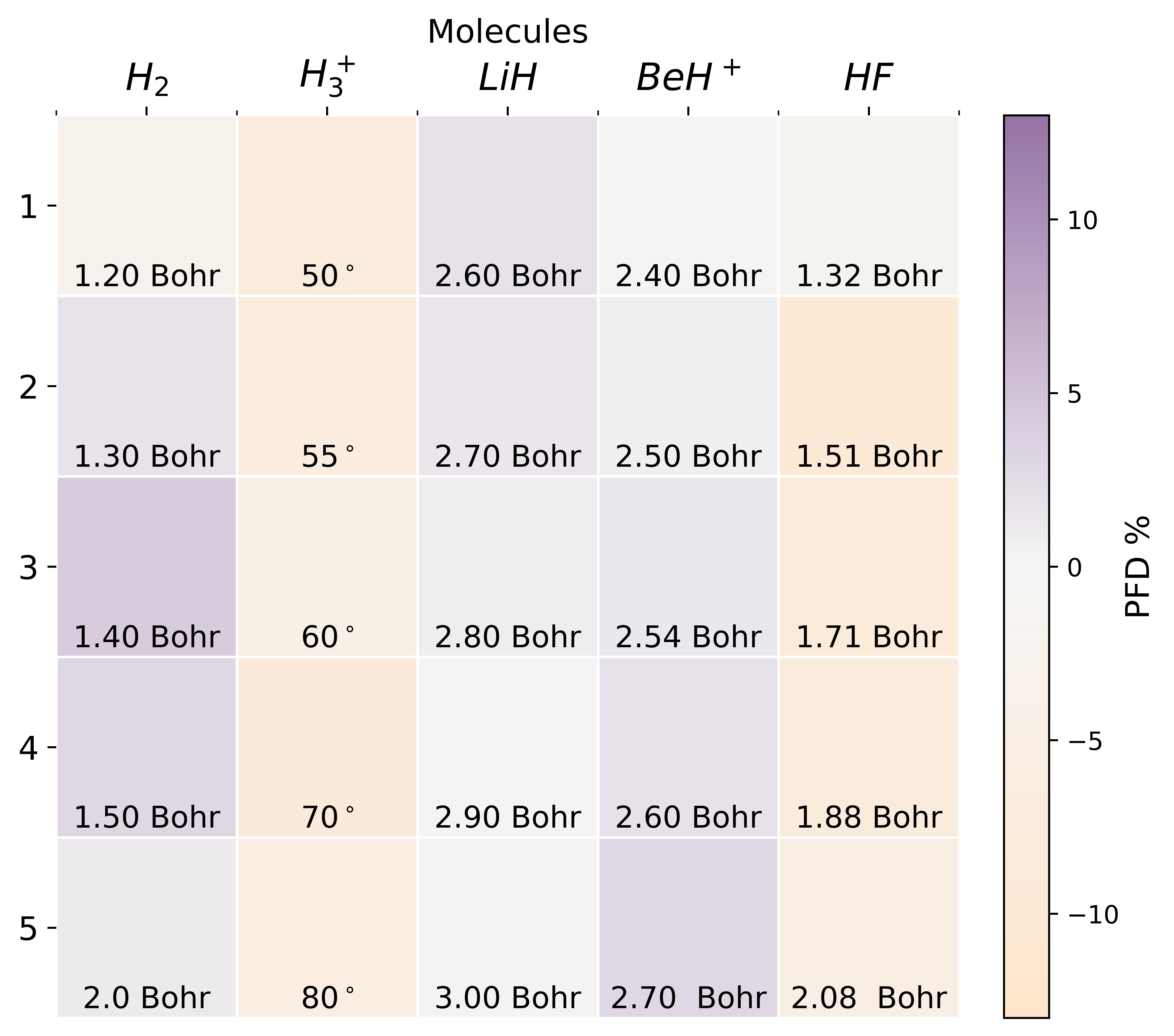} \\
(a) {HHL} & (b) {PerturbedHHL} & (c) {AdaptHHL} \\ \\
\includegraphics[scale=0.40]{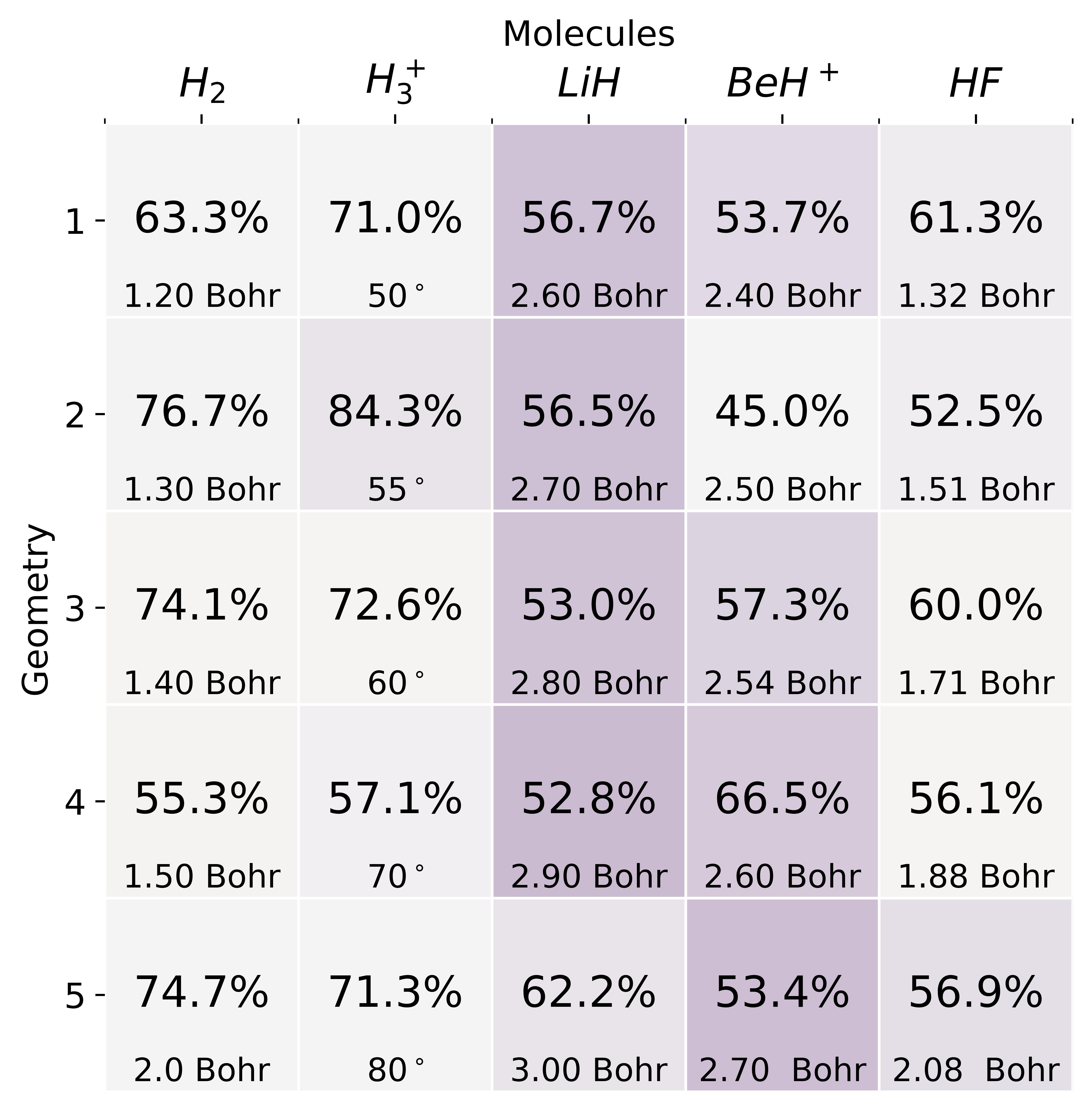} & \includegraphics[scale=0.40]{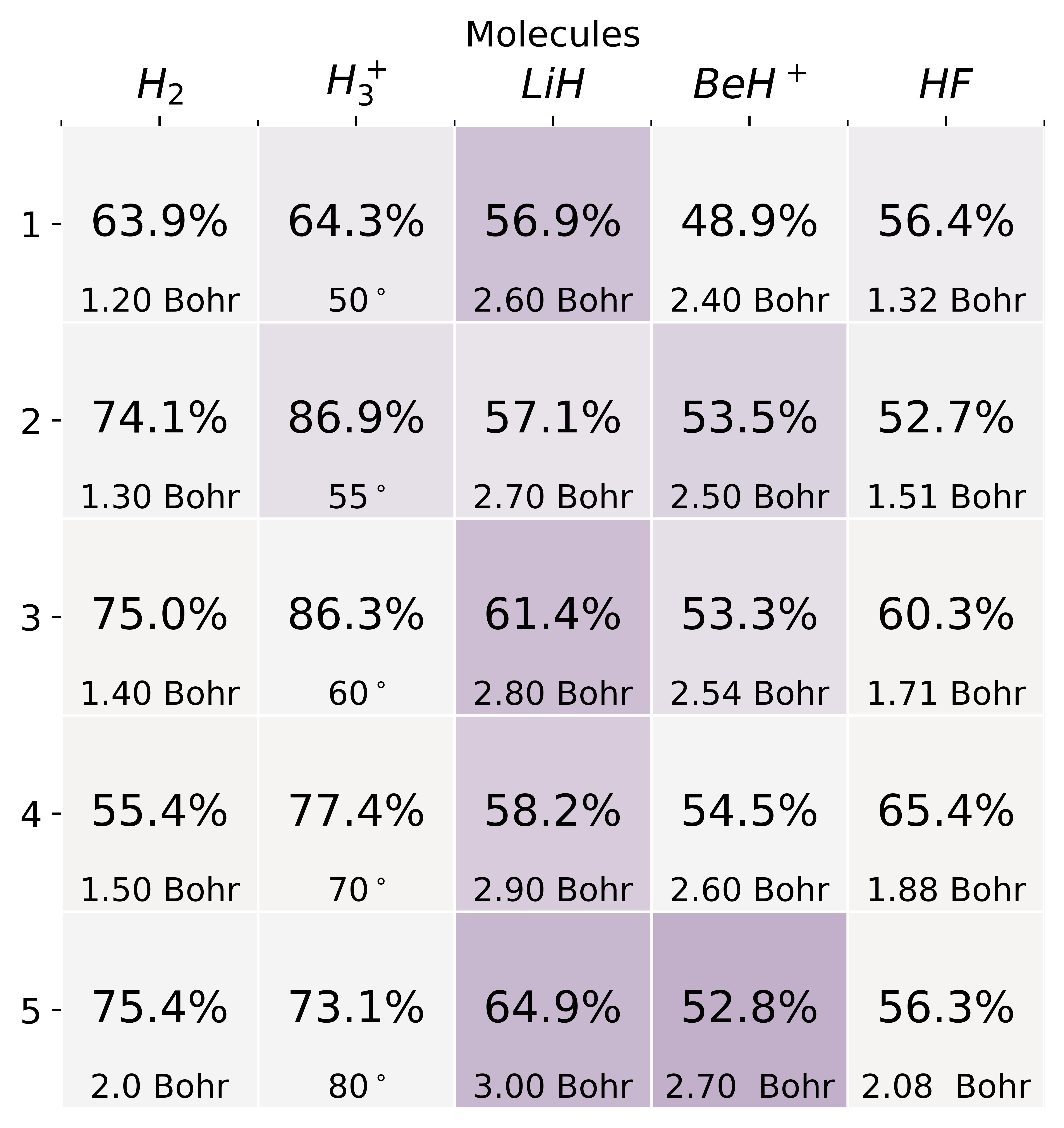} & \includegraphics[scale=0.40]{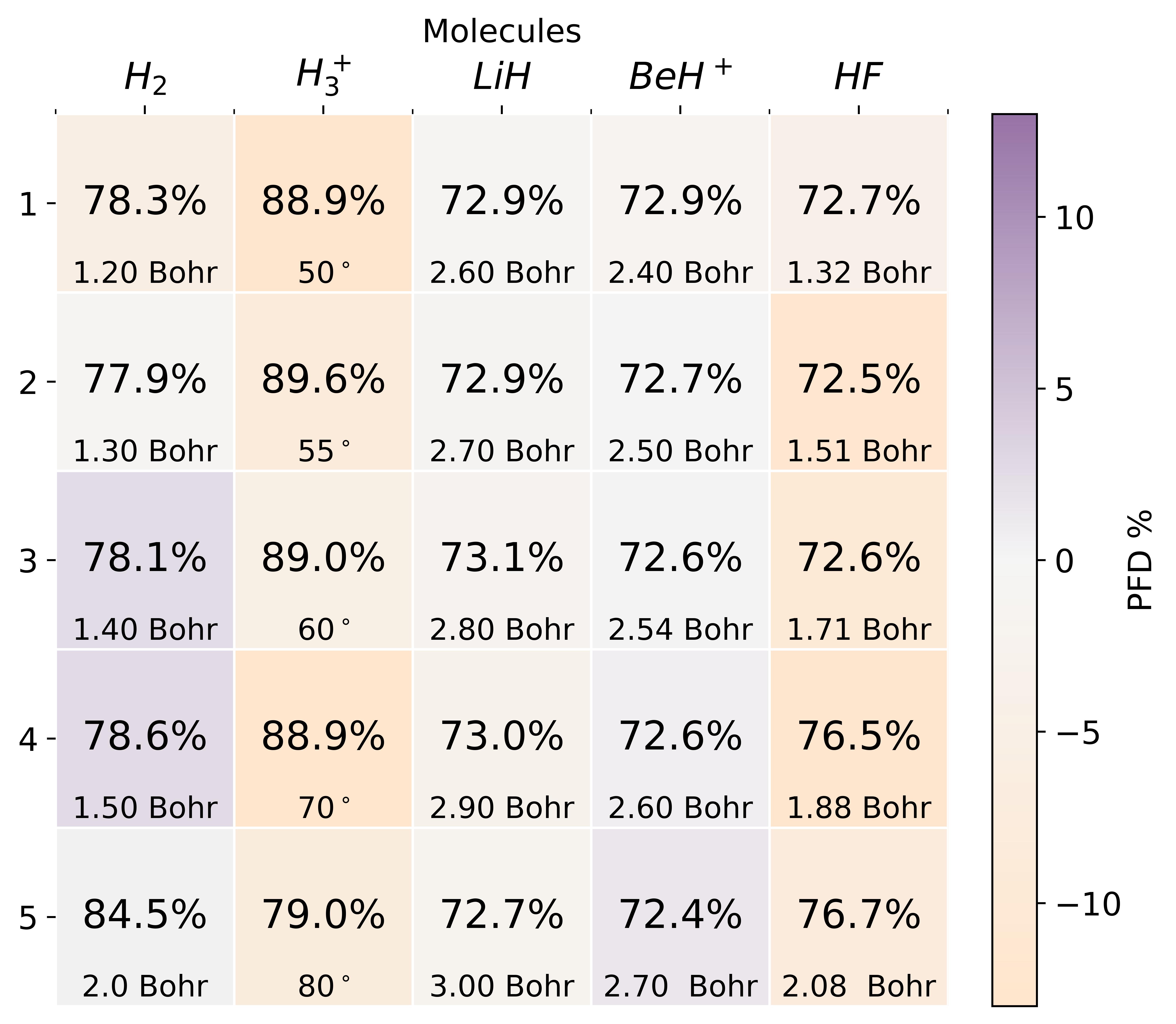} \\
(d) {HHLite} & (e) {PerturbedHHLite} & (f) {AdaptHHLite} \\
\end{tabular}
\caption{\label{fig:FIG5} Heatmap plot showing the circuit depth compression observed in the corresponding HHL variant with respect to quantum circuit depth of the HHL algorithm, as well as the PFD of correlation energies obtained from (a) HHL, (b) PerturbedHHL, (c) AdaptHHL, (d) HHLite, (e) PerturbedHHLite, and (f) AdaptHHLite schemes for the set of molecules considered.  The text inside the grid shows the circuit depth compression in percentage along with the geometry of the molecule. The color gradient represents the PFD with respect to LCC calculation on a traditional computer. We note that for all of our numerical results for AdaptHHL and AdaptHHLite, we set $\tilde{d}_{min}$ to $d_{min}$, as mentioned in Section \ref{theory-secB2}(a). } 
\end{figure*} 

\section{\label{results} Results and Discussion} 

\subsection{\label{results-secA}A summary of our results}  

As we are currently in the NISQ era, practical results for correlation energies can realistically only be obtained in the NISQ variants of PerturbedHHLite and AdaptHHLite, with both the approaches utilizing classical multi-qubit fixing and pipeline-based quantum circuit optimization. To that end, we begin with HHL results, against which we compare the performance of PerturbedHHLite and AdaptHHLite (we hereafter do not explicitly mention classical multi-qubit fixing and pipeline-based quantum circuit optimization, as it is implied in the rest of the results) algorithms. Finally, we present our HHL results on the IonQ hardware for a $(2 \times 2)$ size matrix, followed by AdaptHHLite results for a larger $(4 \times 4)$ case with all the qubits fixed as well as for a $(4 \times 4)$ case with all but one qubit fixing. \\

\subsection{\label{results-secB}Computational details} 

Before discussing our results, we provide an overview of the computational aspects involved in our work prior to executing the (Adapt/Perturbed)HHL(ite) algorithm. For a given molecule, we generate the matrix \textit{A} and the vector \textbf{b} using a traditional computer. We obtain the relevant one- and two-body Hamiltonian integrals (the Fock and Coulomb integrals, respectively) using the GAMESS-2014~\cite{GAMESS} package. It is worth noting that we leverage molecular point group symmetry in our matrix generation procedure ($D_{2h}$ for $H_2$, and $C_{2v}$ for the rest of the considered molecules). We also calculate the correlation energy with the LCC approach using our in-house program, by employing a traditional computer. The precision of $E_{corr}$ predicted by (Adapt/Perturbed)HHL(ite) approaches is assessed by comparing them with the result obtained from the traditional computation. 
We consider the simple cases of one-hole orbital and multiple particle orbitals or one particle orbital and multiple hole orbitals for the systems considered in this work. Thus, under a singles and doubles truncation scheme, the two-body spin-free excitation operators are of the type $\{\hat{e}_{ii}^{ab}\}$ or $\{\hat{e}_{ij}^{aa}\}$ and the operator manifold becomes intrinsically orthogonal. No further manipulation of the operators would be required to render the matrix $\textit{A}$  Hermitian. \\ 

The dimension of $\textit{A}$ is equal to the square of the total number of excitations taken in the manifold, while that of vector $\textbf{b}$ is equal to the total number of excitations. For example, in LCCD (LCC with only double excitations, and the approximation that we almost always employ in this work), the dimension of $\textbf{b}$ is $n_h^2n_p^2$ (with $n_h$ and $n_p$ referring to the number of holes and particles respectively). In the case of LCCSD (which includes singles and doubles), the dimension of $\textbf{b}$ is $n_h^2n_p^2 + n_hn_p$. 
The general expression for the dimension of vector $\textbf{b}$, $N$, of an LCCi or CCi (i=degree of excitation taken) theory would be 
$\sum_i (n_hn_p)^i$ and the dimension of \textit{A} would be $(N \times N)$. For a matrix of size $(N \times N)$, the number of qubits in the third register is $(n_b = \lceil log_2(N)\rceil)$. Our base HHL implementation is a slightly modified version of the HHL program from Vazquez \textit{et al}~\cite{richardson}. \\ 
 
In this work, we consider five closed-shell molecules (including a triatomic molecular ion), each in five geometries. The considered systems as well as the details of their geometries and choice of single particle bases are shown in the first three columns of Tables A1 through A3 of the Appendix. $n_r$ is set to 6 for the $(4 \times 4)$ cases, while it is set to 8 for systems of $(16 \times 16)$ matrix size. Figure A5 presents the variation in PFD with $n_r$ for all of the considered molecules in their equilibrium geometries. We note at this point that all our simulation results were obtained using the statevector backend of the Qiskit software development kit~\cite{qiskit}. \\ 

\subsection{\label{results-secC} Results from simulation} 

The results obtained from our calculations are presented in Tables A1 through A3 of the Appendix. The tables provide data on the correlation energies (in units of milliHartree), as well as the depth of the corresponding quantum circuit, the number of two-qubit gates in that circuit, and $n_f$, along with the PFD with respect to LCC calculations carried out on a traditional computer, for HHL and HHLite (Table A1), PerturbedHHL and PerturbedHHLite (Table A2), and AdaptHHL and AdaptHHLite (Table A3). Figure \ref{fig:FIG5} compactly presents the important data given in the tables by considering two metrics for all of the considered molecules and geometries, the PFD with respect to LCC correlation energy obtained on a classical computer denoted by colour in a heat map, and the circuit depth compression in each of our variants relative to the depth of the HHL circuit denoted as percentages. \\ 

We now proceed to comment on the overall observed trends from Figure \ref{fig:FIG5} and Tables A1 through A3, in the subsequent paragraphs. It is important to comment at this juncture that the quantity of interest to us is the correlation energy (presented in units of milliHartree), and not the total energy (which is a substantially larger quantity relative to correlation energy). Therefore, all the PFDs indicate the precision with which we obtain the correlation energy in the present work. We note here that the smallest value of $E_{corr}$ is about 10 milliHartree, while the largest is about 44 milliHartree. \\ 

The data in the table shows, that while results from HHL in general agree well with the LCC calculation carried out on a traditional computer, it leads to deep circuits, with the deepest one being containing over 18,700 two-qubit gates (with a depth of almost 100,000). On the other hand, the worst-case scenario for AdaptHHLite is about 9400 two-qubit gates (depth of almost 27,000). On this note, it is worth noting that AdaptHHLite accommodates a large $n_f$ for our $P_{\rm th}$ of 0.8, including all-qubit fixing for four out of five geometries for $H_3^+$ and at least an $n_f$ of 6 for the three larger molecules ($(16 \times 16)$ cases). Figure A6 of the Appendix shows the reduction in depth with increasing $n_f$ for our HHLite, PerturbedHHLite, and AdaptHHLite. We note that this particular analysis was carried out by hard-coding $n_f$, and not via a probability threshold. Turning our attention to Figure \ref{fig:FIG5}, we find that HHLite leads to a depth compression of at least 45.0 percent, and can lead to a compression as much as 84.3 percent. AdaptHHLite takes it further and leads to a minimum and maximum compression of 72.4 and 89.6 percent respectively. Note that the molecular systems where we find the minimum and maximum compression need not be the same for HHLite and AdaptHHLite. As the figure shows, PerturbedHHLite offers a noticeable improvement in terms of depth reduction over HHLite for the worst and best cases but does marginally worse relative to AdaptHHLite. However, if we compare the three approaches, namely HHLite, PerturbedHHLite, and AdaptHHLite, for each of the considered molecules, we observe that while PerturbedHHLite sometimes can give a slightly deeper circuit than HHLite, but AdaptHHLite always provides a noticeable reduction in depth relative to HHLite. \\ 

We also observe the unsurprising implication of resource reduction from both Figure~\ref{fig:FIG5} as well as the tables A1 through A3, namely the trade-off with precision. To that end, we devote the rest of this paragraph for reporting our PFDs along with the energy difference with respect to the LCC value for correlation energy obtained on a traditional computer, for our considered variants. For HHL, the best and worst case PFDs are $-$0.02 and 3.04 respectively, with the energy differences themselves being below a milliHartree level throughout. The worst-case energy difference is observed to be about 0.45 milliHartree. In the case of HHLite, the PFDs range from $-$0.04 to 5.78, with the worst-case energy difference now being about 1.3 milliHartree. For the PerturbedHHL algorithm, the worst case PFD goes to 7.28 with the corresponding energy difference being about 1.306 milliHartree, while for the PerturbedHHLite variant, it is 6.95 with the associated energy difference of about 1.6 milliHartree. Finally, we see that the AdaptHHL and the AdaptHHLite variants yield worst-case PFDs of 9.2 and 13.7 respectively, with the worst-case energy difference being about 4 milliHartree for the latter. We reiterate at this point that the precision can be improved to some extent by tuning the probability threshold but at the cost of incurring more depth. \\ 

We now discuss the sources of errors in all our numerical simulations. Note that in the HHL algorithm, the inadequacy in the number of QPE clock register qubits ($n_r$) could limit precision. In the AdaptHHL variant, the choice of $\tilde{d}_{min}$ and $n_r$ has a direct impact on the precision. We recall that we choose $\tilde{d}_{min} = d_{min}$ for all our numerical results. The problem of finding the optimal $\tilde{d}_{min}$ in a resource-efficient manner is deferred to a future study. However, we present our results from a preliminary analysis on the possibility of narrowing down the search range in Figure A3 of the Appendix. In the AdaptHHLite framework, the multi-qubit fixing scheme of the Lite module involves elimination of the non-dominant outcome probabilities from the QPE module via a probability threshold, which in turn contributes to the error budget. In the pipeline optimization module, the optimized unitary is slightly different from the original one. This minute difference still reflects as an error in correlation energies. \\ 

We finally comment very briefly on the efficacy of our quantum multi-qubit fixing procedure described in Section~\ref{theory-secB1}(d). We found that in the AdaptHHLite framework, for twenty-four out of the twenty-five molecular systems that we considered, we obtain the same values for $n_f$ as in the equivalent classical multi-qubit fixing procedure, and for the remaining case, we obtained a slightly better value of $n_f$ with quantum multi-qubit fixing. \\ 


\subsection{\label{results-secD}Proof-of-principle hardware results}

We carry out two proof-of-principle hardware experiments. The first computation serves as a pilot demonstration of HHL being able to predict LCC correlation energies. To that end, we consider a trivial case of $H_2$ in the STO-6G basis (the \textit{A} matrix size is $(2 \times 2)$), and carry out an HHL computation in the LCCSD approximation to construct the potential energy curve for the molecule in the neighbourhood of the equilibrium bond length. The second experiment involves two sets of calculations for obtaining the potential energy curves of $LiH$ and $HF$, to demonstrate the performance of our AdaptHHLite algorithm. Both of these cases involve $(4 \times 4)$ size \textit{A} matrices, but since we obtained all-qubit fixing for our $P_{\rm th}$ of 0.8, the computation vastly simplifies. All of our results are presented in Figure \ref{fig:FIG6}. \\ 

\subsubsection{\label{results-secD1}HHL hardware results}

Before carrying out our hardware experiments, we need to compute the required number of shots needed to achieve reasonable precision. In Figure A1, we demonstrate the behaviour of the correlation energy of \textit{\ce{H2}} in the STO-6G basis (chosen as a representative system for this purpose) with the number of shots, such that each computation with a given number of shots is repeated 200 times, on the Qiskit noiseless QASM simulator. We observe that the mean correlation energy for these repetitions lies between the full configuration interaction (FCI) and LCCSD classical computation value and is closer to the latter. Furthermore, beyond 1000 shots, we observe that the spread in the correlation energy starts falling below 0.25 milliHartree, around the LCCSD (classical) value. Therefore, we set the number of shots to 1000 while performing our hardware experiments. \\ 

\begin{figure*} 
\begin{tabular}{c c c}
\includegraphics[scale=0.58]{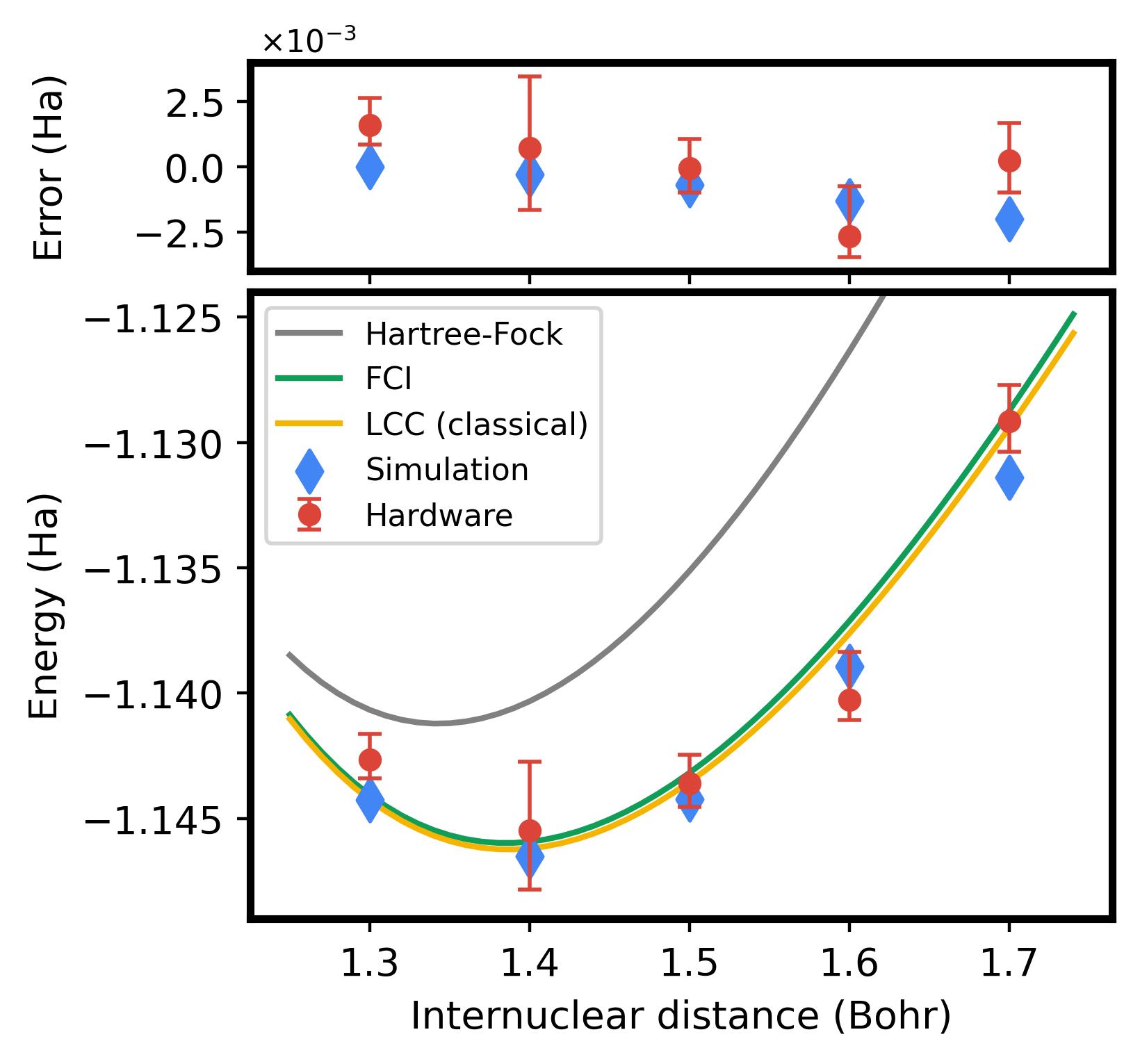} & \includegraphics[scale=0.58]{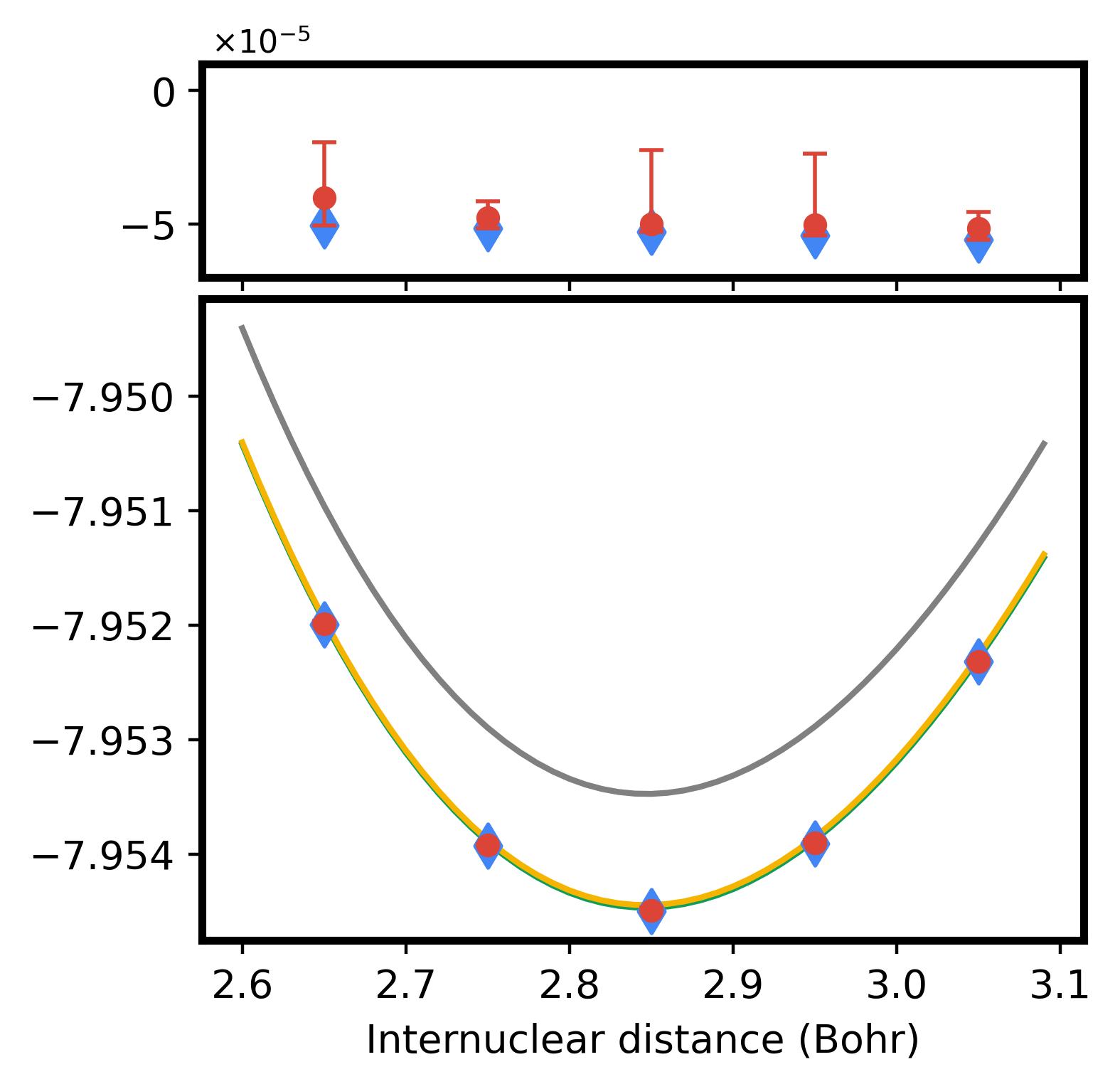}  & \includegraphics[scale=0.58]{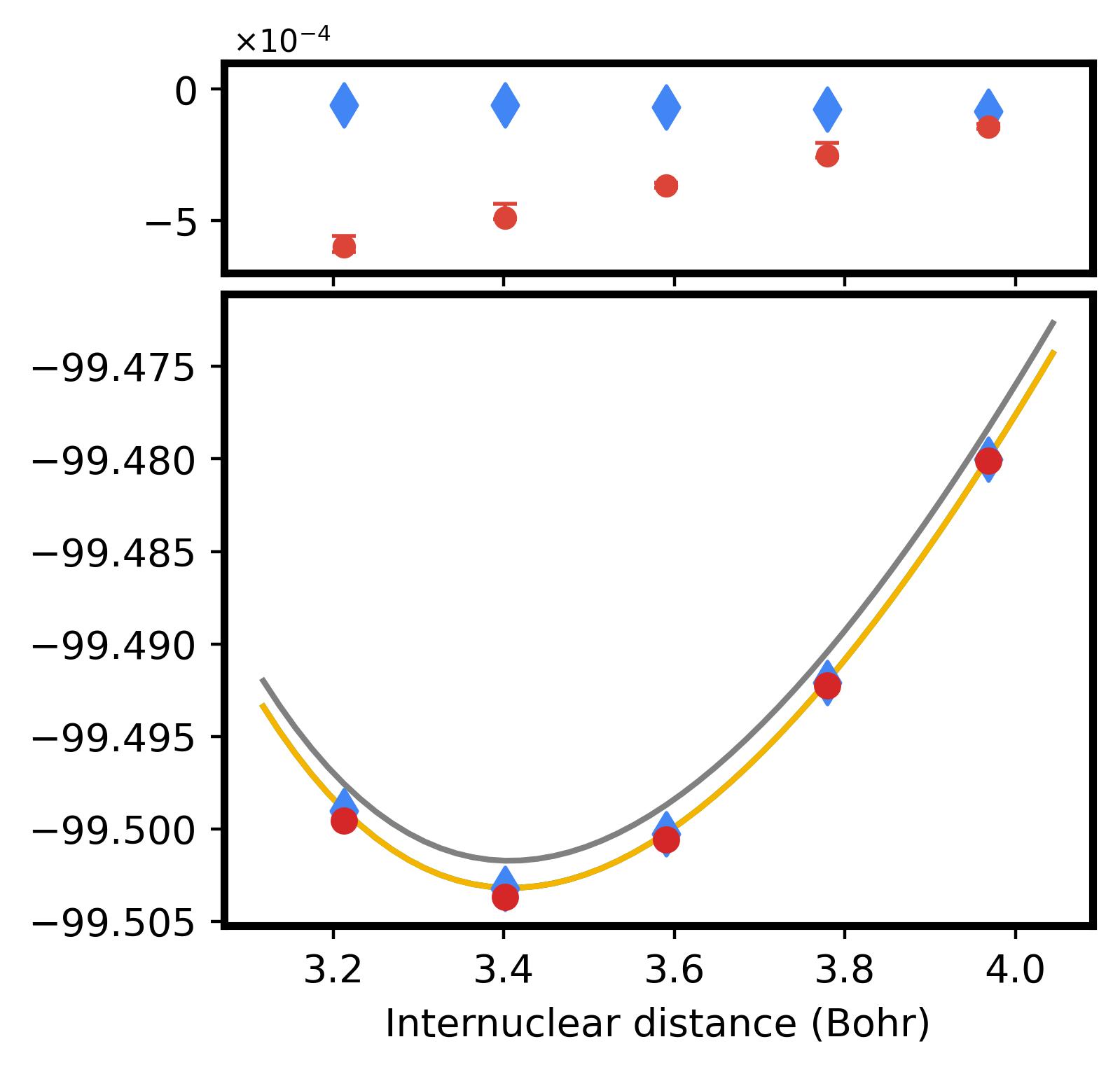}\\
(a) $H_2$ & (b) $LiH$ & (c) $HF$ \\
\end{tabular}
\caption{\label{fig:FIG6} (a) The potential energy curve of \textit{\ce{H2}} in the STO-6G basis obtained using the HHL algorithm (simulation and hardware) compared with results obtained using LCC, FCI, and the Hartree-Fock levels of theory on traditional computers. The grey line represents the Hartree-Fock energy, and serves as a visual reference that shows the size of correlation effects for each of the considered geometries. To avoid an excessively zoomed-out panel due to the diverse range of correlation energies across the geometries that we considered, the Hartree-Fock energy curve is shifted down vertically by a fixed quantity C$_1$(= -0.015 Ha), so that all the important components of computation are seen in the same graph. The FCI and LCC energy curves obtained from computations on a traditional computer are represented in the green and yellow curves respectively. The correlation energies obtained from hardware~(red circles with its error bars) and simulation~(blue diamonds) are summed with corresponding Hartree-Fock energies, yielding total energies, which we show in the plot. We also present the energy differences with respect to LCC energies from classical computers for each of the methods on the top panel. Potential energy curves obtained from AdaptHHLite (all-qubit fixing cases) of $LiH$ and $HF$ are presented in subfigures (b) and (c) respectively.} 
\end{figure*}

We now compute the correlation energies for \textit{\ce{H2}} in the STO-6G basis with 1000 shots and 10 repetitions for various values of the internuclear distances on the 11-qubit IonQ trapped-ion quantum hardware available via AWS Braket. The benchmark FCI values have been generated using the GAMESS-2014~\cite{GAMESS} software. Figure~\ref{fig:FIG6}(a) gives the potential energy curve of \textit{\ce{H2}} plotted around the vicinity of the equilibrium region. The HHL simulation and hardware experiments are computed by setting $n_r = 2$ keeping in mind the limitations of the current state-of-art quantum hardware. Increasing the number of qubits in the clock register improves the accuracy of $E_{corr}$ but translates to an increased number of gates and depth, thereby accumulating more hardware noise to the final result. For example, consider the case of \textit{\ce{H2}} in STO-6G basis with a bond length of 1.70 Bohr. Increasing $n_r$ from 2 to 4 improves the correlation energy difference with respect to LCC result on a traditional computer from 2 to 0.1 milliHartree. Since our hardware runs are proof-of-principle in nature, we make a deliberate trade-off between precision with depth. The  mean results for the correlation energies obtained using hardware experiments for various internuclear distances lie within $\sim$3 milliHartree difference with respect to LCCSD on a traditional computer and are presented in Figure~\ref{fig:FIG6}(a). The well-known lower bound nature of LCC energies with respect to configuration interaction~\cite{MUKHERJEE1981AnoteCPMET} is also validated in our results. Table A4 illustrates the mean correlation energies and standard deviation with respect to the mean for the results obtained on the trapped-ion device. The value of the standard deviation is at most 1.4 milliHartree over the multiple repetitions on hardware. \\ 

\subsubsection{\label{results-secD2}AdaptHHLite (NISQ variant) hardware results}

We now turn our attention to AdaptHHLite. For this purpose, we choose $LiH$ and $HF$ molecules, in the STO-6G basis. We carry out LCCD computations with one-hole and two-particle orbitals and two holes and one-particle orbitals for $LiH$ and $HF$ respectively. We chose five geometries in the neighbourhood of the equilibrium bond lengths for both systems. With a probability threshold of 0.8, we obtain all-qubit fixing for all of the geometries considered. The immediate consequence of all-qubit fixing as discussed in Section \ref{theory-secB2}b is that the calculation simplifies extensively. Therefore, the hardware results we obtain are of very high quality and agree with the LCC calculations carried out on classical computers to within $-$6 percent. Our results are presented in panels (b) and (c) of Figure~\ref{fig:FIG6}. The hardware computation is performed with 100 shots and 10 repetitions for each of the considered data points. \\ 

The preceding paragraph presented our AdaptHHLite results for three molecules, all of whose respective matrices were of $(4\times 4)$ size, with all-qubit fixing, on the IonQ Harmony device. We now briefly discuss our results for a $(4\times 4)$ case once again, but with an all-but-one multi-qubit fixing example. The reason for carrying out the exercise is to demonstrate the performance of our AdaptHHLite algorithm (the NISQ era variant) for a non-trivial case where all-qubit fixing is not possible. For this purpose, we choose the $H_2$ molecule in the 6-31G basis and at 2 Bohr bond length, where we achieve all-but-one fixing with a $P_{\rm th}=$0.8. The resulting AdaptHHLite quantum circuit has 171 single-qubit gates ($GPi$ and $GPi2$ gates) and 41 two-qubit $MS$ gates. We observed that the error rate was a staggering 50 percent with even 3000 shots on the Harmony machine, on which we had reported all our earlier results. We therefore executed these computations on the next generation Aria machine, and we found the error to be 26 percent with 5000 shots. After mitigating the dominant source of error in our results due to imperfect two-qubit gates using the zero noise extrapolation approach~\cite{zne_prl}, the error reduced to about 13 percent as presented in Figure~A4. \\

\subsection{On re-usability of clock register qubits} 

We briefly comment on a possible alternative to introducing more qubits to HHL algorithm via the Hong-Ou-Mandel module. An interesting offshoot of multi-qubit fixing is the observation that the associated clock register qubits that are fixed to 0 have no gates acting on them, although the fact that they are fixed to 0 influences the operations on the state register and the HHL ancilla register. Therefore, as long as the number of clock register qubits that are fixed to 0, $n_{(f,0)}$, is greater than or equal to $n_b$, we could reuse $n_b$ number of clock register qubits by preparing $|b\rangle$ for the Hong-Ou-Mandel module. However, one may not expect the number of clock register qubits to scale strongly with system size, as they are only related to the precision in eigenvalues. For large system sizes, when $n_b > n_{(f,0)}$, we could resort to using the other clock register qubits that are fixed to 1. Note that after the AdaptHHLite execution, the clock register qubits are not strictly $|0\rangle^{\otimes n_r}$. Therefore, while using those clock register qubits that are qubit-fixed to 0 is ideal, reusing those fixed to 1 is not, and introduces an error to the resulting correlation energy. In the extreme case of requiring more $n_b$-register qubits than there are multi-qubit fixed clock register qubits, additional qubits need to be introduced. Furthermore, if the quantum hardware permits mid-circuit qubit reinitialization, then one could reuse the $n_r$ appropriately by reinitializing them to $\ket{0}^{\otimes n_r}$ and encode $\ket{b}$ for Hong-Ou-Mandel module computation. \\ 

\subsection{A comparison between QPE-CASCI and HHL-LCCSD: coexistence across scales} 

Finally, we compare HHL-LCC and QPE-CASCI (with CASCI standing for complete active space CI, where one does full CI within an active space, and which is much more commonplace in practical CI calculations than FCI), both on the grounds of cost and accuracy. \\ 

Although HHL involves two QPE routines, thus appearing seemingly more expensive relative to QPE, this is true if both the algorithms have the same qubit count. One can see that for the same molecular system in the same single particle basis and the same active space, HHL-LCCSD (we choose SD as an example) incurs fewer qubits, and hence lower cost than QPE-CASCI. The number of qubits in the state register for QPE-CASCI is the number of spinorbitals, whereas in the case of HHL-LCC, it is the $\lceil log_2 \rceil$ of the number of excitations, which are defined in our work in terms of spatial orbital substitutions. This exponential suppression is precisely why the number of state register qubits grow slower in HHL-LCC relative to QPE-CASCI. \\

We consider now a few examples for number of qubits incurred on state register for QPE-CASCI versus HHL-LCCSD (with the number of excitations chosen to be $\sim n_h^2n_p^2$), in a given basis set to illustrate the growth pattern of the qubit number: (a) \textit{LiH} STO-3G: 12 and 6 for an all-electron-all virtual (ae-av) computation, and with an active space of 2 occupied spinorbitals and 4 virtuals (denoted hereafter as [2,4]), it is 6 and 2, (b) \textit{LiH} cc-pVDZ: 38 and 11 for an ae-av computation, and 12 and 5 for active space choice of [2,10], (c) \textit{RbH} Sapporo dzp basis: 80 and 18 for an ae-av calculation, and 28 and 11 for [10,18] active space, and (d) \textit{RbH} Sapporo qzp basis: 224 and 22 for an ae-av calculation. \\ 

We note that we mention `CASCI’ and not truncated CI such as CISD or CISDT while comparing QPE and HHL, as there is no straightforward way that we are aware of to carry out truncated CI using QPE. Note that this aspect indicates that the HHL algorithm offers flexibility in terms of number of qubits (and therefore depth) for truncated LCC computations, whereas QPE-CASCI does not. \\

We now turn our attention to the accuracy aspect, the crux of which is that QPE and HHL are not to be viewed as competitors, but rather algorithms that excel at different system sizes, due to the fewer qubits involved on HHL-LCC. Therefore, while LCCSD is inferior to CASCI but is still superior to the HF method, it can always be employed to capture correlation effects for larger system sizes than QPE-CASCI can, for a given number of qubits that some era's hardware can at most accommodate. For example, if at some point in the future, a quantum hardware can accommodate 1000 qubits reasonably (with sufficiently good two-qubit gate fidelity, etc), QPE-CASCI allows one to explore systems where the number of spinorbitals is a few hundreds whereas HHL-LCCSD offers the scope to probe correlation effects in larger systems with substantially more spinorbitals but with lesser accuracy than CASCI. In this sense, HHL-LCCSD and QPE-CASCI can be thought of as being analogous to density functional theory and CCSD respectively on classical hardware. It is also important to note that the CC method (as well as its linearized counterpart) scales correctly with increase in system size. That is, it is rigorously size extensive due to the presence of completely connected terms, while a truncated CI will not be so (where some future quantum algorithms can handle truncated CI efficiently). This is also why CC is preferred over CI in the classical computing scenario. On a related note, the well-known interesting lower bound feature for the energy of a system using LCC means that the theory does possess seemingly good predictive capability, and can yield energies even lower than those predicted by CISD. \\

Lastly, we comment on future scope for extensions of our work stemming from our pilot study. We note that our work opens a new direction for quantum chemistry on quantum computers, outside the scope of VQE and QPE. Since this is a pilot study, we envisage that there will be future works in this direction that improve accuracy and go beyond linearized CC by including non-linear terms in a systematic manner. \\ 

\section{\label{conclusion}Conclusion} 

In summary, we recognize that the HHL algorithm can be utilized in quantum many-body theory (in this work, the linearized coupled cluster method) to compute ground state correlation energies of molecular systems. We customize the HHL algorithm to cater to the needs of different quantum computing eras-- NISQ, late-NISQ, early fault-tolerant, and fault-tolerant. In particular, our HHL variants integrate two aspects--- (a) given a Hermitian matrix, we demonstrate how to scale down the matrix efficiently without precomputing its eigenvalues and choose the coefficient $c$ in Eq.~\eqref{cr} conveniently with a reasonably low loss in precision, and (b) we then rigorously aim at reducing the quantum circuit depth and tailor the HHL modules to suit various timelines of quantum computing. Central to this work is the variant called AdaptHHLite, with the `Adapt' part consistently integrated into the algorithm in all of the eras. The `Lite' component assumes different forms in the NISQ, late-NISQ, and early fault-tolerant timelines while becoming obsolete in the fault-tolerant era. \\ 

The `Adapt' part introduces a novel scaling scheme that in turn decides the controlled-rotation angles through a coefficient $c$ in the HHL algorithm. Instead of having to choose the so-called coefficient $c$ in Eq.~\eqref{cr} by trial and error, we lay down a more stringent yet powerful condition that allows one to pick $c$ suitably, resulting in correlation energies with reasonably high precision. Our simple yet subtle modification altogether eliminates the classical overhead of $\mathcal{O}(N^3)$ required for calculating eigenvalues of the matrix, $A$, thus making the `Adapt' part suitable for all quantum computing eras. On the other hand, the `Lite' component draws upon classical multi-qubit fixing and pipeline-based quantum circuit optimization to reduce quantum resources in the NISQ era. For the late NISQ period, the classical fixing can be superseded by quantum multi-qubit fixing with/without circuit optimization schemes. For the early fault-tolerant era, we recommend executing an LMR module-based quantum algorithm in place of quantum multi-qubit fixing to achieve the desired resource reduction. We note at this juncture that the scaling strategy that we introduce in the AdaptHHL framework broadens the reach of the algorithm to encompass problems across various domains. \\ 

We calculated correlation energies of several light molecular systems (five molecules, each in five geometries) using the HHL, HHLite, PerturbedHHL, PerturbedHHLite (with the Perturbed variants being less powerful cousins of the Adapt versions), AdaptHHL, and AdaptHHLite algorithms (NISQ version), in order to compare each variant’s performance in balancing precision with resource reduction. We add a Hong-Ou-Mandel module at the end of all of these implementations to extract the correlation energy. For the molecules considered in our work, we deal with matrix (which we call $A$) sizes that are $(2\times2)$, $(4\times4)$, and $(16\times16)$, resulting in computations on 5, 11, and 17 qubits, respectively. Our findings demonstrate that HHLite can yield a depth compression of as much as 84.3 percent while incurring a precision loss of 0.4 milliHartree in $\sim$25 milliHartree correlation energy. On the other hand, AdaptHHLite can lead to a substantial compression of at most 89.6 percent, while compromising on precision by about 2 milliHartree in a total of 27.5 milliHartree. We observe that, more generally, our scaling approach in AdaptHHLite aids in substantial circuit depth reduction yet yields results with reasonable loss in precision. Our hardware results on the 11-qubit IonQ Harmony device show that (a) the HHL algorithm predicts correlation energies of the $H_2$ molecule ($(2\times2)$ size matrix for $A$) in the 0.03 to 16.59 percent precision band, and (b) the AdaptHHLite (NISQ variant) algorithm, in the event of an all qubit-fixing case, for the $LiH$ and $HF$ molecule ($(4\times4)$ size matrix for $A$), yields precise results to within $-$6 percent trivially. \\ 

In conclusion, we note that our algorithmic enhancements could be applicable across multiple domains going beyond quantum chemistry, and across different quantum computing eras. Furthermore, our study has the potential to unveil new possibilities for exploring quantum many-body methods in the field of quantum chemistry without the restriction of unitarity in existing quantum algorithms. \\ 

\begin{acknowledgments}
The authors thank Prof. Bhanu Pratap Das for conceptual discussions on the LCC aspects of the work. S.R. thanks Prof. Ian R. Petersen for confirming a property of positive definite matrices. N.B. thanks Dr. Olivia Di Matteo for useful discussions on IonQ native gate set. The classical computations were done on Rudra (SankhyaSutra Labs) supercomputers. The AWS Braket platform was used for cloud access to IonQ hardware through the credits provided by the MeitY QCAL project (N-21/17/2020-NeGD, 2022-2024). K.S. acknowledges support from JST PRESTO “Quantum Software” project (Grant No. JPMJPR1914), Japan, KAKENHI Scientific Research C (21K03407) and Transformative Research Area B (23H03819) from JSPS, Japan, Center of Innovations for Sustainable Quantum AI (Grant No. JPMJPF2221) from JST, Japan, and Quantum Leap Flagship Program (JPMXS0120319794) from the MEXT, Japan. 
\end{acknowledgments}

\bibliography{references}

\newpage

\appendix

\section*{Appendix}

\setcounter{figure}{0}
\renewcommand\theequation{A\arabic{equation}}
\setcounter{equation}{0}
\renewcommand\thesubsection{A\arabic{subsection}}

\renewcommand{\thetable}{A\arabic{table}}
\renewcommand{\thefigure}{A\arabic{figure}}
\renewcommand{\theHfigure}{A\arabic{figure}}

\subsection{\label{TCC}Traditional coupled cluster theory}

In this subsection, we present a pedagogical yet brief description of the coupled cluster (CC) theory, beginning by introducing the spin-free linearized CC (LCC) framework that we work with, followed by recasting the LCC equations as a system of linear equations, and finally expressing the quantity of interest to us, namely the correlation energy in the required form to implement in a quantum algorithm. \\ 

Substituting the CC wavefunction ansatz from {Eq.~(1) of the main text, that is, $\ket{\Psi} = \text{e}^{\hat{T}} \ket{\phi_0}$, onto the Schr\"{o}dinger equation gives 
\begin{equation}\label{SRCC ansatz in schrod}
    \hat{H} \text{e}^{\hat{T}}\ket{\Phi_0}=E \text{e}^{\hat{T}}\ket{\Phi_0}. 
\end{equation}
By writing the Hamiltonian operator in a normal ordered form with respect to the Hartree-Fock function, one may subtract the vacuum energy from Eq. \eqref{SRCC ansatz in schrod} to get 
\begin{equation}\label{normal ordered SRCC}
    \{\hat{H}\} \text{e}^{\hat{T}}\ket{\Phi_0}=E_{corr} \text{e}^{\hat{T}}\ket{\Phi_0}, 
\end{equation}
where $E_{\text{corr}}$ denotes the electron correlation energy and $\{\hat{H}\}$ denotes the normal ordered portion of the Hamiltonian.\\

Left multiplying with e$^{-{\hat{T}}}$ on both sides of Eq. \eqref{normal ordered SRCC} we get,
\begin{align}
    \text{e}^{-{\hat{T}}}\{\hat{H}\}\text{e}^{\hat{T}}\ket{\Phi_0}=E_{corr}\ket{\Phi_0} \\
    \{\overline{H}\}\ket{\Phi_0}=E_{corr}\ket{\Phi_0}. \label{SRCC eqn}
\end{align}
$\overline{H}$ is the similarity transformed Hamiltonian which can be expressed in terms of commutators following the Baker-Campbell-Hausdorff (BCH) expansion,
\begin{align}
    \overline{H}=\hat{H}+[\hat{H},{\hat{T}}]+\frac{1}{2!}[[\hat{H},{\hat{T}}],{\hat{T}}]+\frac{1}{3!}[[[\hat{H},{\hat{T}}],{\hat{T}}],{\hat{T}}]+\cdots.
\end{align}
The commutators are evaluated using the Wick's theorem and a commutator $[\hat{H},\hat{T}]$ can also be represented as a Wick contraction $\contraction{}{\hat{H}}{}{\hat{T}}{\hat{H} \hat{T}} - \contraction{}{\hat{T}}{}{\hat{H}}{\hat{T} \hat{H}}$. Since all the cluster operators are normal ordered with respect to $\ket{\Phi_0}$, different components of the cluster operators commute among themselves and they can only contract with $\hat{H}$ from the right side resulting in a single term $\contraction{}{\hat{H}}{}{\hat{T}}{\hat{H} \hat{T}}$. The Hamiltonian consists up to two-body terms and hence the BCH expansion of the similarity transformed Hamiltonian terminates at the quartic power of T.\\

Continuing from Eq. (\ref{SRCC eqn}) in the LCC approximation, we find that the LCC equations are 

\begin{equation}\label{LCCopeqn}
    \{(\hat{H}+\contraction{}{\hat{H}}{}{\hat{T}}{\hat{H} \hat{T}})\}\ket{\Phi_0}=E_{corr}\ket{\Phi_0}.
\end{equation}
We now write the operators involved in Eq. \eqref{LCCopeqn} in terms of a set of composite indices \textit{P,Q,$\cdots$}, where each such index represents the orbital substitutions involved in a specific kind of excitation, i.e. single excitations, double excitations, etc. 
The equation for the set of cluster amplitudes \textit{t$_P$} is obtained by left-projecting Eq. \eqref{LCCopeqn} by a set of excited functions, say \textit{$\chi_P$}, $\forall$ P. Here the excited functions are defined as  $ \ket{\chi_P} = \{\hat{e}_P\}\ket{\Phi_0}$, and after projection, we have 
\begin{equation} \label{LCCtampeqn}
    \bra{\chi_P}\hat{h}\ket{\Phi_0} + \sum_Q \bra{\chi_P} \hat{h} \{\hat{e}_Q\} \ket{\Phi_0}t_Q=0,\ \  \forall P. 
\end{equation}
In the above equation, $\hat{h}$ refers to the Hamiltonian operator. The above equation can be rearranged to give 
\begin{equation}
    \sum_Q\bra{\chi_P} \hat{h}\ket{\chi_Q}t_Q=-\bra{\chi_P}\hat{h}\ket{\Phi_0}. 
\end{equation}
\\ 


We finally discuss the expression for the LCC correlation energy. It is found by left-projecting Eq. \eqref{LCCopeqn} by $\ket{\Phi_0}$. Since the first term on the right hand side of Eq. \eqref{LCCopeqn} is the operator part of the Hamiltonian, it vanishes upon projection by $\ket{\Phi_0}$ and we are left with 
\begin{eqnarray}
    \bra{\Phi_0} \hat{h} \{\hat{e}_P\} \ket{\Phi_0}t_P&=&E_{corr} \nonumber \\
    \bra{\Phi_0} \hat{h} \ket{\chi_P}t_P&=&E_{corr} . \label{LCCecorreqn}
\end{eqnarray}

\begin{figure}[h]
\includegraphics[width = \columnwidth]{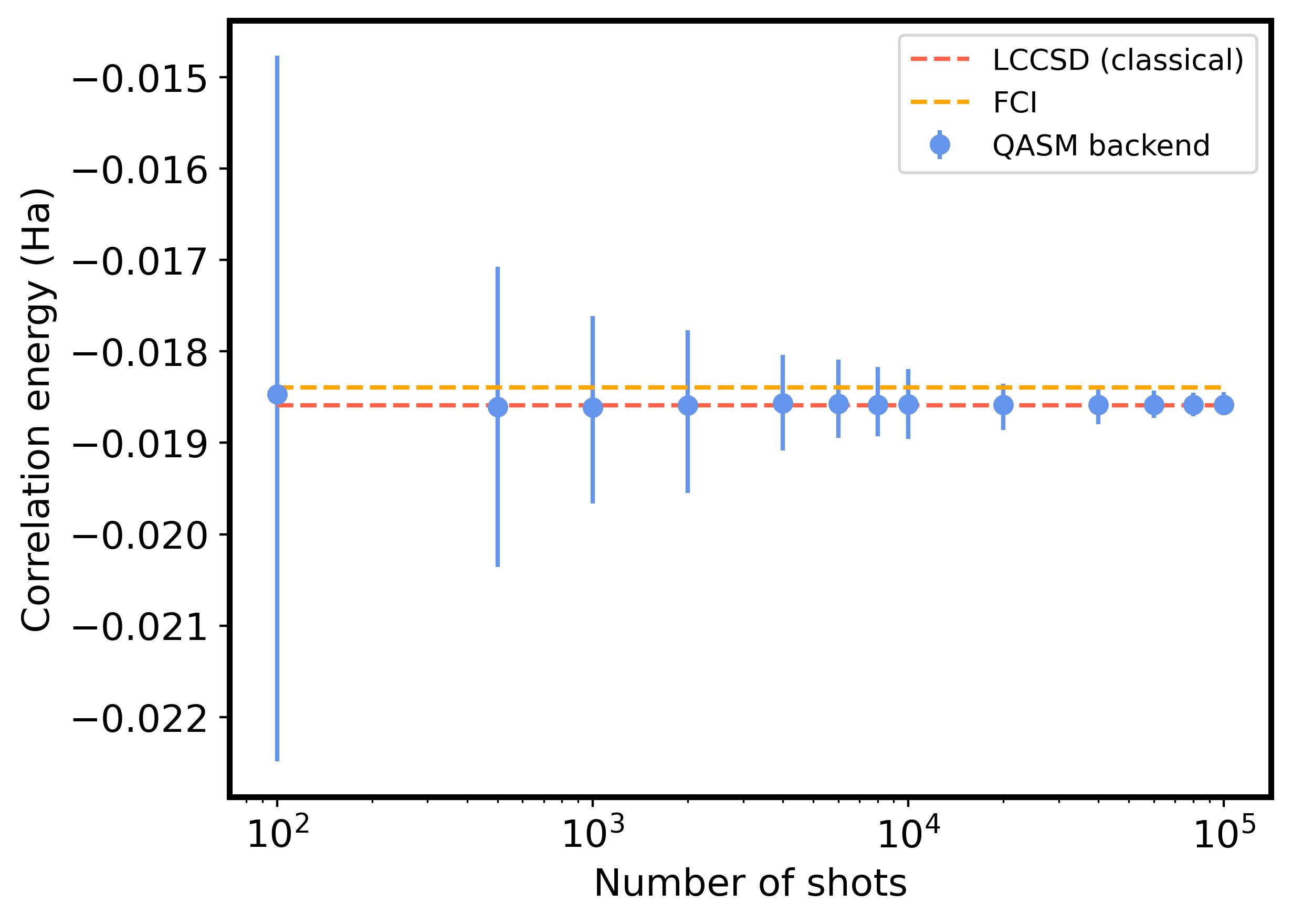}
\caption{\label{fig:figs4} Correlation energy versus number of shots for \textit{\ce{H2}} molecule in 1.30 Bohr geometry and in the STO-6G basis, with each data point repeated $200$ times. The orange and red dashed lines signify the correlation energies obtained using FCI and LCCSD approaches respectively. The mean $E_{corr}$ values are represented by blue circles and their respective error bars are given by the blue vertical lines. }
\end{figure}

\begin{figure*}
\begin{tabular}{c c}
    \quad \quad \includegraphics[scale=0.5]{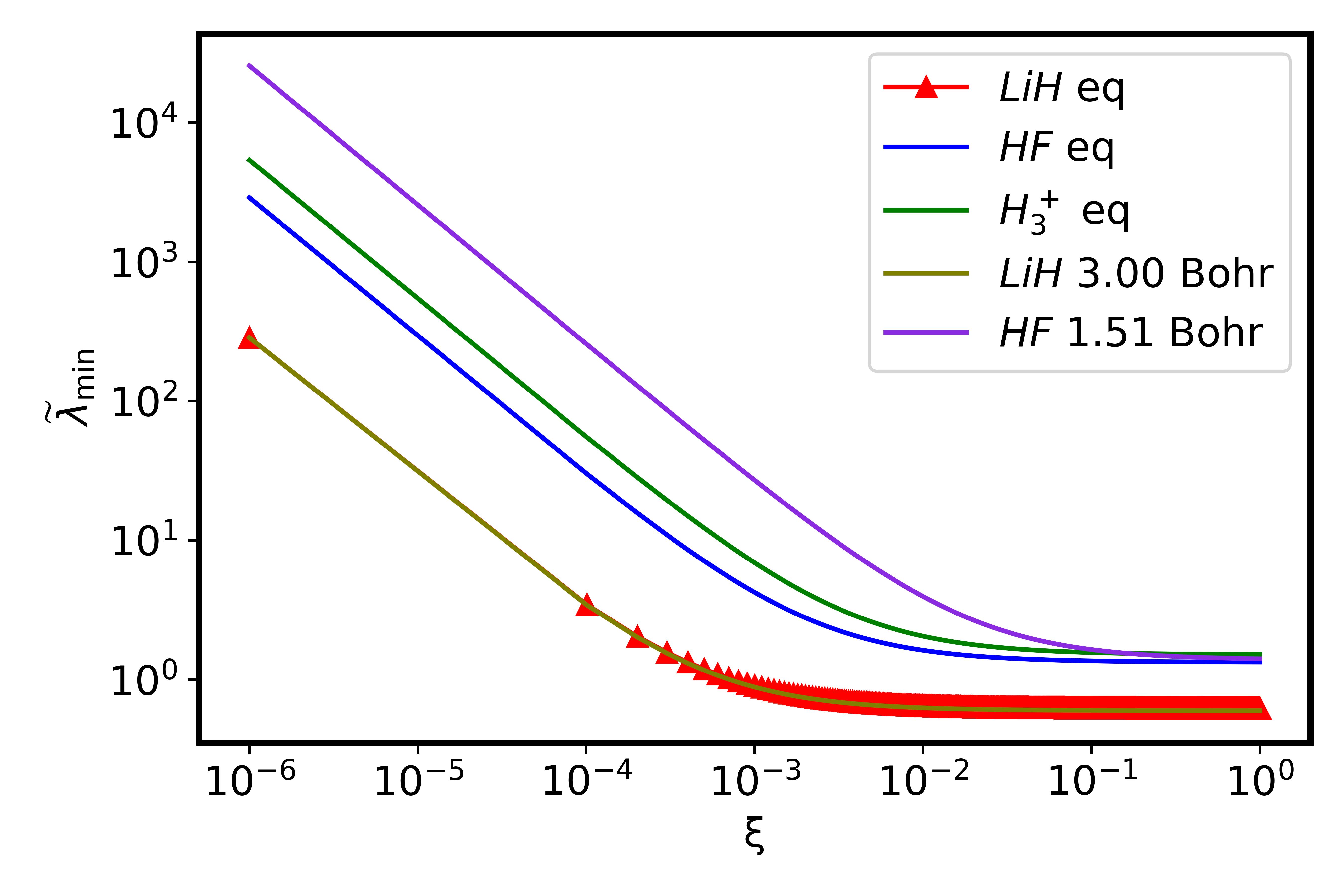} \quad \quad & \quad \quad \includegraphics[scale=0.5]{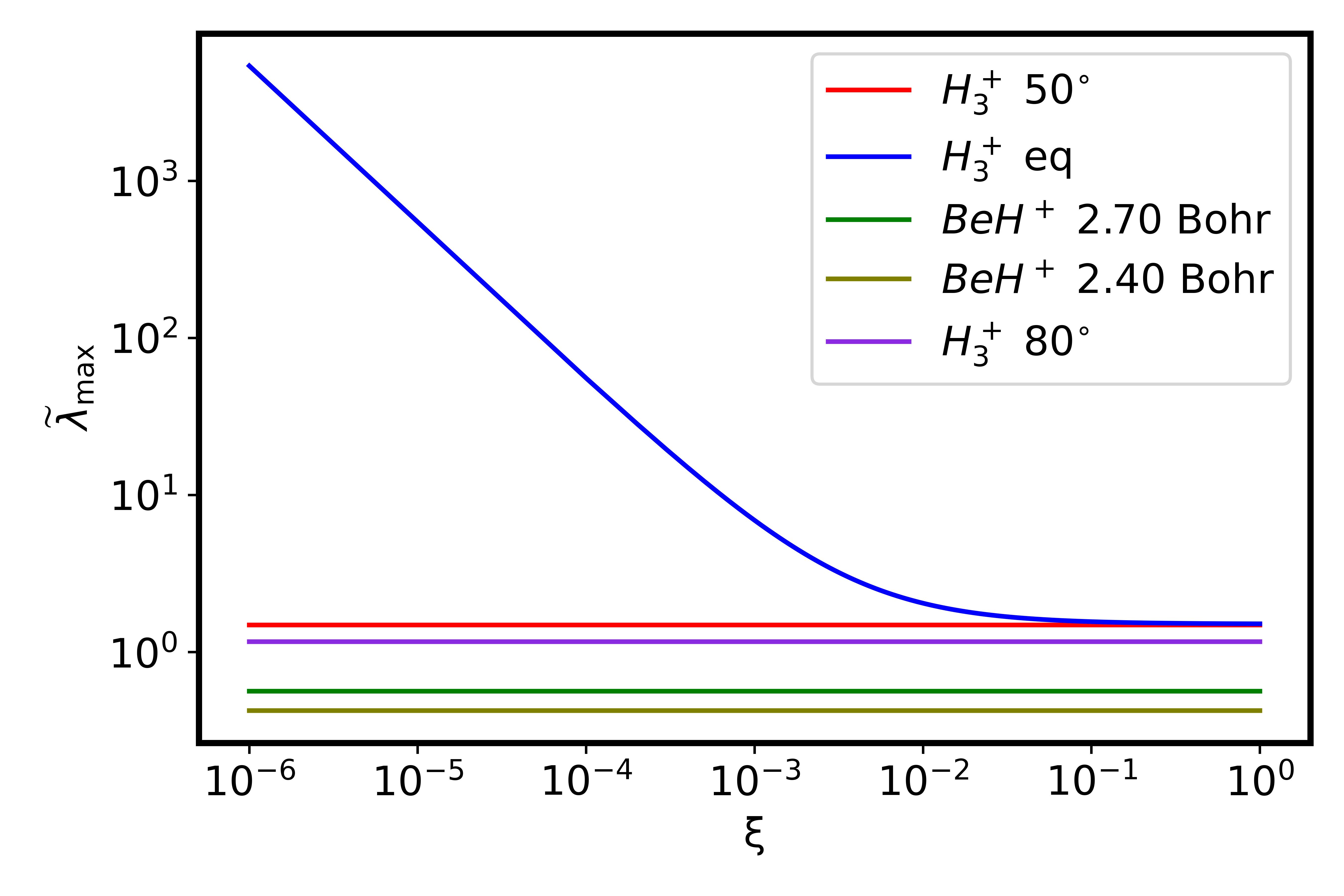} \quad \quad \\
    (a) & (b) \\
\end{tabular}
    \caption{\label{fig:figs3}{Variation of (a) $\tilde{\lambda}_{min}$  (b) $\tilde{\lambda}_{max}$ with the choice of $\xi$ for five representative molecular geometries, in the context of the PerturbedHHL and the PerturbedHHLite algorithms. In the legend, `Eq' refers to the equilibrium geometry for a given molecule. }}
\end{figure*}
\begin{figure*} 
\begin{tabular}{c c c}
\includegraphics[scale=0.40]{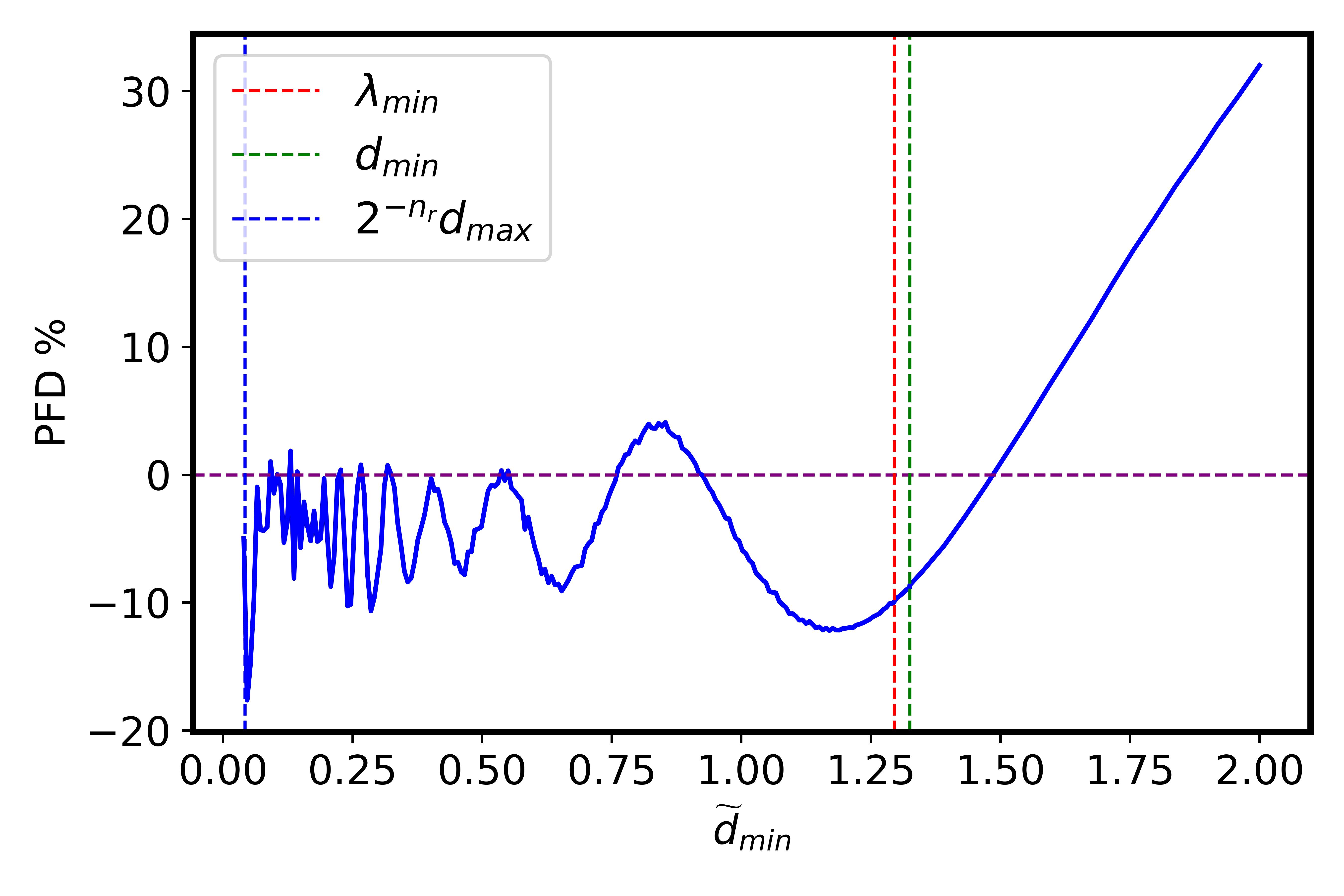} & \includegraphics[scale=0.40]{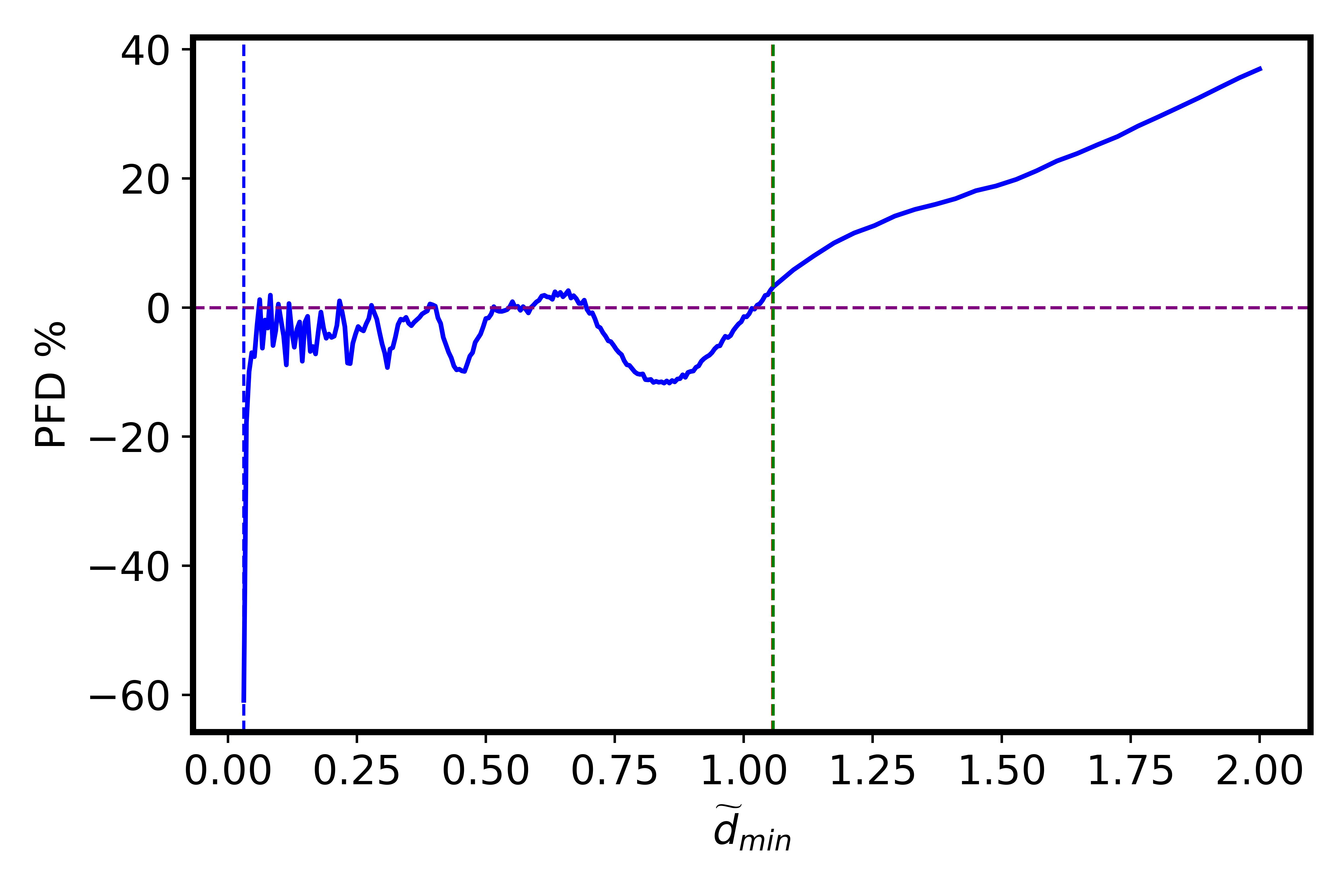} & \includegraphics[scale=0.40]{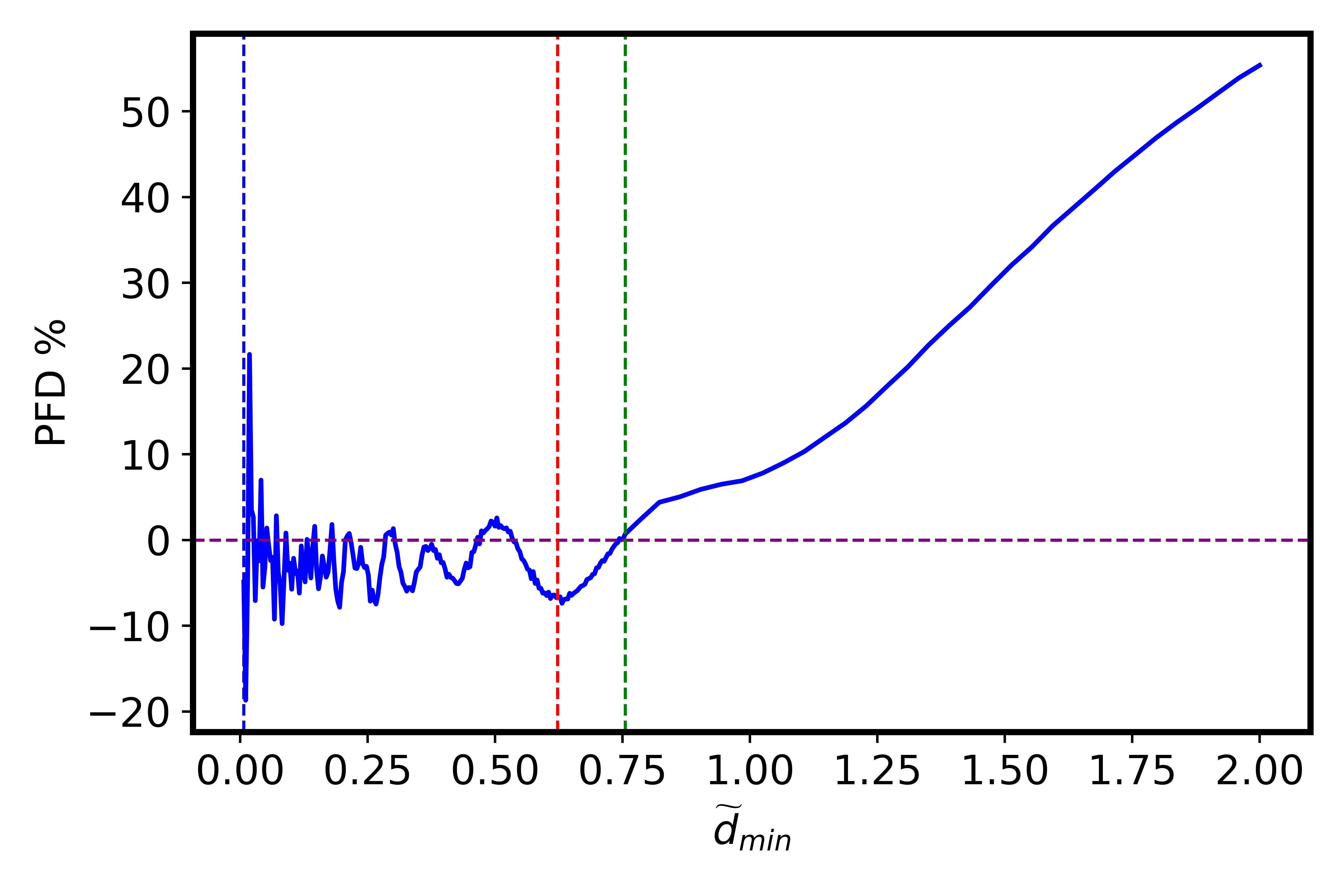} \\
(a) \textit{\ce{H3+}} 70$^{\circ}$ & (b) \textit{\ce{H2}} 1.50 Bohr & (c) \textit{\ce{BeH+}} 2.50 Bohr \\ \\
\includegraphics[scale=0.40]{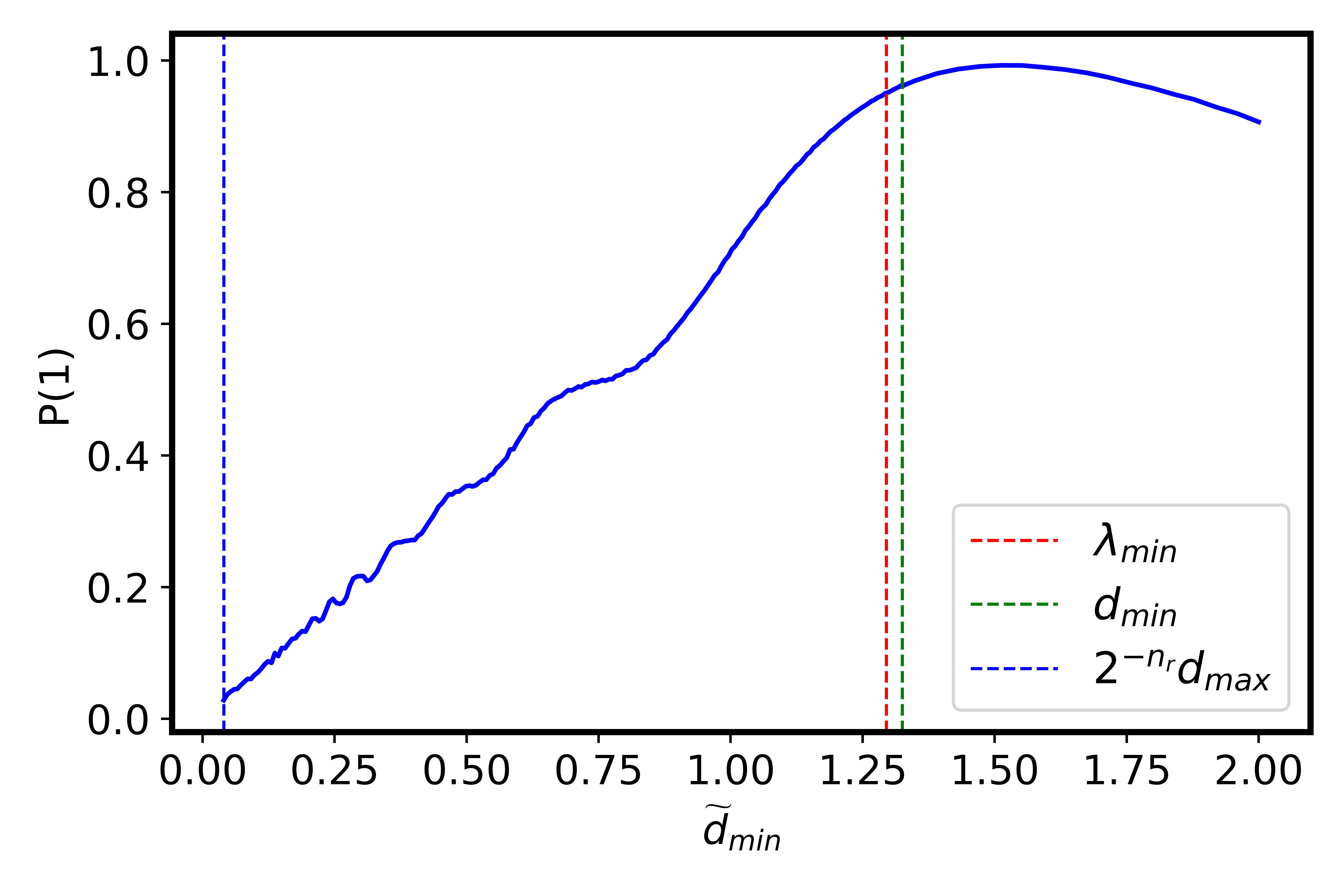} & \includegraphics[scale=0.40]{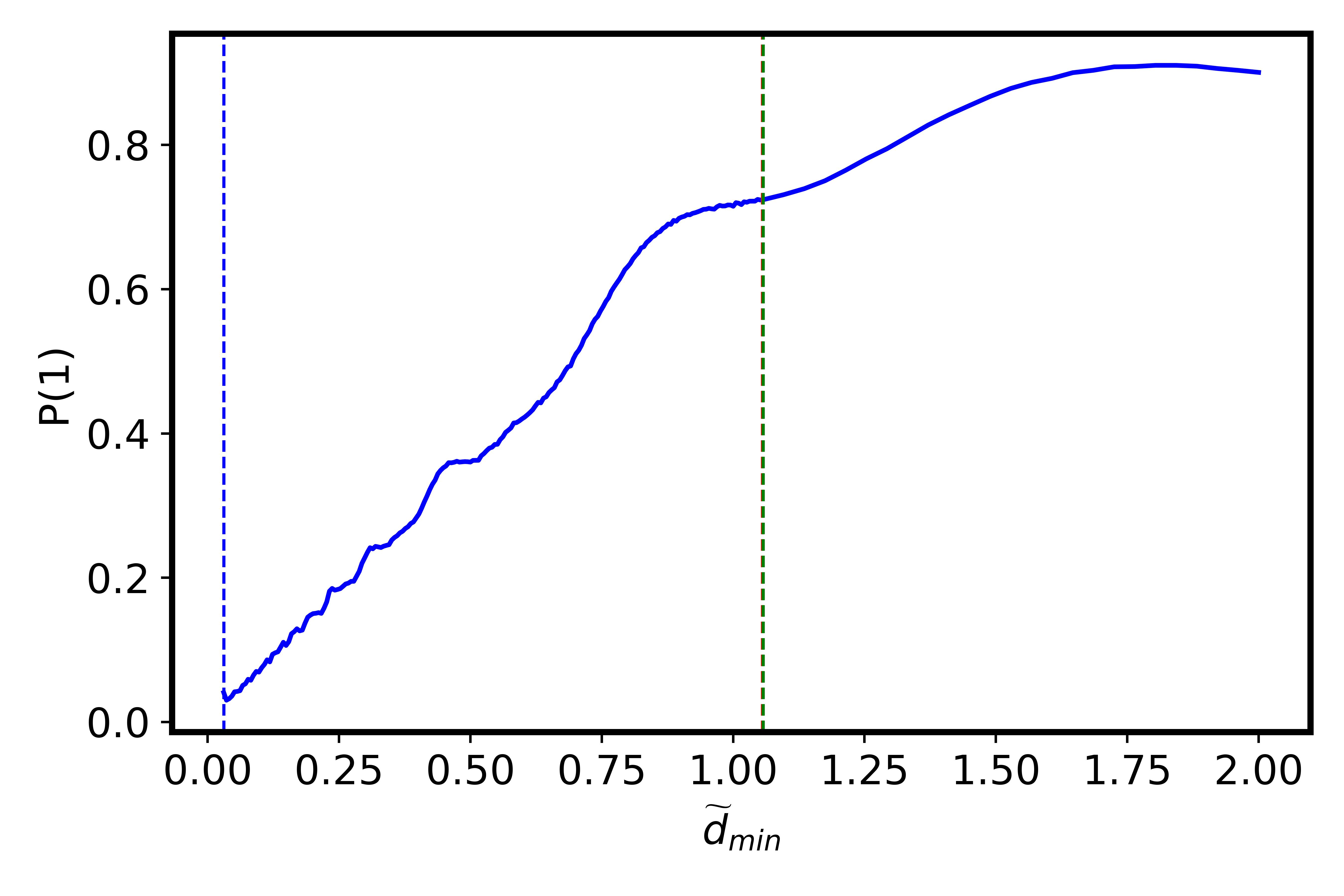} & \includegraphics[scale=0.40]{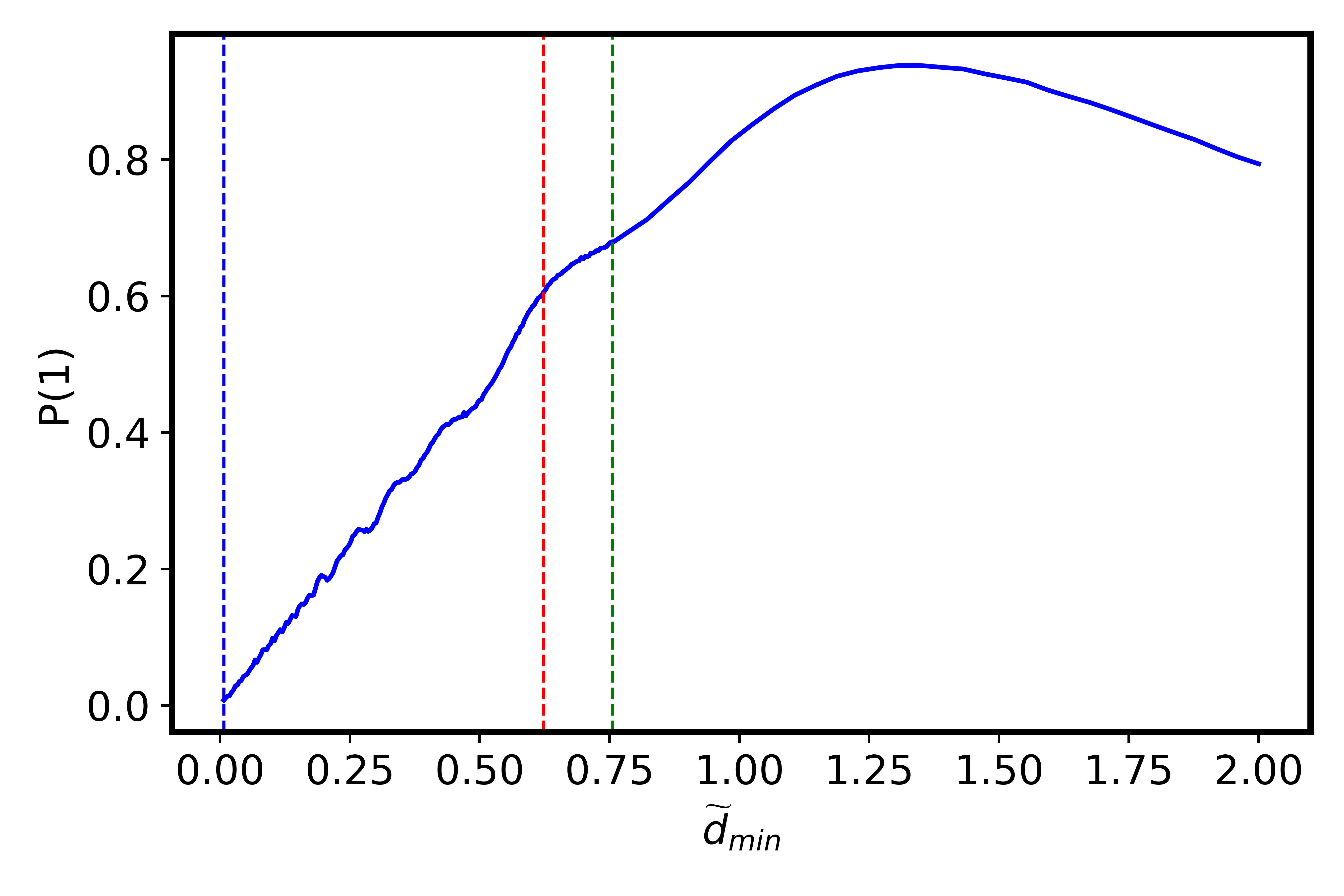} \\
(d) \textit{\ce{H3+}} 70$^{\circ}$ & (e) \textit{\ce{H2}} 1.50 Bohr & (f) \textit{\ce{BeH+}} 2.50 Bohr \\
\end{tabular}
\caption{\label{fig:figs5} Variation of PFD and prescaled $\norm{x}$, which is $P(1)$, with $\tilde{d}_{min}$ for three representative molecules, \textit{\ce{H3+}} at 70$^{\circ}$ (subfigures (a) and (d)), \textit{\ce{H2}} at 1.50 Bohr bond length (subfigures (b) and (e)), and \textit{\ce{BeH+}} at 2.50 Bohr bond length (subfigures (c) and (f)), which yield the worst, the in-between, and the best PFDs respectively in our numerical results with $\tilde{d}_{min}=d_{min}$. The blue and the green lines show the bounds that we set in Eq.~(21) in Section IIB2 of the main text and therefore the relevant search range, while the red line shows $\lambda_{min}$. One can see from the top panels that the relevant search range could be narrowed down by ignoring the part of the curve corresponding to `high frequency' oscillations and instead carry out a coarse grained (and hence `low-cost') search in the `low frequency' regime. To that end, a threshold at approximately midway between the bounds can be set. It is important to stress that the analysis is very preliminary in nature, and further analysis on a resource-efficient choice of $\tilde{d}_{min}$ without having to rely on a priori knowledge of LCC results from traditional computing is deferred to a future work. }
\end{figure*} 


\begin{figure}
\begin{tabular}{c}
  \includegraphics[width = \columnwidth]{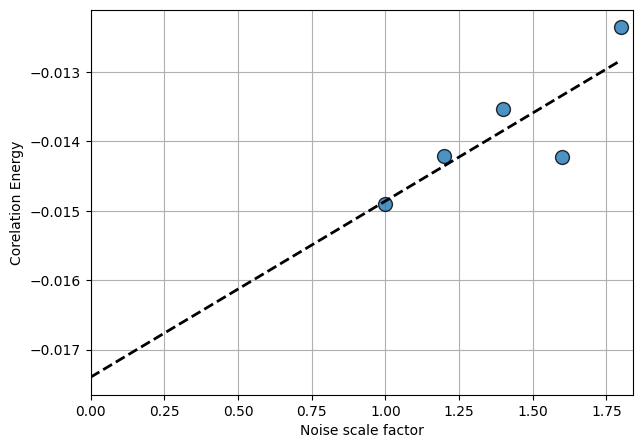} \\
\end{tabular}
\caption{\label{fig:figs200}{Our zero noise extrapolated (ZNE) result for the $H_2$ molecule in the 6-31G basis and at 2 Bohr bond length. We achieve all-but-one fixing with a $P_{\rm th}=$0.8. For ZNE, we choose local folding of only the two-qubit gates (MS gates) from the left, with non-integer scale factors (1, 1.2, 1.4, 1.6, and 1.8), and a linear fit to our data points. Further, each data point presented in the figure is an average over three repetitions, each with 5000 shots. We note that the computations were carried out on the IonQ Aria device, unlike the other less complicated calculations (with all-qubit fixing) for which we used the older Harmony machine. This is in view of the former yielding a PFD of 25.71 for scale factor 1 for 5000 shots, whereas the latter gave about 38 even with 10000 shots. After ZNE, the PFD improved from 25.71 to 13.24. We forecast that as state-of-the-art quantum devices improve in the coming years, much better PFDs can be achieved with these mitigation techniques. }}
\end{figure}

\begin{table*}[t]
\caption{\label{table1} The table presents the relevant data associated with our HHL and HHLite computations for the chosen molecules. The first two columns provide the molecule along with the choice of single-particle basis as well as the information on molecular geometries. $^*$ represents the equilibrium geometry. $N_t(= 2\times n_b + n_r + 1)$ is a shorthand for the total number of qubits, while $E^{clc}_{corr}$ refers to the correlation energy obtained from LCC calculation on a traditional computer. This serves as the benchmark for our computations. `2Q' is the number of two-qubit gates involved in a computation, and PFD ($ = \frac{E_{corr}^{clc}-E_{corr}^X}{E_{corr}^{clc}} \times 100$) is a percentage fraction difference, where $X \in$ \{HHL, HHLite\}. $n_f$ is the total number of qubit fixings for a given calculation. The correlation energies are given in units of milliHartree, and are rounded off to 3 decimal places. }
 \begin{tabular}{ c c c c| c c c c| c c c c c }
    \hline \hline
        Molecule  & Geometry & $N_t$ & $E^{clc}_{corr}$  &\multicolumn{4}{ c |}{HHL}  & \multicolumn{5}{ c }{HHLite}  \\
        (Basis)& & ($n_b$, $n_r$) &  &  Depth & \quad 2Q \quad & \quad $E_{corr}$ \quad & \quad $E_{diff}$(PFD)\quad  &  Depth  & \quad 2Q \quad & \quad $E_{corr}$ \quad & \quad $E_{diff}$(PFD) \quad & \quad $n_f$ \quad \\ 
    \hline
    
        \textit{\ce{H2}} & 1.20 & 11(2,6) & -10.207 & 2452  & 536   & -10.001 & -0.206 (2.02) & 899   & 339   & -10.235 & 0.028 (-0.28) & 2 \\
        (6-31G) & 1.30 & 11(2,6) & -10.813 & 2452  & 536   & -10.485 & -0.328 (3.04) & 572   & 199   & -10.793 & -0.020 (0.19) & 4 \\
                & 1.40$^*$ & 11(2,6) & -11.546 & 2452  & 536   & -11.440 & -0.106 (0.92) & 635   & 240   & -11.606 & 0.060 (-0.52) & 3 \\
                & 1.50 & 11(2,6) & -12.437 & 2452  & 536   & -12.241 & -0.196 (1.57) & 1095  & 399   & -12.556 & 0.119 (-0.96) & 1 \\
                & 2.00 & 11(2,6) & -20.053 & 2452  & 536   & -19.929 & -0.124 (0.62) & 621   & 211   & -20.037 & -0.016 (0.08) & 4 \\ \hline
        \textit{\ce{H3+}} & 1.727, 50$^{\circ}$ & 11(2,6) & -24.425 & 2452  & 536   & -24.133 & -0.292 (1.19) & 710   & 243   & -24.492 & 0.067 (-0.27) & 4 \\
         (STO-6G)& 1.727, 55$^{\circ}$ & 11(2,6) & -25.394 & 2452  & 536   & -25.367 & -0.027 (0.11) & 384   & 134   & -24.972 & -0.422 (1.66) & 6 \\
                 & 1.727, 60$^{\circ *}$ & 11(2,6) & -26.197 & 2452  & 536   & -25.850 & -0.347 (1.33) & 672   & 240   & -26.326 & 0.129 (-0.49) & 4 \\
                 & 1.727, 70$^{\circ}$ & 11(2,6) & -27.423 & 2452  & 536   & -27.215 & -0.208 (0.76) & 1052  & 368   & -27.282 & -0.141 (0.51) & 2 \\
                 & 1.727, 80$^{\circ}$ & 11(2,6) & -28.315 & 2452  & 536   & -28.345 & 0.030 (-0.11) & 704   & 236   & -28.342 & 0.027 (-0.09) & 4 \\  \hline
        \textit{\ce{LiH}} & 2.60 & 17(4,8) & -17.464 & 97656 & 18644 & -17.016 & -0.448 (2.56) & 42283 & 14970 & -16.579 & -0.885 (5.06) & 3 \\
        (STO-6G) & 2.70 & 17(4,8) & -17.948 & 97557 & 18616 & -17.648 & -0.300 (1.67) & 42464 & 15012 & -16.984 & -0.964 (5.37) & 3 \\
                 & 2.80$^*$ & 17(4,8) & -18.536 & 98013 & 18698 & -18.282 & -0.254 (1.37) & 46096 & 13273 & -17.615 & -0.921 (4.97) & 2 \\
                 & 2.90 & 17(4,8) & -19.227 & 97990 & 18704 & -19.147 & -0.080 (0.42) & 46204 & 16394 & -18.116 & -1.111 (5.78) & 2 \\
                 & 3.00 & 17(4,8) & -20.024 & 97479 & 18590 & -19.986 & -0.038 (0.19) & 36861 & 13070 & -19.691 & -0.333 (1.66) & 4 \\ \hline
        \textit{\ce{BeH+}} & 2.40 & 17(4,8) & -18.963 & 98189 & 18728 & -18.966 & 0.003 (-0.02) & 45431 & 16134 & -18.452 & -0.511 (2.70) & 2 \\
        (STO-6G)  & 2.50 & 17(4,8) & -20.278 & 98025 & 18708 & -20.286 & 0.008 (-0.04) & 53895 & 19130 & -20.285 & 0.007 (-0.04) & 0 \\
                  & 2.54$^*$ & 17(4,8) & -20.848 & 97444 & 18606 & -20.833 & -0.015 (0.07) & 41596 & 14764 & -20.149 & -0.699 (3.35) & 3 \\
                  & 2.60 & 17(4,8) & -21.752 & 97950 & 18690 & -21.761 & 0.009 (-0.05) & 32823 & 11632 & -20.810 & -0.942 (4.33) & 5 \\
                  & 2.70 & 17(4,8) & -23.391 & 96675 & 18476 & -23.395 & 0.004 (-0.02) & 45044 & 16102 & -22.115 & -1.276 (5.45) & 2 \\ \hline
        \textit{\ce{HF}} & 1.32 & 17(4,8) & -13.139 & 95807 & 18298 & -13.100 & -0.039 (0.29) & 37040 & 13184 & -13.025 & -0.114 (0.87) & 4 \\
        (STO-6G)& 1.51 & 17(4,8) & -18.224 & 96005 & 18342 & -18.172 & -0.052 (0.28) & 45642 & 16285 & -18.082 & -0.142 (0.78) & 2 \\
                & 1.71$^*$  & 17(4,8) & -25.555 & 96033 & 18348 & -25.439 & -0.116 (0.46) & 38369 & 13587 & -25.830 & 0.275 (-1.07) & 4 \\
                & 1.88 & 17(4,8) & -33.238 & 96012 & 18336 & -33.193 & -0.045 (0.13) & 42147 & 14984 & -33.491 & 0.253 (-0.76) & 3 \\
                & 2.08 & 17(4,8) & -43.947 & 95783 & 18316 & -43.800 & -0.147 (0.34) & 41260 & 14712 & -42.967 & -0.980 (2.23) & 3 \\
    \hline \hline
\end{tabular}
\end{table*}

\begin{center}
\begin{table*}[t]
\caption{\label{table2} The table presents the relevant data associated with our PerturbedHHL and PerturbedHHLite computations for the chosen molecules using $\tilde{\lambda}_{min}$ and $\tilde{\lambda}_{max}$ . The first two columns provide the molecule along with the choice of single-particle basis as well as the information on molecular geometries. $^*$ represents the equilibrium geometry. $N_t(= 2\times n_b + n_r + 1)$ is a shorthand for the total number of qubits, while $E_{corr}^{clc}$ refers to the correlation energy obtained from LCC calculation on a traditional computer. This serves as the benchmark for our computations. `2Q count' is the number of two-qubit gates involved in a computation, and PFD ($ = \frac{E_{corr}^{clc}-E_{corr}^X}{E_{corr}^{clc}} \times 100$) is a percentage fraction difference, where $X \in$ \{PerturbedHHL, PerturbedHHLite\}. $n_f$ is the total number of qubit fixings for a given calculation. The correlation energies are given in units of milliHartree, and are rounded off to five decimal places. }
 \begin{tabular}{ c c c c| c c c c| c c c c c }
    \hline \hline
        Molecule  & Geometry & $N_t$ & $E^{clc}_{corr}$  &\multicolumn{4}{ c |}{PerturbedHHL}  & \multicolumn{5}{ c }{PerturbedHHLite}  \\
        (Basis)& & ($n_b$, $n_r$) &  &  Depth & \quad 2Q \quad & \quad $E_{corr}$ \quad & \quad $E_{diff}$(PFD)\quad  &  Depth  & \quad 2Q \quad & \quad $E_{corr}$ \quad & \quad $E_{diff}$(PFD) \quad & \quad $n_f$ \quad \\ 
    \hline
        \textit{\ce{H2}} & 1.20 & 11(2,6) & -10.207 & 2452  & 536   & -10.001 & -0.206 (2.02) & 884   & 334   & -10.233 & 0.026 (-0.25) & 2 \\
        (6-31G) & 1.30 & 11(2,6) & -10.813 & 2452  & 536   & -10.479 & -0.334 (3.09) & 636   & 213   & -10.795 & -0.018 (0.17) & 4 \\
                & 1.40$^*$ & 11(2,6) & -11.546 & 2452  & 536   & -11.440 & -0.106 (0.92) & 613   & 232   & -11.607 & 0.061 (-0.53) & 3 \\
                & 1.50 & 11(2,6) & -12.437 & 2452  & 536   & -12.238 & -0.199 (1.59) & 1093  & 395   & -12.553 & 0.116 (-0.93) & 1 \\
                & 2.00 & 11(2,6) & -20.053 & 2452  & 536   & -19.924 & -0.129 (0.64) & 604   & 204   & -20.034 & -0.019 (0.09) & 4 \\\hline
        \textit{\ce{H3+}} & 1.727, 50$^{\circ}$ & 11(2,6) & -24.425 & 2452  & 536   & -24.199 & -0.226 (0.92) & 876   & 318   & -24.157 & -0.268 (1.10) & 2 \\
        (STO-6G) & 1.727, 55$^{\circ}$ & 11(2,6) & -25.394 & 2452  & 536   & -25.143 & -0.251 (0.99) & 321   & 115   & -24.848 & -0.546 (2.15) & 6 \\
                 & 1.727, 60$^{\circ *}$ & 11(2,6) & -26.197 & 2452  & 534   & -26.221 & 0.024 (-0.09) & 337   & 120   & -26.233 & 0.036 (-0.14) & 6 \\
                 & 1.727, 70$^{\circ}$ & 11(2,6) & -27.423 & 2452  & 536   & -27.384 & -0.039 (0.14) & 554   & 194   & -27.642 & 0.219 (-0.80) & 4 \\
                 & 1.727, 80$^{\circ}$ & 11(2,6) & -28.315 & 2452  & 536   & -28.354 & 0.039 (-0.14) & 659   & 225   & -28.360 & 0.045 (-0.16) & 4 \\ \hline
        \textit{\ce{LiH}} & 2.60 & 17(4,8) & -17.464 & 97493 & 18608 & -16.650 & -0.814 (4.66) & 42043 & 14934 & -16.561 & -0.903 (5.17) & 3 \\
        (STO-6G) & 2.70 & 17(4,8) & -17.948 & 97672 & 18636 & -16.642 & -1.306 (7.28) & 41804 & 14800 & -17.653 & -0.295 (1.65) & 3 \\
                 & 2.80$^*$ & 17(4,8) & -18.536 & 97837 & 18652 & -17.604 & -0.932 (5.02) & 37825 & 13356 & -17.524 & -1.012 (5.46) & 4 \\
                 & 2.90 & 17(4,8) & -19.227 & 98065 & 18702 & -18.135 & -1.092 (5.68) & 40983 & 14595 & -18.449 & -0.778 (4.05) & 3 \\
                 & 3.00 & 17(4,8) & -20.024 & 97998 & 18706 & -18.610 & -1.414 (7.06) & 34239 & 12011 & -18.786 & -1.238 (6.18) & 5 \\ \hline
        \textit{\ce{BeH+}} & 2.40 & 17(4,8) & -18.963 & 98065 & 18698 & -18.974 & 0.011 (-0.06) & 50140 & 17796 & -19.001 & 0.038 (-0.20) & 1 \\
         (STO-6G) & 2.50 & 17(4,8) & -20.278 & 98680 & 18826 & -20.290 & 0.012 (-0.06) & 45584 & 16167 & -19.570 & -0.708 (3.49) & 2 \\
                  & 2.54$^*$ & 17(4,8) & -20.848 & 97475 & 18586 & -20.832 & -0.016 (0.08) & 45469 & 16167 & -20.412 & -0.436 (2.09) & 2 \\
                  & 2.60 & 17(4,8) & -21.752 & 97624 & 18620 & -21.768 & 0.016 (-0.07) & 44576 & 15915 & -21.753 & 0.001 (-0.01) & 2 \\
                  & 2.70 & 17(4,8) & -23.391 & 98089 & 18684 & -23.179 & -0.212 (0.91) & 45626 & 16208 & -21.765 & -1.626 (6.95) & 2 \\ \hline
        \textit{\ce{HF}} & 1.32 & 17(4,8) & -13.139 & 95807 & 18302 & -12.936 & -0.203 (1.54) & 41809 & 14888 & -13.028 & -0.111 (0.85) & 3 \\
        (STO-6G) & 1.51 & 17(4,8) & -18.224 & 95715 & 18292 & -18.127 & -0.097 (0.53) & 45450 & 16260 & -18.164 & -0.060 (0.33) & 2 \\
                & 1.71$^*$  & 17(4,8) & -25.555 & 95999 & 18332 & -25.437 & -0.118 (0.46) & 38117 & 13527 & -25.820 & 0.265 (-1.04) & 4 \\
                & 1.88 & 17(4,8) & -33.238 & 95417 & 18236 & -33.186 & -0.052 (0.16) & 33201 & 11742 & -33.500 & 0.262 (-0.79) & 5 \\
                & 2.08 & 17(4,8) & -43.947 & 95583 & 18274 & -43.776 & -0.171 (0.39) & 41882 & 14903 & -44.191 & 0.244 (-0.55) & 3 \\
    \hline \hline
\end{tabular}
\end{table*}
\end{center}

\begin{center}
\begin{table*}[t]
\caption{\label{table3} The table presents the relevant data associated with our AdaptHHL and AdaptHHLite computations for the chosen molecules without using any $\lambda_{min}$ and $\lambda_{max}$ information. The first two columns provide the molecule along with the choice of single-particle basis as well as the information on molecular geometries. $^*$ represents the equilibrium geometry. $N_t(= 2\times n_b + n_r + 1)$ is a shorthand for the total number of qubits, while $E_{corr}^{clc}$ refers to the correlation energy obtained from LCC calculation on a traditional computer. This serves as the benchmark for our computations. `2Q count' is the number of two-qubit gates involved in a computation, and PFD ($ = \frac{E_{corr}^{clc}-E_{corr}^X}{E_{corr}^{clc}} \times 100$) is a percentage fraction difference, where $X \in$ \{AdaptHHL, AdaptHHLite\}. $n_f$ is the total number of qubit fixings for a given calculation. The correlation energies are given in units of milliHartree, and are rounded off to five decimal places. }
 \begin{tabular}{ c c c c| c c c c| c c c c c }
    \hline \hline
        Molecule  & Geometry & $N_t$ & $E^{clc}_{corr}$  &\multicolumn{4}{ c |}{AdaptHHL}  & \multicolumn{5}{ c }{AdaptHHLite}  \\
        (Basis)& & ($n_b$, $n_r$) &  &  Depth & \quad 2Q \quad & \quad $E_{corr}$ \quad & \quad $E_{diff}$(PFD)\quad  &  Depth  & \quad 2Q \quad & \quad $E_{corr}$ \quad & \quad $E_{diff}$(PFD) \quad & \quad $n_f$ \quad \\ 
    \hline
        \textit{\ce{H2}} & 1.20 & 11(2,6) & -10.207 & 2452  & 536   & -10.489 & 0.282 (-2.76) & 532   & 183  & -10.742 & 0.535 (-5.24)  & 4 \\
        (6-31G) & 1.30 & 11(2,6) & -10.813 & 2452  & 536   & -10.631 & -0.182 (1.68) & 543   & 185  & -10.884 & 0.071 (-0.66)  & 4 \\
                & 1.40$^*$ & 11(2,6) & -11.546 & 2452  & 536   & -11.081 & -0.465 (4.02) & 537   & 182  & -11.262 & -0.284 (2.46)  & 4 \\
                & 1.50 & 11(2,6) & -12.437 & 2452  & 536   & -12.065 & -0.372 (2.99) & 525   & 183  & -12.124 & -0.313 (2.52)  & 4 \\
                & 2.00 & 11(2,6) & -20.053 & 2452  & 536   & -19.839 & -0.214 (1.07) & 379   & 136  & -19.979 & -0.074 (0.37)  & 5 \\\hline
        \textit{\ce{H3+}} & 1.727, 50$^{\circ}$ & 11(2,6) & -24.425 & 2452  & 536   & -26.274 & 1.849 (-7.57) & 273   & 100  & -27.388 & 2.963 (-12.13) & 6 \\
        (STO-6G) & 1.727, 55$^{\circ}$ & 11(2,6) & -25.394 & 2452  & 536   & -27.120 & 1.726 (-6.80) & 256   & 97   & -27.533 & 2.139 (-8.42)  & 6 \\
                 & 1.727, 60$^{\circ *}$ & 11(2,6) & -26.197 & 2452  & 536   & -27.375 & 1.178 (-4.50) & 270   & 98   & -27.472 & 1.275 (-4.87)  & 6 \\
                 & 1.727, 70$^{\circ}$ & 11(2,6) & -27.423 & 2452  & 536   & -29.809 & 2.386 (-8.70) & 272   & 100  & -31.189 & 3.766 (-13.73) & 6 \\
                 & 1.727, 80$^{\circ}$ & 11(2,6) & -28.315 & 2452  & 536   & -30.016 & 1.701 (-6.01) & 514   & 177  & -30.463 & 2.148 (-7.59)  & 4 \\ \hline
        \textit{\ce{LiH}} & 2.60 & 17(4,8) & -17.464 & 80663 & 16374 & -17.136 & -0.328 (1.87) & 26458 & 9456 & -17.527 & 0.063 (-0.37)  & 6 \\
        (STO-6G) & 2.70 & 17(4,8) & -17.948 & 80779 & 16394 & -17.690 & -0.258 (1.44) & 26441 & 9440 & -18.124 & 0.176 (-0.98)  & 6 \\
                 & 2.80$^*$ & 17(4,8) & -18.536 & 80476 & 16338 & -18.399 & -0.137 (0.74) & 26348 & 9412 & -18.874 & 0.338 (-1.83)  & 6 \\
                 & 2.90 & 17(4,8) & -19.227 & 80543 & 16350 & -19.259 & 0.032 (-0.17) & 26425 & 9429 & -19.771 & 0.544 (-2.83)  & 6 \\
                 & 3.00 & 17(4,8) & -20.024 & 80591 & 16356 & -20.255 & 0.231 (-1.15) & 26574 & 9465 & -20.438 & 0.414 (-2.06)  & 6 \\ \hline
        \textit{\ce{BeH+}} & 2.40 & 17(4,8) & -18.963 & 80827 & 16392 & -19.071 & 0.108 (-0.57) & 26580 & 9440 & -19.224 & 0.261 (-1.37)  & 6 \\
         (STO-6G) & 2.50 & 17(4,8) & -20.278 & 81031 & 16422 & -20.134 & -0.144 (0.71) & 26783 & 9493 & -20.335 & 0.057 (-0.28)  & 6 \\
                  & 2.54$^*$ & 17(4,8) & -20.848 & 80724 & 16372 & -20.591 & -0.257 (1.23) & 26654 & 9456 & -20.814 & -0.034 (0.16)  & 6 \\
                  & 2.60 & 17(4,8) & -21.752 & 80699 & 16368 & -21.344 & -0.408 (1.88) & 26807 & 9482 & -21.613 & -0.139 (0.64)  & 6 \\
                  & 2.70 & 17(4,8) & -23.391 & 80748 & 16374 & -22.698 & -0.693 (2.96) & 26712 & 9468 & -23.048 & -0.343 (1.47)  & 6 \\ \hline
        \textit{\ce{HF}} & 1.32 & 17(4,8) & -13.139 & 80238 & 16292 & -13.308 & 0.169 (-1.29) & 26190 & 9396 & -13.669 & 0.530 (-4.03)  & 6 \\
        (STO-6G) & 1.51 & 17(4,8) & -18.224 & 80098 & 16266 & -19.901 & 1.677 (-9.20) & 26388 & 9427 & -20.276 & 2.052 (-11.26) & 6 \\
                & 1.71$^*$  & 17(4,8) & -25.555 & 80254 & 16292 & -27.744 & 2.189 (-8.57) & 26314 & 9428 & -27.995 & 2.440 (-9.55)  & 6 \\
                & 1.88 & 17(4,8) & -33.238 & 80256 & 16294 & -35.680 & 2.442 (-7.35) & 22589 & 8055 & -37.295 & 4.057 (-12.21) & 7 \\
                & 2.08 & 17(4,8) & -43.947 & 79769 & 16224 & -46.323 & 2.376 (-5.41) & 22326 & 8012 & -47.394 & 3.447 (-7.84)  & 7 \\
    \hline \hline
\end{tabular}
\end{table*}
\end{center}


\begin{figure*}[]
\begin{tabular}{c}
\includegraphics[scale=0.40]{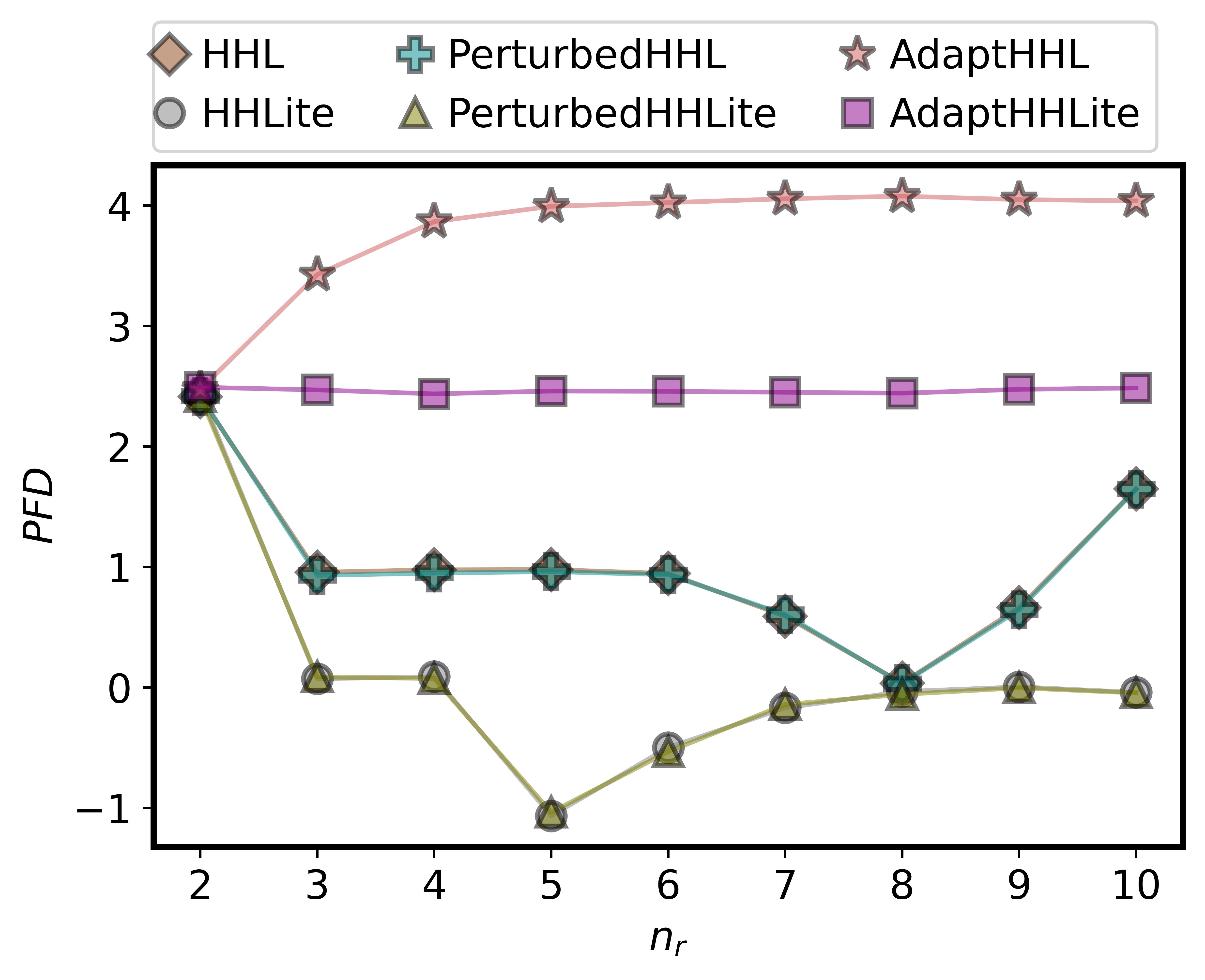}  \\
(a) {$H_2$} \\
\includegraphics[scale=0.40]{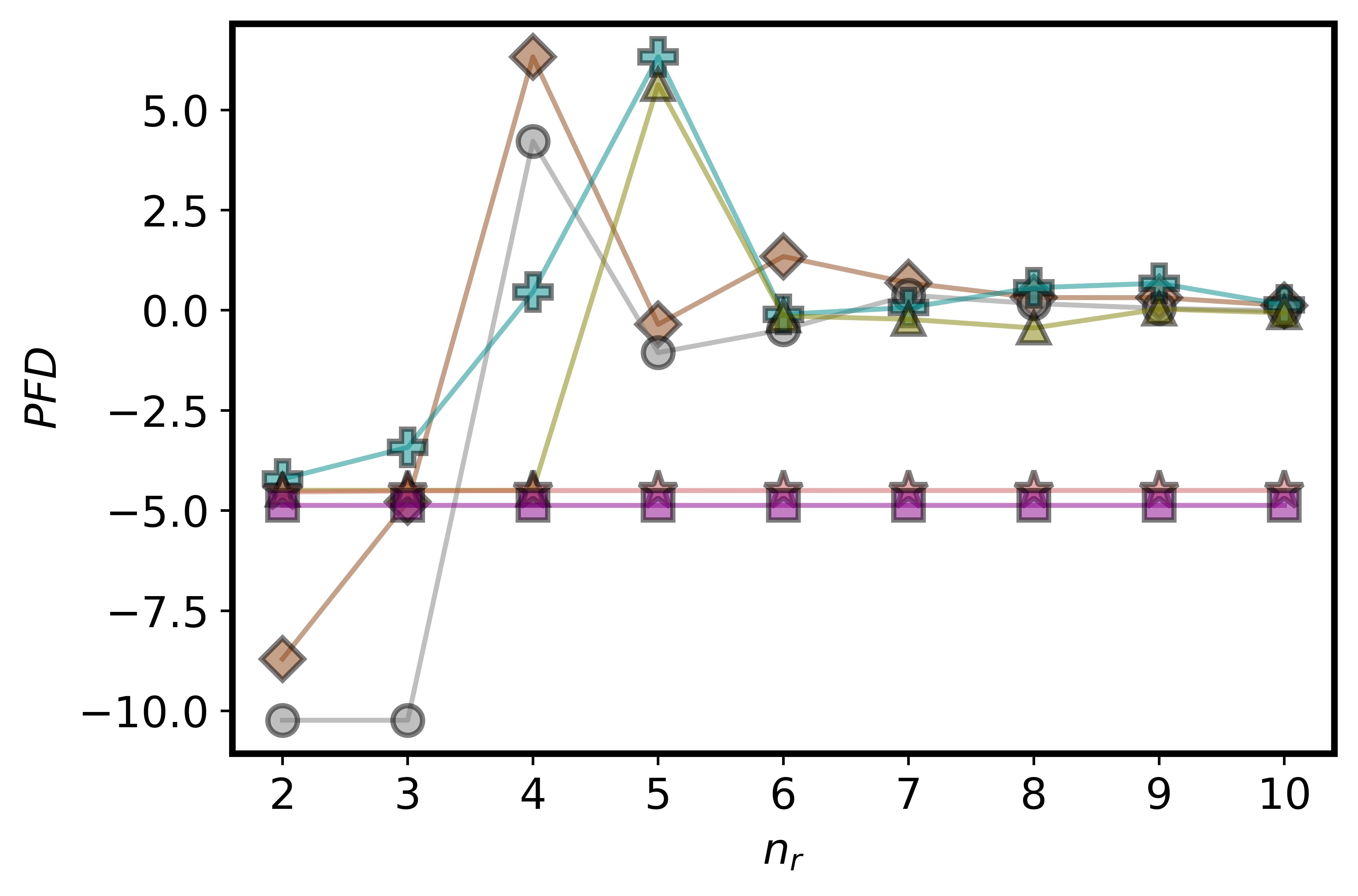}   \\
 (b) {$H_3^+$} \\ 
\includegraphics[scale=0.40]{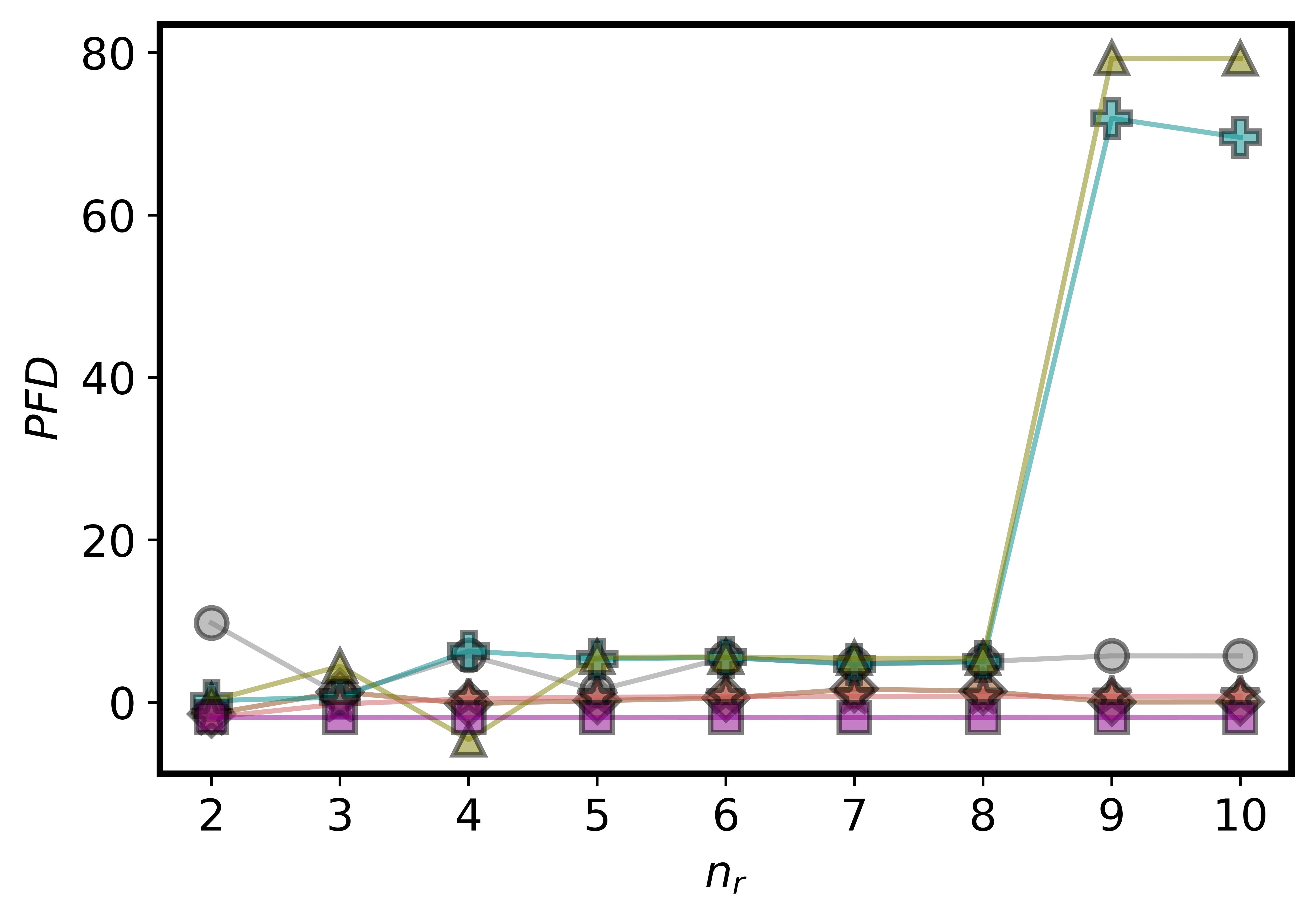} \\
(c) {$LiH$} \\
\includegraphics[scale=0.40]{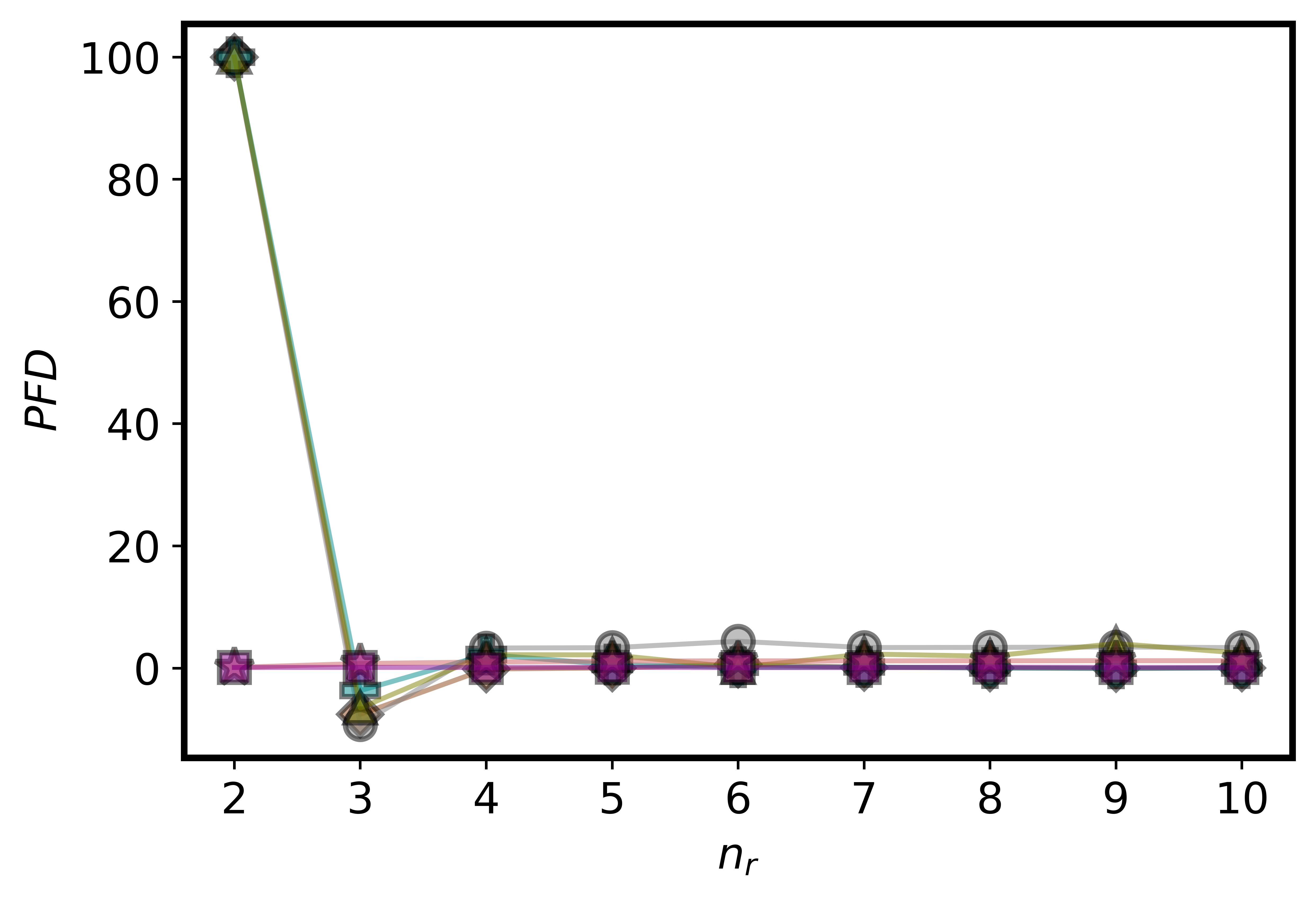} \\
(d){$BeH^+$}\\
\includegraphics[scale=0.40]{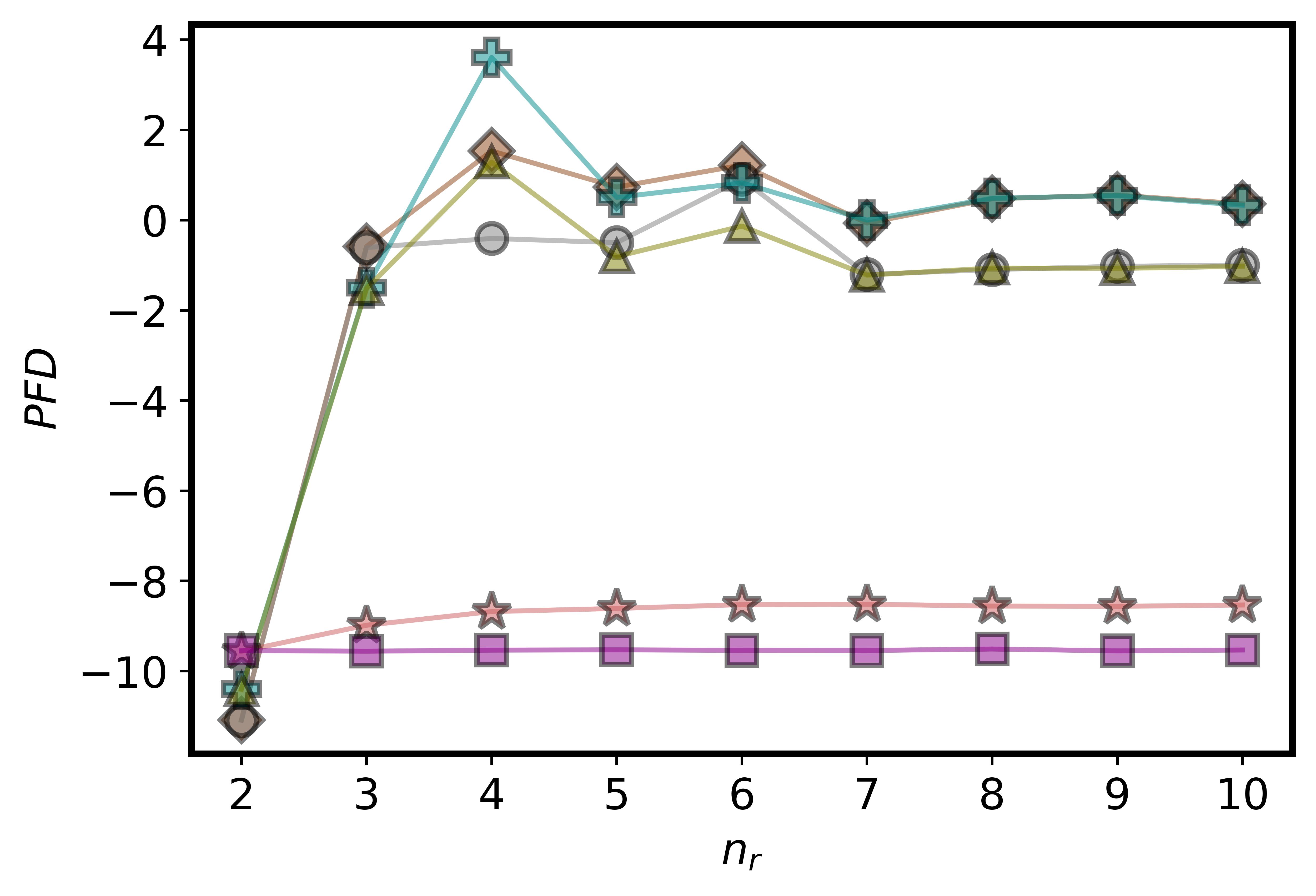} \\
(e) {$HF$}\\
\end{tabular}
\caption{\label{fig:figs1} Figure showing PFD with varying number of ancilla qubits ($n_r$) with $P_{\rm th}$=0.8, for HHL, PerturbedHHL, AdaptHHL, HHLite, PerturbedHHLite and AdaptHHLite for $H_2$, $H_3^+$, $LiH$, $BeH^+$ and $HF$. Each PFD is calculated with respect to the classically obtained LCC value of correlation energy. All of the data correspond to molecules in their equilibrium geometries. } 
\end{figure*}

\begin{figure*}[htbp]
\begin{tabular}{c}
\includegraphics[scale=0.38]{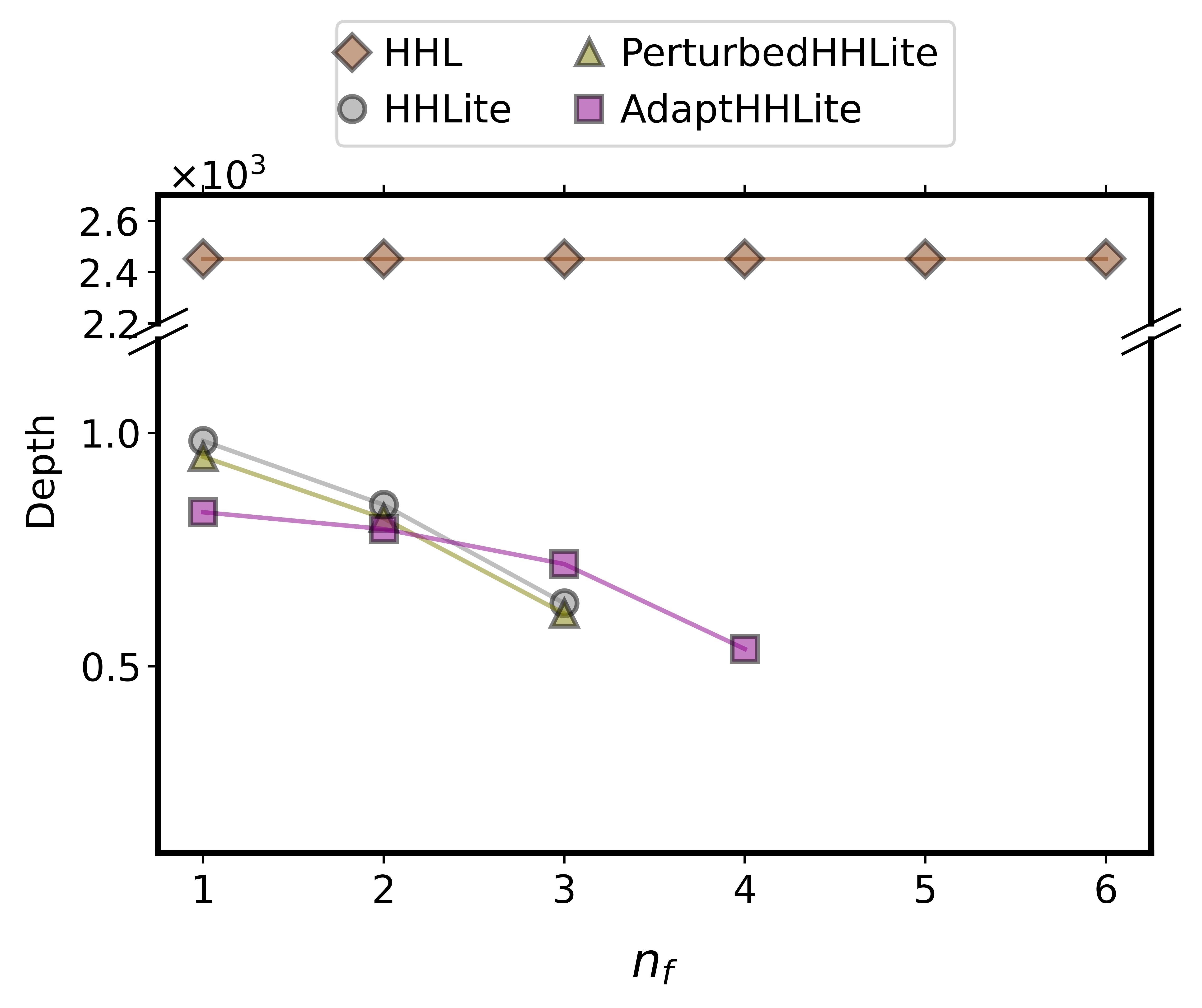} \\
(a) {$H_2$} \\
\includegraphics[scale=0.38]{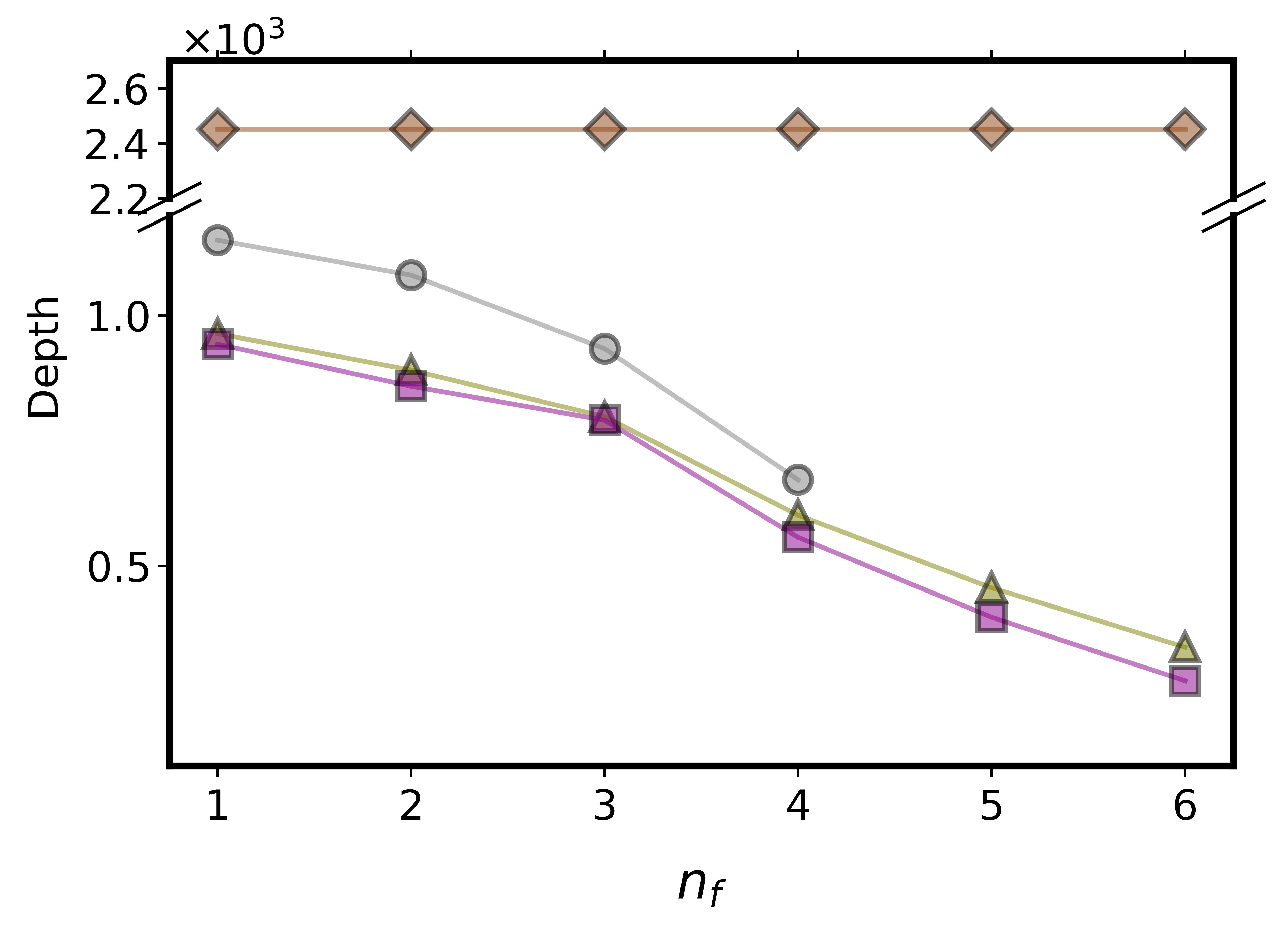} \\
(b) {$H_3^+$} \\

\includegraphics[scale=0.38]{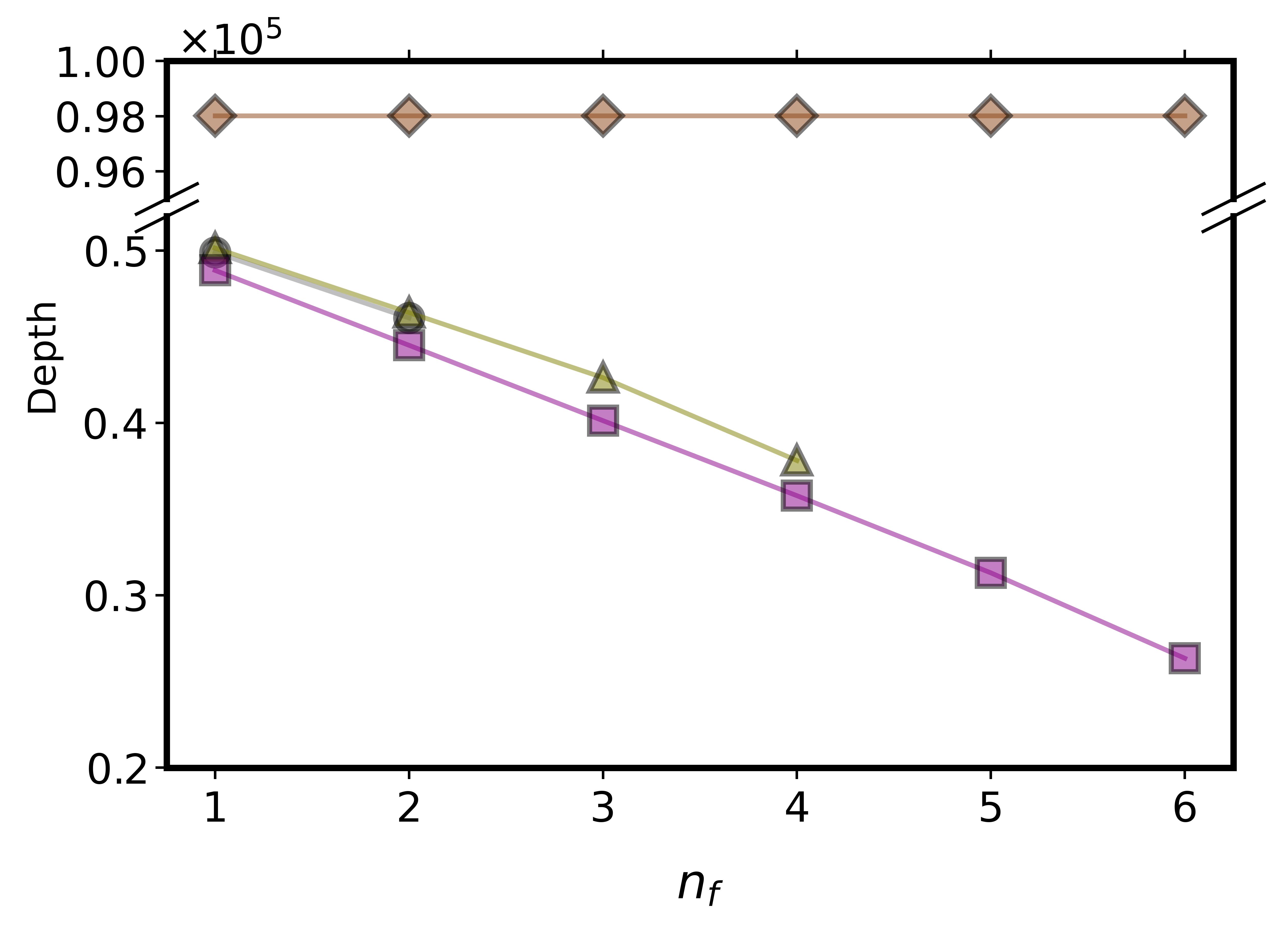} \\
(c) {$LiH$} \\
\includegraphics[scale=0.38]{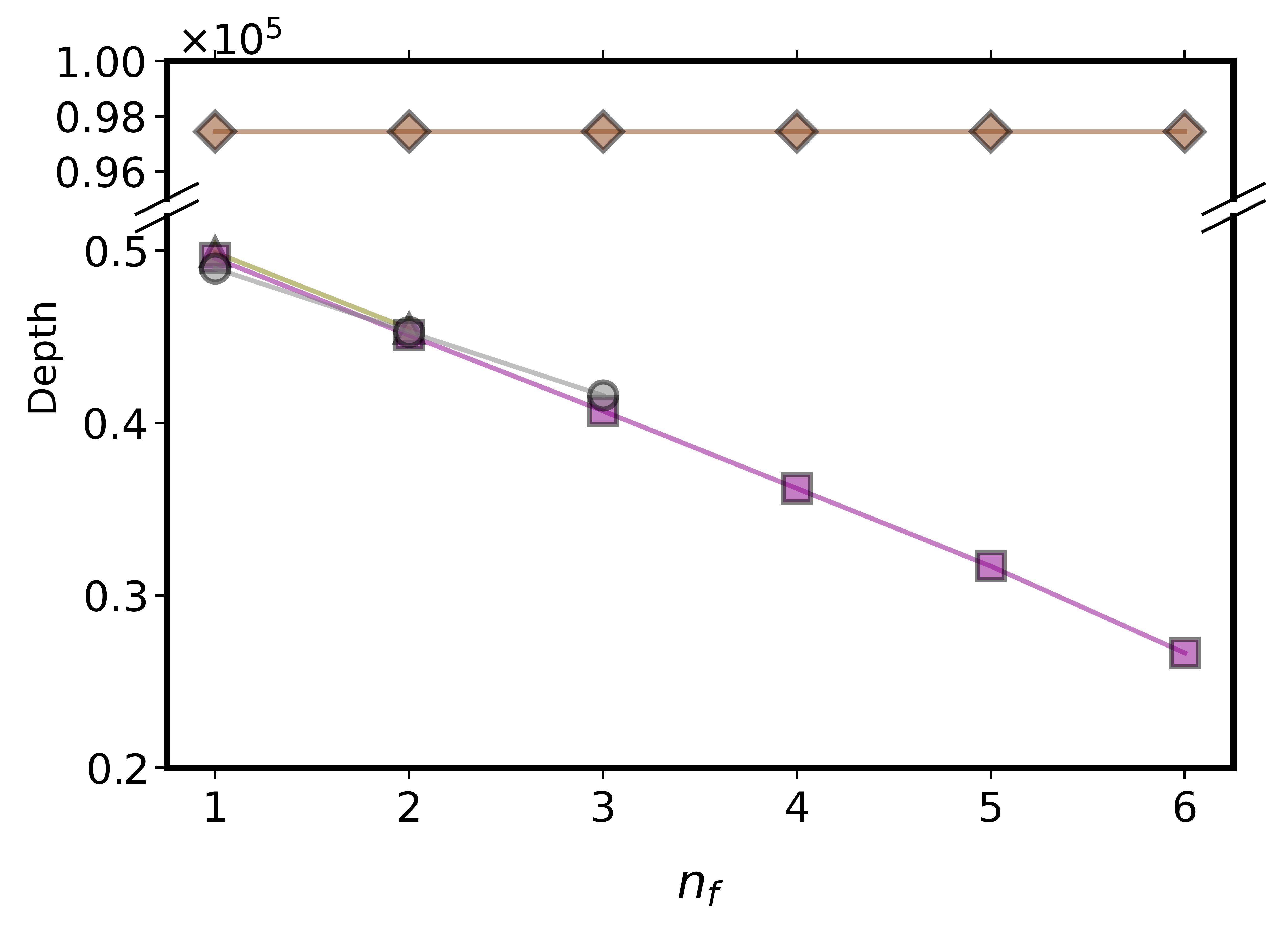} \\
(d){$BeH^+$}\\
\includegraphics[scale=0.38]{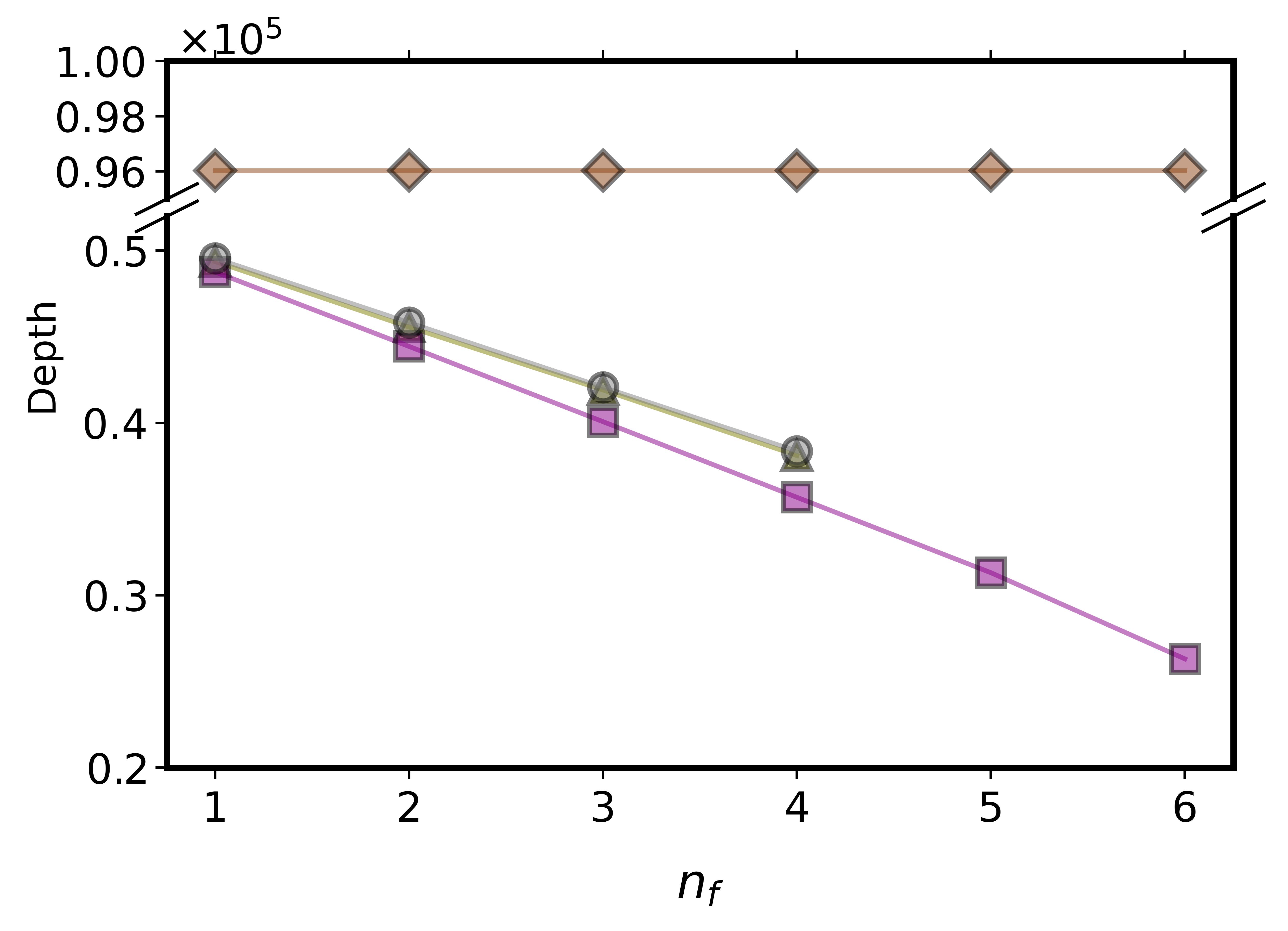} \\
(e) {$HF$}\\
\end{tabular}
\caption{\label{fig:figs2} Depth of the circuit with varying number of fixings ($n_f$) with $P_{\rm th}$=0.8, for HHL, HHLite, PerturbedHHLite and AdaptHHLite for $H_2$, $H_3^+$, $LiH$, $BeH^+$ and $HF$. All of the data is for molecules at their equilibrium geometries. } 
\end{figure*}

\begin{center}
\begin{table*}
\label{HybParam}
\caption{\label{table_statistics} The mean and standard deviation with respect to the mean of $E_{corr}$ of multiple hardware runs for $H_2$ molecule in the STO-6G basis. All units involving correlation energy are in milliHartree.  }
\begin{tabular}{ c c c c }

    \hline \hline
        Internuclear  & $E_{corr}$  & Mean    & Standard Deviation  \\
         distance (Bohr) &  (classical) &  ($E_{corr}$)  &   ($E_{corr}$) \\
    \hline
    1.3 & -18.5920 & -16.991 & 0.652 \\
    1.4 & -20.8738 & -20.145 & 1.393 \\
    1.5 & -23.4209 & -23.463 & 0.610 \\
    1.6 & -26.2622 & -28.928 & 0.768 \\
    1.7 & -29.4289 & -29.183 & 0.883 \\        
    \hline \hline
\end{tabular}
\end{table*}
\end{center}

\end{document}